\documentclass[11pt,a4paper]{article}
\usepackage{epsfig,amsmath,amssymb}
\usepackage{url,color}
\usepackage[english]{babel}
\usepackage{natbib}
\usepackage{authblk}
\usepackage{hyperref}
\usepackage{geometry}
\usepackage{comment}

\def\indep{\perp\kern-0.50em \perp}
\def\var{\mbox{var}}
\def\E{\mbox{E}}

\title{Two-step estimation of latent trait models}

\author[1]{Jouni Kuha}
\affil[1]{Department of Statistics, London School of Economics and
Political Science, London, UK. \texttt{j.kuha@lse.ac.uk}}

\author[2]{Zsuzsa Bakk}
\affil[2]{Department of Methodology and Statistics, Leiden University,
Leiden, The~Netherlands. \texttt{z.bakk@fsw.leidenuniv.nl}}

\begin{document}

%\begin{center}
%{\LARGE{Two-step estimation of latent trait models}}
%\end{center}

%\vspace*{40ex}
%\vspace*{-4ex}

\maketitle

%\vspace*{-4ex}

\begin{abstract}
We consider likelihood-based two-step estimation of latent variable models, in which just
the measurement model is estimated in the first step and the measurement
parameters are then fixed at their estimated values in the second step
where the structural model is estimated. We show how this approach can
be implemented for latent trait models (item response theory models)
where the latent variables are continuous and their measurement
indicators are categorical variables. The properties of two-step
estimators are examined using simulation studies and applied examples.
They perform well, and have attractive practical and conceptual
properties compared to the alternative one-step and three-step
approaches. These results are in line with expectations from theory and
with previous findings for other
families of latent variable models. This provides strong evidence that
two-step estimation is a flexible and useful general method of
estimation for different types of latent variable models.
\end{abstract}

%\vspace*{2ex}
\vspace*{-2ex}

\section{Introduction}

General latent variable models have two main parts: a
\emph{measurement model} which describes how the latent variables that
appear in the model are measured by observed indicators of them, and a
\emph{structural model} which describes the associations among the
latent variables and any observed explanatory and response variables
which are not treated as measurement indicators. For instance, in the
illustrative example considered in Section \ref{s_example} of this paper
the structural model specifies how individuals' extrinsic and intrinsic
work value orientations are associated with characteristics of the
individuals, and the measurement model specifies how these value orientations are
measured by a set of survey questions. Here, as in many applications,
the structural model is the focus of substantive interest, but the
measurement model also needs to be included in order for
the structural model to be estimable.

Estimation of these elements can be organised in different ways. In
joint or \emph{one-step estimation}, both parts of the model are
estimated together. When this is done by maximizing the joint
likelihood, one-step estimates are the maximum likelihood (ML) estimates
of the model parameters, also known in this context as full-information
ML (FIML) estimates.. In contrast, ``stepwise'' approaches divide the
estimation into separate steps. The most familiar of them is
\emph{three-step estimation}. In its first step, the measurement model
is estimated from a specification which omits all or most of the
structural model. In the second step, this estimated measurement model
is used to assign predicted values of the latent variables to the units
of analysis, such as factor scores for continuous latent variables. In
the third step, these scores are used in place of the latent variables
to estimate the structural model. There are ``naive'' and ``adjusted''
versions of three-step estimation, depending on whether it attempts to
account for the measurement error that results from using the predicted
values.

The focus of this paper is on a different stepwise approach,
\emph{two-step estimation} of latent variable models. Its first step is
estimation of the measurement model, as in the three-step method. In the
second step, instead of being used to calculate explicit predictions of
the latent variables, the parameters of the measurement model are simply
fixed at their estimated values. In other words, the second step takes
the same form as one-step estimation, except that all the measurement
parameters are treated as known numbers rather than unknown estimands.
Likelihood-based two-step estimation derives its justification and
properties from the general theory of pseudo maximum likelihood
estimation. Its large-sample properties are very similar to those of
one-step ML estimation.

Two-step estimates avoid the measurement error bias of naive three-step
estimates, and are typically comparable in performance but more
generally usable and practically simpler than adjusted three-step estimates. Compared to
one-step estimation, the two-step approach has in principle attractive
practical and conceptual advantages. Because it is split into two steps,
both of them will be computationally less demanding than the single step
of one-step estimation. It is also easy to estimate the measurement and
structural models using fully or partially different sets of data, which
is useful in some applications. The conceptual advantage arises from
an inherent difference in how the two methods use the available data.
Because the one-step approach fits all parts of the model at once, the
estimated measurement model is informed not only by the measurement
indicators but also by observed covariates and responses in the
structural model. This can cause what \citeauthor{burt73}
(\citeyear{burt73}, \citeyear{burt76}) terms ``interpretational
confounding'', a situation where the implied definition of a latent
variable is partly determined by variables that should be
conceptually separate from it. In other words, if we change the
specificaton of the structural model, this also changes the estimated
measurement model and hence in effect changes the definition of the
latent variables that appear in the structural model. This circularity is
avoided by the two-step approach because it estimates the measurement
model only once and using only the measurement indicators.

Two-step estimation is a general idea, and one of the points that we
want to emphasise is that it can in principle be applied to any
latent variable models. However, there may be differences in performance
of the estimates, ease of implementation, or other points to consider
when it is used for different broad classes of models. In the literature
so far, two-step estimation has accordingly been described separately
for different types of models. It was proposed for latent class
analysis, that is for models where both the latent variables and their
indicators are categorical, by \cite{bandeenrocheetal97},
\cite{xue+bandeen-roche02} and \cite{bakk-kuha}, and has since been
extended to further versions of them (e.g.\ multilevel latent class
analysis in \citealt{dimarietal23}). For structural equation models
(SEMs), where the latent variables and indicators are both continuous,
the idea was introduced already by \citeauthor{burt73}
(\citeyear{burt73}, \citeyear{burt76}), but detailed exploration of it
is much more recent. In particular, \cite{rosseel+loh24} proposed a
general two-step approach for SEMs, with the title of
``structural-after-measurement'' (SAM) estimation. They describe two
(usually equivalent) forms of it: ``local SAM'' which summarises the
first step in the form of the estimated means and covariance matrix of
the variables in the structural model, and ``global'' SAM which is
analogous to the approach that we describe in this paper for
latent trait models. A method for SEMs is also proposed by
\cite{levy23}, using Bayesian (MCMC) estimation for both steps, and
\cite{levy+mcneish25} extended it to the kinds of models with
categorical indicators that we will also consider here. Another closely
related paper is \cite{skrondal+kuha12}, who use two-step estimation to
correct for covariate measurement error in regression models. Two-step
methods for these classes of models are now also being implemented in
general-purpose software for latent variable modelling, in Latent Gold
for latent class analysis \citep{vermunt+magidson21B} and in the R
package \emph{lavaan} for SAM \citep{rosseel12, rosseel+loh24}. A
comprehensive recent review of two-step and other stepwise methods is
given by \cite{vermunt25}.

In this paper we propose and examine two-step estimation for another
family of latent variable models, one where the latent variables are
continuous but their indicators are categorical. These are commonly
known as \emph{Item Response Theory (IRT) models}, and they can also be
placed in a more general context as instances of generalised linear
latent and mixed models (GLLAMMs; \citealt{skrondal+rabehesketh04}) or
as structural models combined with general linear latent variable
measurement models (GLLVMs; \citealt{bartholomewetal11}). We refer to
them as \emph{latent trait models}. They resemble SEMs in having
continuous latent variables, and standard latent class models in having
categorical indicators. Like SEMs, they are often used with complex
structural models which include multiple latent variables, whereas
latent class analysis more often includes only one latent class
variable. Computationally, latent trait models are typically the most
demanding of these families, at least for ML estimation,
because their log likelihood cannot be expressed in a closed form.

We describe the theory and implementation of two-step estimation of
latent trait models. The presentation of this broadly parallels that of
\cite{bakk-kuha} for latent class models, but with the changes that are
needed now that the latent variables are continuous. We then use
simulation studies and an application example to explore the properties
of the method in this context. The focus of comparisons is between
two-step and one-step estimates, because generally applicable adjusted
three-step estimates are not available for latent trait models. One
question of interest is then whether the relative performance of the
two-step method is similar to what it is for previously considered types
of models. The conclusion is that it is, with some differences of
emphasis (these findings are discussed in more detail in Sections
\ref{s_simulation}--\ref{s_discussion}). Two-step estimates of the
structural model again perform very well. In complex models their
computational advantage over one-step estimates is even larger than for
latent class models and SEMs. On the other hand, interpretational
confounding of one-step estimates may be less worrying here than for
latent class models, because shifts in interpretation tend to be less
abrupt for continuous latent variables than for categorical ones.

The methods of estimation that we consider throughout are
likelihood-based. The starting point is thus that one-step FIML
estimation would be considered appropriate, but we look for the kinds of
further advantages that stepwise modifications of it can offer. The
merits of likelihood-based estimation itself are not at issue here. Such
a focus is standard when these models are approached from an IRT
perspective or as general statistical models with latent variables.
Other methods of estimation are also available. In particular, in
structural equation modelling literature, where these models would be
seen as SEMs with categorical indicators, it is also common to use
limited-information methods such as different forms of weighted least
squares estimation. This choice is distinct from that of the number of
steps in the procedure. In other words, both one-step and two-step
methods can also be combined with different methods of estimation, with
appropriate modifications to their implementation and theoretical
justification. We discuss this topic briefly further in Section
\ref{ss_estimation_LS}.

The latent trait models that we consider are introduced in Section
\ref{s_models}. The definition, implementation and properties of
two-step estimation for them are described in Section~\ref{s_estimation}.
Simulation studies are reported in Section \ref{s_simulation} and the
real-data example in Section \ref{s_example}, and concluding discussion
is given in Section \ref{s_discussion}. Extended tables of some of the results
from the simulations and data analysis are given in a supplementary
appendix. Computer code for the estimation in R and Mplus (both of
which are needed) and replication code for the applied example are also
provided as supplementary materials.

\section{Models and variables}
\label{s_models}

To begin with a general formulation of a latent variable model, let
$\boldsymbol{\eta}_{i}$ be a vector of latent variables, and
$\mathbf{Y}_{i}$ and $\mathbf{X}_{i}$ vectors of observed variables, for
a unit of analysis~$i$. Consider a model of the form
$p(\mathbf{Y}_{i},\boldsymbol{\eta}_{i}| \mathbf{X}_{i};
\boldsymbol{\theta}) = p(\mathbf{Y}_{i}|\boldsymbol{\eta}_{i},
\mathbf{X}_{i}; \boldsymbol{\theta}_{1})\,
p(\boldsymbol{\eta}_{i}|\mathbf{X}_{i}; \boldsymbol{\theta}_{2})$, where
$p(\cdot|\cdot)$ denotes a conditional distribution and
$\boldsymbol{\theta}=(\boldsymbol{\theta}_{1}',\boldsymbol{\theta}_{2}')'$
are parameters. Here $p(\mathbf{Y}_{i}|\boldsymbol{\eta}_{i},
\mathbf{X}_{i}; \boldsymbol{\theta}_{1})$ is the \emph{measurement
model} for $\boldsymbol{\eta}_{i}$ and
$p(\boldsymbol{\eta}_{i}|\mathbf{X}_{i}; \boldsymbol{\theta}_{2})$ the
\emph{structural model}. The observed variables $\mathbf{Y}_{i}$ (the
measurement \emph{items}) serve as measures of the latent
$\boldsymbol{\eta}_{i}$, and $\mathbf{X}_{i}$ are exogenous observed
explanatory variables. Endogenous observed variables can also be
included, by representing them as elements of $\boldsymbol{\eta}_{i}$
which are perfectly measured by single items in $\mathbf{Y}_{i}$. The
parameters in $\boldsymbol{\theta}_{1}$ determine the measurement model,
and we refer to them as the \emph{measurement parameters}, and
$\boldsymbol{\theta}_{2}$ (\emph{structural parameters}) determine the
structural model. It is assumed that $\boldsymbol{\theta}_{1}$ and
$\boldsymbol{\theta}_{2}$ are distinct, in the sense that the joint
parameter space for $\boldsymbol{\theta}$ is the product of the spaces
for $\boldsymbol{\theta}_{1}$ and $\boldsymbol{\theta}_{2}$.

In two-step estimation, $\boldsymbol{\theta}_{1}$ are estimated first
and then fixed at their estimated values in the second step where
$\boldsymbol{\theta}_{2}$ are estimated. This approach can be used for
any instance of this general model. For specificity, however, in most of
this paper we focus on its use for a particular class of latent variable
models. Here the items in $\mathbf{Y}_{i}$ are categorical variables,
$\boldsymbol{\eta}_{i}$ are continuous and normally distributed, and the
measurement model is of a certain common form. This specification is
defined in this section. All of its elements can be further relaxed, as
discussed in Section~\ref{ss_estimation_extensions} below.

Consider first the structural model
$p(\boldsymbol{\eta}_{i}|\mathbf{X}_{i};\boldsymbol{\theta}_{2})$. A
general expression of it for the class of models that we focus on is
\begin{equation}
\boldsymbol{\eta}_{i} = \mathbf{B}_{0} +
\mathbf{B}_{\eta}\boldsymbol{\eta}_{i}
+\mathbf{B}_{x}\mathbf{X}_{i}+\boldsymbol{\zeta}_{i},
\label{structural_general}
\end{equation}
where $\mathbf{B}_{0}$, $\mathbf{B}_{\eta}$ and $\mathbf{B}_{x}$ are
matrices of parameters, and $\boldsymbol{\zeta}_{i}\sim
N(\mathbf{0},\boldsymbol{\Psi})$ are normally distributed model
residuals. This is a variant of a formulation that is commonly used to
specify structural equation models in a compact form. For our purposes,
however, it is more convenient to write the model in a different way, in
terms of a chain of conditional distributions. Suppose that
$\boldsymbol{\eta}_{i}=(\eta_{i1},\dots,\eta_{iK})'$ includes $K\ge 1$
variables, and that the model specification further groups them into
$1\le M\le K$ blocks $\boldsymbol{\eta}_{i(m)}$, so that we can also
write $\boldsymbol{\eta}_{i}=
(\boldsymbol{\eta}_{i(1)}',\dots,\boldsymbol{\eta}_{i(M)}')'$ in this
pre-specified order.
We assume that
(\ref{structural_general}) is recursive, i.e.\ that $\mathbf{B}_{\eta}$
is lower triangular and~$\boldsymbol{\Psi}$ is block diagonal in the
order of the blocks in $\boldsymbol{\eta}_{i}$.
The structural model can then be written as
\begin{eqnarray}
%\lefteqn{
p(\boldsymbol{\eta}_{i}|\mathbf{X}_{i};\boldsymbol{\theta}_{2})
&=&
p(\boldsymbol{\eta}_{i(1)}|\mathbf{X}_{i};\boldsymbol{\theta}_{21})\,
p(\boldsymbol{\eta}_{i(2)}|\mathbf{X}_{i},\boldsymbol{\eta}_{i(1)};\boldsymbol{\theta}_{22})
%}
\nonumber
\\
\hspace*{5em}
&
%\hspace*{-1em}
%\hspace*{-3em}
&
%\hspace*{-1em}
\hspace*{1em}\times
\cdots
\times
p(\boldsymbol{\eta}_{i(M)}|\mathbf{X}_{i},\boldsymbol{\eta}_{i(1)},
\dots,\boldsymbol{\eta}_{i(M-1)};\boldsymbol{\theta}_{2M})
\label{structural_model}
\end{eqnarray}
where the structural parameters are partitioned correspondingly as
$\boldsymbol{\theta}_{2}=(\boldsymbol{\theta}_{21}',\dots,\boldsymbol{\theta}_{2M}')'$.
The conditional
distributions are all
(multivariate or univariate) normal and specified by
the linear models
\begin{eqnarray}
p(\boldsymbol{\eta}_{i(1)}|\mathbf{X}_{i};\boldsymbol{\theta}_{21})
&\sim& N(
\boldsymbol{\beta}_{10}
+\boldsymbol{\beta}_{1x}\mathbf{X}_{i}, \, \boldsymbol{\Psi}_{1}
)
\label{eta_model1}
\\
p(\boldsymbol{\eta}_{i(m)}
|\mathbf{X}_{i},\boldsymbol{\eta}_{i(1)},\dots,\boldsymbol{\eta}_{i(m-1)};\boldsymbol{\theta}_{2m})
&\sim&\hspace*{-.5em} N(
\boldsymbol{\beta}_{m0}
+\boldsymbol{\beta}_{mx}\mathbf{X}_{i}
+\sum_{l=1}^{m-1}\boldsymbol{\beta}_{ml}\boldsymbol{\eta}_{i(l)}, \,
\boldsymbol{\Psi}_{m}
)
%\hspace*{.2em}\text{ for }\hspace*{.2em} m=2,\dots,M,
\label{eta_model2}
\end{eqnarray}
for $m=2,\dots,M$. Each of
$\boldsymbol{\theta}_{21},\dots,\boldsymbol{\theta}_{2M}$ thus consists of
the corresponding $\boldsymbol{\beta}$ and $\boldsymbol{\Psi}$
parameters. Constraints on elements of $\boldsymbol{\theta}_{2}$ can be
included, most often zero constraints on the $\boldsymbol{\beta}$
parameters which  omit some of the explanatory variables from some parts
of the model and thus yield different non-saturated choices for the
overall structural model~(\ref{structural_model}). If observed
covariates $\mathbf{X}_{i}$ are not present, they are omitted from
(\ref{structural_model})--(\ref{eta_model2}). If any
$\boldsymbol{\eta}_{i(m)}$ are actually observed (rather than latent)
variables, it would also be straightfoward to replace the linear model with
some other model for them. For example, if $\boldsymbol{\eta}_{i(M)}$ is
a single binary observed variable, we could specify a binary logistic
model for it given $\mathbf{X}_{i}$ and
$\boldsymbol{\eta}_{i(1)},\dots,\boldsymbol{\eta}_{i(M-1)}$.

Consider then the measurement model. We
assume that
$p(\mathbf{Y}_{i}|\boldsymbol{\eta}_{i}, \mathbf{X}_{i};
\boldsymbol{\theta}_{1}) =p(\mathbf{Y}_{i}|\boldsymbol{\eta}_{i};
\boldsymbol{\theta}_{1})$, so that there
is no non-equivalence of measurement (differential item functioning) with
respect to observed covariates $\mathbf{X}_{i}$.
The overall model can then be written as
\begin{equation}
p(\mathbf{Y}_{i},\boldsymbol{\eta}_{i}|\mathbf{X}_{i}; \boldsymbol{\theta}) =
p(\mathbf{Y}_{i}|\boldsymbol{\eta}_{i};
\boldsymbol{\theta}_{1})\,
p(\boldsymbol{\eta}_{i}|\mathbf{X}_{i}; \boldsymbol{\theta}_{2}),
\label{general_model1}
\end{equation}
which also implies a model for the observed variables as
\begin{equation}
p(\mathbf{Y}_{i}|\mathbf{X}_{i}; \boldsymbol{\theta}) =
\int \, p(\mathbf{Y}_{i}|\boldsymbol{\eta}_{i};
\boldsymbol{\theta}_{1})\,
p(\boldsymbol{\eta}_{i}|\mathbf{X}_{i}; \boldsymbol{\theta}_{2})\,
d\boldsymbol{\eta}_{i}.
\label{general_model2}
\end{equation}
We make the common assumption that the items in
$\mathbf{Y}_{i}$ are conditionally independent of each other given
$\boldsymbol{\eta}_{i}$.
Typically each latent
variable in $\boldsymbol{\eta}_{i}$ is measured by only a subset of the
items in $\mathbf{Y}_{i}$, and the simplest structure is obtained when each
item measures exactly one variable. In that case
$\mathbf{Y}_{i}=(\mathbf{Y}_{i1}',\dots,\mathbf{Y}_{iK}')'$ can be
partitioned corresponding to
$\boldsymbol{\eta}_{i}=(\eta_{i1},\dots,\eta_{iK})'$, so that only items
in $\mathbf{Y}_{ik}= (Y_{ik1},\dots,Y_{ikp_{k}})'$ are measures of
$\eta_{ik}$, for each $k=1,\dots,K$. The measurement model can then be
written as
\begin{equation}
p(\mathbf{Y}_{i}|\boldsymbol{\eta}_{i};
\boldsymbol{\theta}_{1}) = \prod_{k=1}^{K} \,
p(\mathbf{Y}_{ik}|\eta_{ik}; \boldsymbol{\theta}_{1k})
=\prod_{k=1}^{K}\prod_{j=1}^{p_{k}} \, p(Y_{ikj}|\eta_{ik};
\boldsymbol{\theta}_{1k})
\label{measurement_model1}
\end{equation}
so that $\boldsymbol{\theta}_{1} =
(\boldsymbol{\theta}_{11}',\dots,\boldsymbol{\theta}_{1K}')'$, with the
different $\boldsymbol{\theta}_{1k}$ taken to be distinct from each
other. We consider measurement models of this form in most of our
specific examples for simplicity, but two-step estimation is not limited
to this case (in the simulation studies Section~\ref{s_simulation} we
also include models where some items measure more than one latent trait).

Specification of the measurement models for individual items $Y_{ijk}$
depends on the forms of the items. We consider situations where each item is a
categorical variable with a finite number of possible values. In most of
our specific examples the items
are binary. Coding their values as 0 and 1, in that case we specify the
 model for each item in (\ref{measurement_model1}) as the
logit model
\begin{equation}
\text{logit}[P(Y_{ikj}=1|\eta_{ik};
\boldsymbol{\theta}_{1k})] = \tau_{kj}+\lambda_{kj}\eta_{ik}
\label{logit}
\end{equation}
so that $\boldsymbol{\theta}_{1k}= (\tau_{k1},\dots,\tau_{kp_{k}},
\lambda_{k1},\dots, \lambda_{kp_{k}})'$. In the language of IRT
modelling, this is a two-parameter logistic (2-PL)
model. There can be parameter constraints within
$\boldsymbol{\theta}_{1k}$, such as taking all $\lambda_{kj}$ for the
same $k$ to be equal. Other model specifications will be used if an item
has $L>2$ categories. If they are regarded as ordered, the binary model
can be replaced by, for example, the
ordinal logistic model $\text{logit}[P(Y_{ikj}\le l|\eta_{ik};
\boldsymbol{\theta}_{1k})] = \tau_{kjl}-\lambda_{kj}\eta_{ik}$
for categories $l=1,\dots,L-1$. For items with unordered categories, we
can use the multinomial logistic model
$\log[
P(Y_{ikj}=l|\eta_{ik};\boldsymbol{\theta}_{1k})/
P(Y_{ikj}=1|\eta_{ik};\boldsymbol{\theta}_{1k})
]= \tau_{kjl}+\lambda_{kjl}\eta_{ik}$ for $l=2,\dots,L$.

%If these are all taken to be normally distributed continuous
%variables, $p(Y_{ikj}|\eta_{ik};\boldsymbol{\theta}_{1k})$ can be
%specified as linear (factor analysis) models. The joint model
%(\ref{general_model1}) is then a linear structural equation model with
%continuous items, for which two-step estimation has been described by
%\cite{rosseel+loh22}.

\section{Two-step estimation of the structural model}
\label{s_estimation}

\subsection{Two-step point estimates}
\label{ss_estimation_twostep}

Suppose that we observe $(\mathbf{Y}_{i}, \mathbf{X}_{i})$
for units $i=1,\dots,n$, assumed to be independent of each other.
We use likelihood-based estimation in a parametric framework.
The log-likelihood function for the class of models models
that we consider is
\begin{eqnarray}
\lefteqn{\ell(\boldsymbol{\theta}) =
\sum_{i=1}^{n}
\, \log p(\mathbf{Y}_{i}|\mathbf{X}_{i};
\boldsymbol{\theta})
=
\sum_{i=1}^{n}\, \log \,
\int p(\mathbf{Y}_{i}|\boldsymbol{\eta}_{i};
\boldsymbol{\theta}_{1})\,
p(\boldsymbol{\eta}_{i}|\mathbf{X}_{i};
\boldsymbol{\theta}_{2}) \, d\boldsymbol{\eta}_{i}}
\nonumber \\
&=&
\sum_{i=1}^{n}\, \log \int \,
\left[
\prod_{k=1}^{K}
\prod_{j=1}^{p_{k}}
p(Y_{ikj}|\eta_{ik}; \boldsymbol{\theta}_{1k})
\right] \,
p(\boldsymbol{\eta}_{i}|\mathbf{X}_{i};
\boldsymbol{\theta}_{2}) \, d\boldsymbol{\eta}_{i}
\label{loglik1}
\end{eqnarray}
where the structural and measurement models are specified as
described in Section \ref{s_models}.
The measurement model on the second line of (\ref{loglik1})
is for the
simple measurement structure where
each item measures exactly one latent variable $\eta_{ik}$, and it would
be modified appropriately if this was not the case.
Maximizing (\ref{loglik1}) with respect to
$\boldsymbol{\theta}=(\boldsymbol{\theta}_{1}',\boldsymbol{\theta}_{2}')'$
would give the overall (full-information) maximum likelihood estimates of
$\boldsymbol{\theta}$, i.e.\ the one-step estimates.

In step 1 of two-step estimation, the measurement parameters
$\boldsymbol{\theta}_{1}$ are estimated. This can be done using any
specification that allows consistent estimation of
$\boldsymbol{\theta}_{1}$. The default would be to use the simplest
model with that property.
When the  measurement structure is simple, this means that the step-1
estimation can be done separately for each
$\eta_{ik}$, using
the log-likelihoods
\begin{equation}
\ell(\boldsymbol{\psi}_{k})=
\sum_{i=1}^{n}\, \log
\int
\left[ \prod_{j=1}^{p_{k}} \,
p(Y_{ikj}|\eta_{ik}; \boldsymbol{\theta}_{1k})
\right]
\,
p(\eta_{ik};
\mu_{k}, \sigma^{2}_{k})\,
d\eta_{ik}
\label{step1loglik}
\end{equation}
for $k=1,\dots,K$, where $p(\eta_{ik}; \mu_{k},\sigma^{2}_{k})\sim
N(\mu_{k},\sigma^{2}_{k})$
and $\boldsymbol{\psi}_{k}=(\boldsymbol{\theta}_{1k}', \mu_{k},
\sigma^{2}_{k})'$. Maximizing (\ref{step1loglik}) gives
estimate $\tilde{\boldsymbol{\psi}}_{k}$, from which
$\tilde{\boldsymbol{\theta}}_{1k}$ are
the step-1 estimates of $\boldsymbol{\theta}_{1k}$ and
$\tilde{\mu}_{k}$ and $\tilde{\sigma}_{k}^{2}$ are discarded.

In step 2 of two-step estimation, the parameters
$\boldsymbol{\theta}_{2}$ of the structural model are then estimated,
holding $\boldsymbol{\theta}_{1}$ fixed at their estimates
$\tilde{\boldsymbol{\theta}}_{1}$ from step 1. Here the log-likelihood
is $\ell(\tilde{\boldsymbol{\theta}}_{1},\boldsymbol{\theta}_{2})$,
which is of the same form as the one-step log-likelihood
$\ell(\boldsymbol{\theta})=\ell(\boldsymbol{\theta}_{1},\boldsymbol{\theta}_{2})$
in (\ref{loglik1}) but with the fixed values
$\tilde{\boldsymbol{\theta}}_{1}$ substituted for
$\boldsymbol{\theta}_{1}$. Maximizing this with respect to
$\boldsymbol{\theta}_{2}$ gives the two-step estimate of the parameters
of the structural model, which we denote by
$\tilde{\boldsymbol{\theta}}_{2}$.

The same step-1 estimate $\tilde{\boldsymbol{\theta}}_{1}$ can be used
for any structural models for the same $\boldsymbol{\eta}_{i}$ in step
2, for example if we want to compare models with different choices of
the covariates $\mathbf{X}_{i}$ or with some associations omitted in the
structural model. Note also that for simplicity the notation
in~(\ref{loglik1}) and (\ref{step1loglik}) assumes that
$\ell(\boldsymbol{\theta})$ and all $\ell(\boldsymbol{\psi}_{k})$ use
the same set of $n$ observations. This is not necessary in general.
Two-step estimation also allows partially or completely different sets
of data to be used for estimating $\tilde{\boldsymbol{\theta}}_{1}$ in
step 1 and $\tilde{\boldsymbol{\theta}}_{2}$ in step 2, or for
estimating $\tilde{\boldsymbol{\theta}}_{1k}$ for different $k$ in step
1, as long as we can assume that the same measurement model holds in all
of them.

Compared with latent class models (as discussed in \citealt{bakk-kuha}),
there are some additional considerations here where the latent variables
$\boldsymbol{\eta}_{i}$ are continuous. First,
some parameter constraints are needed
to fix the scales of $\boldsymbol{\eta}_{i}$ and identify the
parameters $\boldsymbol{\theta}$. In two-step estimation they are
imposed in the step 1, and the scale implied by them then carries
over to the step 2 where no further constraints are required. We
identify the scale of each $\eta_{ik}$ by fixing parameters in the
measurement model, for example $\tau_{kj}=0$ and $\lambda_{kj}=1$ for
one $j$ in the logistic model in (\ref{logit}). This anchors the latent
scale in the items $\mathbf{Y}_{i}$, leaving the distribution of
$\boldsymbol{\eta}_{i}$ free. An alternative first-step constraint would
be to set $\mu_{k}=0$ and $\sigma^{2}_{k}=1$ for all $k$, implying that
each latent variable is on a scale where its marginal distribution  is
standard normal.

Another difference to latent class models is that in
applications of latent trait models it is more common for the latent
$\boldsymbol{\eta}_{i}$ to be multivariate. The measurement model can
then be correspondingly more complex. When it has simple
structure, we recomment that step-1 estimation is carried out for
the single-trait models (\ref{step1loglik}) for each latent trait
separately. In other cases this is generalised only as far as is needed.
For example, suppose that the only deviation from a simple measurement
structure is that some items measure both $\eta_{i1}$ and $\eta_{i2}$.
Then the measurement parameters for them would be estimated from a
two-trait model with a bivariate normal distribution for
($\eta_{i1},\eta_{i2}$), and the rest of $\boldsymbol{\theta}_{1}$ from
one-trait models for any other traits. An alternative would be to
estimate the whole measurement model at once, from a $K$-trait model
for a $K$-variate normal $\boldsymbol{\eta}_{i}$. In other words, this
would be obtained by omitting just $\mathbf{X}_{i}$ from the structural
model, and ignoring any parameter constraints in it. This, however,
would make the first step more complex, with no clear benefit.
\cite{rosseel+loh24} give a good discussion of the considerations of how
to organise step 1 in this respect. They too recommend estimating the
measurement model of each distinct block of latent variables separately
in most cases.

In practice some of the observed variables $\mathbf{X}_{i}$ and
$\mathbf{Y}_{i}$ may be missing. Any missing items in $\mathbf{Y}_{i}$
are taken to be missing at random (MAR), and their contributions are
simply omitted from (\ref{loglik1}) and~(\ref{step1loglik}). In
(\ref{loglik1}) we take $\mathbf{X}_{i}$ to be complete, implying that
units with any missing values in $\mathbf{X}_{i}$ are omitted from
estimation of the structural model. This gives valid estimates for the
model as long as the missingness in $\mathbf{X}_{i}$ does not depend on
the response variables~$\boldsymbol{\eta}_{i}$ (or on~$\mathbf{Y}_{i}$).
Multiple imputation could be used to include also incomplete
observations of $\mathbf{X}_{i}$. These assumptions are conventional,
and their implications are mostly the same for both two-step and
one-step estimation. There are, however, some small differences which
favour the two-step approach here. First, if the MAR assumption fails
only for some items, say those in $\mathbf{Y}_{il}$, this affects only
the estimates of the measurement model they contribute to, i.e.\ only
$\tilde{\boldsymbol{\theta}}_{1l}$. Second, even if we omit some units
with incomplete data on $\mathbf{X}_{i}$ in step 2, we can still use
them in step~1 where $\mathbf{X}_{i}$ are not involved, hence reducing
the loss of information from missing data.

\subsection{Asymptotic properties and variance estimation}
\label{ss_estimation_variance}

The properties of the step-2 estimates $\tilde{\boldsymbol{\theta}}_{2}$
follow from the general theory of pseudo-maximum likelihood (PML)
estimation \citep{gong+samaniego81}. Much of the presentation here is
similar to that of \cite{bakk-kuha} for latent class models, because the
general form of these results is the same irrespective of the details of
the model specification.

PML refers to the general approach where one set of parameters of a
model (here $\boldsymbol{\theta}_{1}$) are estimated first, and their
estimates are then treated as known in a log likelihood (here
$\ell(\tilde{\boldsymbol{\theta}}_{1},\boldsymbol{\theta}_{2})$) that is maximized to estimate a
second set of parameters ($\boldsymbol{\theta}_{2}$).
PML estimators are consistent and asymptotically normally distributed
under very general regularity conditions. In our situation these
require, broadly, that the joint model is such that the one-step ML
estimator $\hat{\boldsymbol{\theta}}$ is consistent for
$\boldsymbol{\theta}$, that the values of $\boldsymbol{\theta}_{1}$ and
$\boldsymbol{\theta}_{2}$ can vary independently of each other, and that
the step-1 estimator $\tilde{\boldsymbol{\theta}}_{1}$ is consistent for
$\boldsymbol{\theta}_{1}$. Here these conditions are satisfied, with the
partial exception of one approximation which is discussed further at the end of
this section.

To get the variance matrix of $\tilde{\boldsymbol{\theta}}_{2}$,
let the Fisher information matrix for $\boldsymbol{\theta}$ in
the full model be
\begin{equation}
\boldsymbol{{\cal I}}(\boldsymbol{\theta}^{*})=
\begin{bmatrix}
\boldsymbol{{\cal I}}_{11} & \\
\boldsymbol{{\cal I}}'_{12} &
\boldsymbol{{\cal I}}_{22}
\end{bmatrix}
\label{Imat}
\end{equation}
where $\boldsymbol{\theta}^{*}$ denotes the true value of
$\boldsymbol{\theta}$ and the partitioning corresponds to
$\boldsymbol{\theta}_{1}$ and $\boldsymbol{\theta}_{2}$.
Let $\mathbf{\Sigma}_{11}/n$
be the asymptotic variance matrix of the step-1 estimator
$\tilde{\boldsymbol{\theta}}_{1}$.
The asymptotic variance matrix of the two-step estimator
$\tilde{\boldsymbol{\theta}}_{2}$ is then
$\var(\tilde{\boldsymbol{\theta}}_{2})=\mathbf{V}/n$, where
\begin{equation}
\mathbf{V} =
\boldsymbol{{\cal I}}_{22}^{-1}
+
\boldsymbol{{\cal I}}_{22}^{-1}\,
\boldsymbol{{\cal I}}_{12}'\,
\mathbf{\Sigma}_{11}\,
\boldsymbol{{\cal I}}_{12}\,
\boldsymbol{{\cal I}}_{22}^{-1}
\equiv \mathbf{V}_{2} + \mathbf{V}_{1}
\label{Vmat}
\end{equation}
(if $\tilde{\boldsymbol{\theta}}_{1}$ was obtained using
$n_{1}\ne n$ of observations, $\mathbf{\Sigma}_{11}$ is
multiplied by ($n/n_{1}$) in this; see \cite{xue+bandeen-roche02} and
\cite{bakk-kuha}). Here $\mathbf{V}_{2}$ describes the variability in
$\tilde{\boldsymbol{\theta}}_{2}$ if $\boldsymbol{\theta}_{1}$ were
known, and $\mathbf{V}_{1}$ the additional variability arising from
estimating $\boldsymbol{\theta}_{1}$ by
$\tilde{\boldsymbol{\theta}}_{1}$.

The estimated variance matrix of $\tilde{\boldsymbol{\theta}}_{2}$ is
$\hat{\var}(\tilde{\boldsymbol{\theta}}_{2})=\hat{\mathbf{V}}/n$, where
$\hat{\mathbf{V}}=\hat{\mathbf{V}}_{2}+\hat{\mathbf{V}}_{1}$, and the
estimates of these matrices are obtained by substituting
$\tilde{\boldsymbol{\theta}}=(\tilde{\boldsymbol{\theta}}_{1}',
\tilde{\boldsymbol{\theta}}_{2}')'$ for $\boldsymbol{\theta}^{*}$ in
$\boldsymbol{{\cal I}}_{22}$ and $\boldsymbol{{\cal I}}_{12}$ and
substituting an estimate of $\boldsymbol{\Sigma}_{11}$. The estimate of
$\mathbf{V}_{2}=\boldsymbol{{\cal I}}_{22}^{-1}$ is obtained from step 2, while
that of $\boldsymbol{{\cal I}}_{12}$ requires some additional
calculation as discussed below.

We also need an estimate of the variance
matrix $\boldsymbol{\Sigma}_{11}/n$ of $\tilde{\boldsymbol{\theta}}_{1}$
from step 1 of the estimation. Here some new questions arise when
$\boldsymbol{\eta}_{i}=(\eta_{i1},\dots,\eta_{iK})'$ is a vector, with
corresponding measurement parameters
$\boldsymbol{\theta}_{1}=(\boldsymbol{\theta}_{11}',\dots,\boldsymbol{\theta}_{1K}')'$.
Suppose that each $\tilde{\boldsymbol{\theta}}_{1k}$ was estimated
separately in step~1. Consider $\boldsymbol{\Sigma}_{11}$ divided into
diagonal blocks $\boldsymbol{\Sigma}_{11(kk)}=n\,
\text{var}(\tilde{\boldsymbol{\theta}}_{1k})$ and off-diagonal blocks
$\boldsymbol{\Sigma}_{11(kl)}=n\, \text{cov}(
\tilde{\boldsymbol{\theta}}_{1k},\tilde{\boldsymbol{\theta}}_{1l})$ for
each $k,l=1,\dots,K$, and denote the Fisher information matrices for
$\boldsymbol{\psi}_{k}$ in (\ref{step1loglik}) by $\boldsymbol{{\cal
I}}_{*11(k)}$. The $\boldsymbol{\Sigma}_{11(kk)}$ are obtained by
extracting the elements corresponding to $\boldsymbol{\theta}_{1k}$ from
$\boldsymbol{{\cal I}}^{-1}_{*11(k)}$, and estimates
$\tilde{\boldsymbol{\Sigma}}_{11(kk)}$ by substituting
$\tilde{\boldsymbol{\psi}}_{k}$ in them. The off-diagonal blocks,
however, cannot be obtained from these step-1 information matrices. We
choose to set $\tilde{\boldsymbol{\Sigma}}_{11(kl)}=\mathbf{0}$
for $k\ne l$. This is a simplifying approximation, because even though
$\tilde{\boldsymbol{\theta}}_{1k}$ and
$\tilde{\boldsymbol{\theta}}_{1l}$ are estimated from separate models,
they can still be correlated because they use the data on
$\mathbf{Y}_{ik}$ and $\mathbf{Y}_{il}$ for the same units $i$. The same
approach is taken by \cite{rosseel+loh24} for their SAM implementation.
It seems quite justified at least in our simulations in Section~\ref{s_simulation}, where it has no effect on the accuracy of the
variance estimation. A way of estimating these
cross-block covariances could be derived from general results on
estimation equations \citep[see][Chapter 5]{cameron+trivedi05}, but some
elements of this would not be easily available from standard software.

The most inconvenient element of (\ref{Vmat}) is the matrix
$\boldsymbol{\mathcal{I}}_{12}$. An estimate
$\widehat{\boldsymbol{\mathcal{I}}}_{12}$ of it should be obtained as
the off-diagonal block of the information matrix
$\boldsymbol{\cal{I}}(\boldsymbol{\theta}^{*})$ in (\ref{Imat}),
evaluated at the final estimates
$\tilde{\boldsymbol{\theta}}=(\tilde{\boldsymbol{\theta}}_{1}',\tilde{\boldsymbol{\theta}}_{2}')'$.
This is not produced by the two-step estimation procedure, because in its second step $\tilde{\boldsymbol{\theta}}_{1}$ are
treated as fixed numbers and omitted from the information matrix. So
calculating $\widehat{\boldsymbol{\mathcal{I}}}_{12}$ requires an
additional step. We have obtained it by a further call to the estimation
software, where one-step estimation is started from
$\tilde{\boldsymbol{\theta}}$ and
$\widehat{\boldsymbol{\mathcal{I}}}_{12}$ is taken from the information
matrix of this after one iteration. This is only approximately correct
because that one iteration means that $\boldsymbol{\mathcal{I}}_{12}$ is
in the end evaluated at a value which differs to some extent from
$\tilde{\boldsymbol{\theta}}$. Even with it, however, the estimated
standard errors perform well in our simulations.

\cite{dimari+kuha25} proposed an alternative, simulation-based way of
estimating $\mathbf{V}$ which avoids the need to evaluate (\ref{Vmat}).
It is based on the decomposition
$
\var(\tilde{\boldsymbol{\theta}}_{2})=
\E_{\tilde{\boldsymbol{\theta}}_{1}}[\var(\tilde{\boldsymbol{\theta}}_{2}\vert
\tilde{\boldsymbol{\theta}}_{1})]
+\var_{\tilde{\boldsymbol{\theta}}_{1}}[\E(\tilde{\boldsymbol{\theta}}_{2}\vert
\tilde{\boldsymbol{\theta}}_{1})]
$
where the outer expectation and variance are over the sampling
distribution of the step-1 estimator $\tilde{\boldsymbol{\theta}}_{1}$.
The first term of this is estimated by the same $\hat{\mathbf{V}}_{2}$
as above. The second term can be estimated by drawing multiple values of
$\boldsymbol{\theta}_{1}$ from the sampling distribution of
$\tilde{\boldsymbol{\theta}}_{1}$, carrying out step-2 estimation given
each of them in turn, and calculating the variance matrix of the
resulting sample of estimates of $\boldsymbol{\theta}_{2}$. This is
asymptotically equivalent with variance estimation from (\ref{Vmat}).
Essentially the same idea is employed by
\cite{levy23} and \cite{levy+mcneish25} in their Bayesian approach,
which uses draws of the parameters from MCMC simulations.

A rather different approach would be to omit
$\mathbf{V}_{1}$ from the variance altogether. This would mean, in effect, that
once step 1 was done we would treat the measurement model obtained
from it, and thus the definition of the latent variables
$\boldsymbol{\eta}_{i}$, as known and fixed rather than as an estimable
characteristic. Estimated structural models obtained from step 2 would
then be models for this definition of the latent variables. This is close in
spirit to the approach that is also almost always adopted in naive
three-step estimation, as discussed further in Section
\ref{sss_estimation_onethree_naive3} below. This may be desirable or
unproblematic for practical data analysis in many applications. When it is not, however, omitting
the contribution from $\mathbf{V}_{1}$ may result in serious
underestimation of the variance of $\tilde{\boldsymbol{\theta}}_{2}$.

Consider, finally, the potential deviation from the conditions of PML
estimation that was mentioned above. It is a distributional
inconsistency which can arise when the latent
variables~$\boldsymbol{\eta}_{i}$ are continuous. We assume throughout
this paper that the conditional distributions for
$\boldsymbol{\eta}_{i}$ are normal, as shown in
(\ref{eta_model1})--(\ref{eta_model2}). This also implies that
$p(\mathbf{\eta}_{ik}|\mathbf{X}_{i})$ is univariate normal for all~$k$.
The marginal distribution of any single latent trait is then
$p(\eta_{ik})=\int p(\eta_{ik}|\mathbf{X}_{i})p(\mathbf{X}_{i})\,
d\mathbf{X}_{i}$ where $p(\mathbf{X}_{i})$ is the joint sample
distribution of $\mathbf{X}_{i}$. This $p(\eta_{ik})$ is not normal
unless $p(\mathbf{X}_{i})$ is also normal. The marginal normal
distribution that we assume in (\ref{step1loglik}) for $\eta_{ik}$ (and
any multivariate normal distributions for multiple traits which may be
used in step 1) is thus misspecified to some extent, if the structural
model includes non-normal covariates $\mathbf{X}_{i}$. Literature on
latent trait models with misspecified distributions of the trait
\citep[see e.g.][and references therein]{manapat+edwards22} indicates
that estimates of the measurement parameters can then be biased, with a
bias that depends on the true distribution of $\eta_{ik}$ and is largest
when it is very skewed. In two-step estimation this would be a concern
if such bias in $\tilde{\boldsymbol{\theta}}_{1}$ from step 1 also
translated into bias in $\tilde{\boldsymbol{\theta}}_{2}$ from step 2.
We examine this situation in the simulations in
Section~\ref{s_simulation}. There we do not observe any meaningful bias
of this kind in $\tilde{\boldsymbol{\theta}}_{2}$, even when the true
trait distribution in step 1 is very non-normal. Taking the latent traits to
be normally distributed also in step 1 thus seems empirically justified.
We note also that the same approximation is routinely made also in
one-step estimation when we consider different choices of
$\mathbf{X}_{i}$ and take $\eta_{ik}$ be normally distributed given any
of them (which cannot be exactly true for all of them).

\subsection{Model selection}
\label{ss_estimation_modelselection}

We have described how two-step estimation will be done for any given model specification.
In practice, however, we first need to decide which specification(s) to use for
substantive analysis. This too is best done in two steps, selecting
first the form of the measurement model and then the structural model
(this is in fact an older definition of ``two-step analysis'' in latent
variable modelling; see \citealt{anderson+gerbing88}). The basic tools
for doing this are the standard methods of likelihood-based model
comparison, such as likelihood ratio tests and the penalised model
selection criteria AIC and BIC.

Measurement models would again be first examined with the structural
model omitted as far as possible, and typically further split up into
smaller pieces. Here, where the substantive interest is on structural
models, it should be rare for us to want to examine the measurement properties of all
the indicators together, in a fully exploratory manner (if that was
still thought necessary, it is likely that the understanding of the
measurement items was not yet good enough to allow meaningful
substantive analysis of models between the latent constructs).
What is then left to examine, if anything, are smaller choices within
the measurement models, such as possible conditional associations
between items, cross-loadings between indicators of closely related
constucts, or aberrant behaviour of individual items.

Once the form of the measurement model is selected and its parameters
fixed at their estimated values, different structural models can be
compared as a routine part of step 2 of two-step estimation. Here we
would, in particular, examine which (latent and observed) explanatory
variables need or need not be included in different parts of the
structural model. These steps are essentially similar as they would be
for standard regression models with all variables observed.

Elsewhere in latent variable modelling, especially in structural
equation modelling, it is also common to carry out ``global''
assessments of a whole model at once, rather than its individual
parameters separately. This can use overall significance tests
(``overidentification tests'') or goodness-of-fit indices (many of which
exist, especially for SEMs). In a two-step context this would be much
less natural, because it would be a retrograde step: global model
assessment again blends measurement and structural models, which the
two-step approach takes pains to separate. We thus do not recommend such
global methods here. In principle, however, they could still be
implemented (see \citealt{rosseel+loh24} for more on this). For
any indices this would be done simply by calculating them as normal
using the two-step parameter estimates, whereas for goodness-of-fit
tests their sampling distributions might need to be modified.

\subsection{Alternatives: One- and three-step estimation}
\label{ss_estimation_onethree}

\subsubsection{One-step estimation}
\label{sss_estimation_onethree_one}

The main alternative to
two-step estimation is \emph{one-step estimation}. It means obtaining maximum likelihood (ML)
estimates
$\hat{\boldsymbol{\theta}}=(\hat{\boldsymbol{\theta}}_{1},\hat{\boldsymbol{\theta}}_{2})$
of the measurement parameters and structural parameters together, by
maximizing the log-likelihood (\ref{loglik1}) with respect to
$\boldsymbol{\theta}$.
If the model is correctly specified, $\hat{\boldsymbol{\theta}}$ has the
standard asymptotic properties of ML estimates. It is consistent for the
true $\boldsymbol{\theta}^{*}$, and its asymptotic variance matrix is
$\boldsymbol{{\cal I}}^{-1}(\boldsymbol{\theta}^{*})/n$. Here it is
interesting to examine the form of the variance matrix of
$\hat{\boldsymbol{\theta}}_{2}$. Denote it by
$\var(\hat{\boldsymbol{\theta}}_{2})=\mathbf{V}_{ML}/n$, where $\mathbf{V}_{ML}$ is the
block of
$\boldsymbol{{\cal I}}^{-1}(\boldsymbol{\theta}^{*})$
corresponding to $\boldsymbol{\theta}_{2}$.
Applying rules for inverting
partitioned matrices to (\ref{Imat}), we can also write
\begin{equation}
\mathbf{V}_{ML} =
\boldsymbol{{\cal I}}_{22}^{-1}
+
\boldsymbol{{\cal I}}_{22}^{-1}\,
\boldsymbol{{\cal I}}_{12}'\,
\boldsymbol{{\cal I}}^{11}\,
\boldsymbol{{\cal I}}_{12}\,
\boldsymbol{{\cal I}}_{22}^{-1}
\label{VmatML}
\end{equation}
where $\boldsymbol{{\cal I}}^{11}= (\boldsymbol{{\cal I}}_{11}-
\boldsymbol{{\cal I}}_{12}\, \boldsymbol{{\cal I}}_{22}^{-1}\,
\boldsymbol{{\cal I}}_{12}')^{-1} $ is the block of $\boldsymbol{{\cal
I}}^{-1}(\boldsymbol{\theta}^{*})$ corresponding to
$\boldsymbol{\theta}_{1}$. Comparing (\ref{VmatML}) with (\ref{Vmat}),
we can see that they differ only in that $\mathbf{V}_{ML}$ has
$\boldsymbol{{\cal I}}^{11}=n\,\var(\hat{\boldsymbol{\theta}}_{1})$
where $\mathbf{V}$ had
$\boldsymbol{\boldsymbol{\Sigma}}_{11}=n\,\var(\tilde{\boldsymbol{\theta}}_{1})$.
In other words, the only difference between the asymptotic variance
matrices of the one-step and two-step estimates of the structural
parameters $\boldsymbol{\theta}_{2}$ is in how much uncertainty they
incorporate about estimates of the measurement parameters
$\boldsymbol{\theta}_{1}$. Two-step (step-1) estimates of the
measurement parameters of each latent variable $\eta_{ik}$ use only
information from the items that measure that $\eta_{ik}$, whereas
one-step estimates also use information from the measurement items of
other latent variables and from any observed covariates
$\mathbf{X}_{i}$. Because the additional information provided by the
latter is typically small, we would expect  the
variances of $\tilde{\boldsymbol{\theta}}_{2}$ and
$\hat{\boldsymbol{\theta}}_{2}$ to be very similar.
That of
$\tilde{\boldsymbol{\theta}}_{2}$ can even be smaller, in situations
where step 1 of two-step estimation is based on more data than is used for
one-step estimation.

\subsubsection{Naive three-step estimation}
\label{sss_estimation_onethree_naive3}

\emph{Three-step estimation} is another stepwise approach. Its first
step is the same as in the two-step method, i.e.\ obtaining estimates
$\tilde{\boldsymbol{\theta}}_{1}$ of the measurement parameters from
simplest possible models. In the second step, these are used to
calculate predicted values $\tilde{\eta}_{ik}$ for the latent variables,
for $i=1,\dots,n$; $k=1,\dots,K$. In the third step, the structural
model is estimated with
$\tilde{\boldsymbol{\eta}}_{i}=(\tilde{\eta}_{i1},\dots,\tilde{\eta}_{iK})'$
substituted for $\boldsymbol{\eta}_{i}$ and treated as observed
variables. For the recursive model
(\ref{structural_model})--(\ref{eta_model2}) this means simply fitting
the $M$ regression models for
$\tilde{\boldsymbol{\eta}}_{i(1)},\dots,\tilde{\boldsymbol{\eta}}_{i(M)}$
separately. Estimated variances of the parameter estimates are also
obtained as in standard regression, in effect treating
$\tilde{\eta}_{ik}$ like any observed variables. We refer to this as
``naive'' three-step estimation.

The predictions (or ``latent trait scores'') $\tilde{\eta}_{ik}$ could
be obtained in different ways. The ones we use
here are the conditional expected values (empirical Bayes predictions)
$\tilde{\eta}_{ik}=\E(\eta_{k}|\mathbf{Y}_{i(k)};
\tilde{\boldsymbol{\psi}}_{(k)})$, where $\mathbf{Y}_{i(k)}$ denotes
those items which were included in step-1 estimation of the measurement
model for $\eta_{ik}$ and $\tilde{\boldsymbol{\psi}}_{(k)}$ the
estimated parameters of this model. These $\tilde{\eta}_{ik}$ require
numerical integration, but they are normally provided by estimation
software.

This approach is ``naive'' because $\tilde{\eta}_{ik}$ are erroneously
measured versions of $\eta_{ik}$. As a result, in most cases it produces
biased estimates of the parameters of the structural model. Trying to
correct for this bias leads to ``adjusted'' versions of three-step
estimation, which we discuss in the next section. Before that, however,
it is worth noting that whether we would want to do that depends on what
we think the goal of three-step estimation is. Measurement error bias is
a problem if we want to treat the step-3 models as estimates of the
regression models (\ref{structural_model})--(\ref{eta_model2}) for the
latent variables $\boldsymbol{\eta}_{i}$. But it is not an issue if we
view them instead as estimates for the same models for the scores
$\tilde{\boldsymbol{\eta}}_{i}$ themselves. In other words, we can
choose to treat $\tilde{\boldsymbol{\eta}}_{i}$ as variables of interest
rather than as poor measures of $\boldsymbol{\eta}_{i}$. In this view,
steps 1 and 2 of the analysis become data reduction steps which are used
only to develop a calculation that reduces $\mathbf{Y}_{i}$ to these
$\tilde{\boldsymbol{\eta}}_{i}$. This also an entirely coherent approach
to data analysis, if it is substantively interesting and justified.
Here, however, we focus on the situation where models for the latent
$\boldsymbol{\eta}_{i}$ are the goal, because it is also what motivates
two-step and one-step estimation.

\subsubsection{Adjusted three-step estimation}
\label{sss_estimation_onethree_adjusted3}

The goal of adjusted three-step estimation is to obtain valid estimates
of the structural model $p(\boldsymbol{\eta}_{i}|\mathbf{X}_{i};
\boldsymbol{\theta}_{2})$ based on a model for
$p(\tilde{\boldsymbol{\eta}}_{i}|\mathbf{X}_{i};
\boldsymbol{\theta}_{2})$, treating
$\tilde{\boldsymbol{\eta}}_{i}$
as a known function of $\mathbf{Y}_{i}$.
Naive three-step estimation is based on the approximation
$p(\tilde{\boldsymbol{\eta}}_{i}|\mathbf{X}_{i};
\boldsymbol{\theta}_{2})\approx p(\boldsymbol{\eta}_{i}|\mathbf{X}_{i};
\boldsymbol{\theta}_{2})$, which is generally incorrect. The correct
distribution is of the form
\begin{equation}
p(\tilde{\boldsymbol{\eta}}_{i}|\mathbf{X}_{i}; \boldsymbol{\theta}_{2})
=
\int \,
p(\tilde{\boldsymbol{\eta}}_{i}|\boldsymbol{\eta})\,
p(\boldsymbol{\eta}|\mathbf{X}_{i}; \boldsymbol{\theta}_{2})
\, d\boldsymbol{\eta}.
\label{p_thetatilde}
\end{equation}
Note that here we deviate slightly from previous notation, to make one
point clearer below. This is that the latent variables are denoted not
as $\boldsymbol{\eta}_{i}$, but as
$\boldsymbol{\eta}=(\eta_{.1},\dots,\eta_{.K})'$ to emphasise the fact
that in (\ref{p_thetatilde}) they are integration variables and not
fixed values for units $i$. Adjusted three-step estimation could be
based on the right-hand side or left-hand side of (\ref{p_thetatilde}),
and this has been done for other families of latent
variable models. For latent trait models, however, it is less
promising, because it is either
redundant or unavailable as a general method.

Consider first the right-hand side of (\ref{p_thetatilde}). This is
still a latent variable model, but with $\tilde{\boldsymbol{\eta}}_{i}$
rather than $\mathbf{Y}_{i}$ as the observed measurement indicators. It
is thus like step 2 of two-step estimation, but applied to
$\tilde{\boldsymbol{\eta}}_{i}$. To implement it, we would first need to
derive the form of the distribution
$p(\tilde{\boldsymbol{\eta}}_{i}|\boldsymbol{\eta})$. This is fairly
easy for latent class models, where $\mathbf{Y}_{i}$,
$\boldsymbol{\eta}$ and $\tilde{\boldsymbol{\eta}}_{i}$ are all
categorical, and estimation based on this approach has been developed
there by \citet{vermunt:10} and \citet{bakketal13}. It could also be
straightforward for normal structural equation models (SEMs), but it has
not been examined in full in that literature (other
three-step methods discussed below are more natural and
straightforward for them). \cite{savalei18} considered it for SEMs in the special
case where $\tilde{\eta}_{ik}$ are equally-weighted sum scores of the
measurement items.

To examine this possibility for latent trait models, suppose for simplicity
that all the items $Y_{ikj}$ are binary and that the measurement model
is (\ref{measurement_model1}), i.e.\ that the different traits
$\eta_{\cdot k}$ are measured by the distinct sets of items
$\mathbf{Y}_{ik}=(Y_{ik1},\dots,Y_{ikp_{k}})'$. We can then write
(\ref{p_thetatilde}) as
\begin{equation}
p(\tilde{\boldsymbol{\eta}}_{i}|\mathbf{X}_{i}; \boldsymbol{\theta}_{2})
=
\int \,
\left[\prod_{k=1}^{K} p(\tilde{\eta}_{ik}|\eta_{\cdot k})\right]\,
p(\boldsymbol{\eta}|\mathbf{X}_{i};\boldsymbol{\theta}_{2})
\, d\boldsymbol{\eta}.
\label{p_thetatilde2}
\end{equation}
Here $\tilde{\eta}_{ik}=\E(\eta_{.k}|\mathbf{Y}_{ik})$ are the empirical
Bayes predictions, which are not available in a closed form. For
purposes of this discussion, however, it is reasonable to approximate
them by the weighted sums $\tilde{\eta}_{ik}\approx
\sum_{j}a_{kj}Y_{ikj}$, where $a_{kj}$ are functions of the step-1
parameter estimates~$\tilde{\boldsymbol{\psi}}_{k}$. As a function of
the random vector $\mathbf{Y}_{ik}$, this $\tilde{\eta}_{ik}
=\tilde{\eta}_{ik}(\mathbf{Y}_{ik})$ is a discrete random variable with
$2^{p_{k}}$ possible values, many of which will have very low
probability for a given $\eta_{.k}$, and with mean $\sum_{j}
a_{kj}\pi_{ikj}(\eta_{.k})$ and variance
$\var(\tilde{\eta}_{ik}|\eta_{.k})=\sum_{j}
a_{kj}^{2}\pi_{ikj}(\eta_{.k})[1-\pi_{ikj}(\eta_{.k})]$ where
$\pi_{ikj}(\eta_{.k})=P(Y_{ikj}=1|\eta_{.k})$ is given by the
measurement model (\ref{logit}). The $p(\tilde{\eta}_{ik}|\eta_{.k})$ in
(\ref{p_thetatilde2}) is the probability function of this distribution,
evaluated at the observed value of $\tilde{\eta}_{ik}$.

Adjusted three-step estimation of this kind has been proposed for latent
trait models by \citet{lai+hsiao22}. They approximate
$p(\tilde{\eta}_{ik}|\eta_{.k})$ by a normal distribution with mean
$\eta_{.k}$ (when $\eta_{.k}$ is taken to have variance 1) and variance
$\hat{\var}(\eta_{.k}|\tilde{\eta}_{ik})$ (which can be obtained from
estimation software). This, however, is not correct in general, because
$\tilde{\eta}_{ik}$ is not normally distributed and because the variance
of $\tilde{\eta}_{ik}$
given $\eta_{.k}$
is not equal to the variance of $\eta_{.k}$ given
$\tilde{\eta}_{ik}$ and not constant as a function of
$\eta_{.k}$.
It is not clear when this approximation would be
adequate in practice. It seems that it would require, at least, that
$\tilde{\eta}_{ik}$ and $\eta_{.k}$ are scaled to have the same marginal
variance and that $\tilde{\eta}_{ik}$ is approximately normally
distributed (which requires at least large $p_{k}$) and has roughly the
same variance given all the values of $\eta_{.k}$ which make a
non-trivial contribution to the integral in (\ref{p_thetatilde2}).

There is, however, a different way of obtaining these three-step
estimates. Consider again binary items,
and suppose that  all estimated measurement models
(\ref{logit}) for different items $Y_{ikj}$ for $\eta_{.k}$ are
different from each other. The weights $a_{kj}$ in the approximation
$\tilde{\eta}_{ik}\approx \sum_{j}a_{kj}Y_{ikj}$ are then
also all different, so that every possible value of $\mathbf{Y}_{ik}$
gives a distinct value of
$\tilde{\eta}_{ik}=\tilde{\eta}_{ik}(\mathbf{Y}_{ik})$ almost surely. When this
is the case,
$p(\tilde{\eta}_{ik}|\eta_{.k})=p(\mathbf{Y}_{ik}|\eta_{.k})$ and
$p(\tilde{\boldsymbol{\eta}}_{i}|\mathbf{X}_{i};\boldsymbol{\theta}_{2})$
in (\ref{p_thetatilde2}) is equal to the
$p(\mathbf{Y}_{i}|\mathbf{X}_{i};\boldsymbol{\theta}_{2})$ that would be
used for step 2 of two-step estimation. It would seem that the same
argument holds also for the exact values of
$\tilde{\boldsymbol{\eta}}_{i}$, and when any items $Y_{ikj}$ are
polytomous. In other words, if all the items have different estimated
measurement parameters, there is almost surely a one-to-one relationship
between distinct values of $\mathbf{Y}_{i}$ and
$\tilde{\boldsymbol{\eta}}_{i}$.
Adjusted three-step estimation based on
the right-hand side of (\ref{p_thetatilde}), if correctly implemented,
would then give the same estimates of $\boldsymbol{\theta}_{2}$ as
two-step estimation which uses the same step-1 estimates of
$\boldsymbol{\theta}_{1}$ --- but would deliver them with more effort
for no gain. If different items were constrained to have identical measurement models, this kind of three-step
estimation would require separate implementation, but this would
again be effectively pointless because it would in the end reflect
essentially the same information that two-step estimation uses more
easily. We do not consider such three-step estimates further in
this paper.

Adjusted three-step estimation based on the left-hand side of
(\ref{p_thetatilde}) means estimating $\boldsymbol{\theta}_{2}$ from a
closed-form model
$p(\tilde{\boldsymbol{\eta}}_{i}|\mathbf{X}_{i};\boldsymbol{\theta}_{2})$
which is obtained by first evaluating the integral in
(\ref{p_thetatilde}). This is again relatively straightforward for
latent class models (see \citealt{Bolck:04}, \citealt{vermunt:10}, and
\citealt{bakketal13}). It is also convenient for linear SEMs, where it
means obtaining closed-form expressions for the mean vector and
covariance matrix of~$\tilde{\boldsymbol{\eta}}_{i}$ and using them to
estimate~$\boldsymbol{\theta}_{2}$ (see \citealt{croon},
\citealt{devlieger+rosseel17}, and references therein).

For latent trait models this would require a sufficiently good
approximation of the integral which would also yield a conveniently
estimable model. That is not generally available. In some cases,
however, a version of it can be derived. Suppose that the
interesting part of the structural model is
$\eta_{2i}=\beta_{20}+\beta_{21}\eta_{1i}+\psi_{2i}$, with no covariates
$\mathbf{X}_{i}$. Suppose further that we use for $\eta_{1i}$ the
Empirical Bayes score $\tilde{\eta}_{1i}=\E(\eta_{.1}|\mathbf{Y}_{i1})$,
and for $\eta_{2i}$ an unbiased score $\breve{\eta}_{2i}$ for which
$\E(\breve{\eta}_{2i}|\eta_{2i})=\eta_{2i}$. Then we have
$\breve{\eta}_{2i}=\beta_{20}+\beta_{21}\tilde{\eta}_{1i}+\psi_{2i}^{*}$,
so that $(\beta_{20},\beta_{21})$ can be estimated from a linear
regression model for $\tilde{\eta}_{2i}$ given $\tilde{\eta}_{1i}$
(which has a heteroscedastic residual $\psi_{2i}^{*}$, but this can be
allowed for in the estimation). For this model and these trait scores,
naive and adjusted three-step estimation are thus the same. This
situation was described for SEMs by \cite{skrondal_01}, using regression
factor scores as $\tilde{\eta}_{1i}$ and Bartlett factor scores as
$\breve{\eta}_{2i}$. \cite{lu+thomas08} showed that the same
construction would work also for latent trait models, but they also noted
that for these models there are no unbiased scores to be used as
$\breve{\eta}_{2i}$ (the best that can be obtained is consistency as the
number of items increases).

This approach would not generalise well to more complex models. For
example, suppose that the structural model includes a further latent
variable $\eta_{3i}$ and the model $\eta_{3i}=\beta_{30}
+\beta_{31}\eta_{1i}+ \beta_{32}\eta_{2i}+ \psi_{3i}$. The scores that
should be assigned for $(\eta_{1i},\eta_{2i})$ to estimate this model
would be $\E[(\eta_{1i},\eta_{2i})|\mathbf{Y}_{1i},\mathbf{Y}_{2i}]$,
i.e.\ predictions given both of their indicators (\citealt{skrondal_01},
\citealt{lu+thomas08}); these should also be conditional on any
covariates $\mathbf{X}_{i}$ in the structural model. To get these, we
would need to fit the step-1 model for $\eta_{1i}$ and $\eta_{2i}$
jointly rather than separately. Furthermore, the appropriate trait
scores to assign for $\eta_{1i}$ and $\eta_{2i}$ would be different for
the model for $\eta_{3i}$ than for the model for $\eta_{2i}$, and
different again for any further parts of the structural model. This
would quickly get very impracticable for larger models. We do not
consider this version of adjusted three-step estimation further in this
paper, except for some examples with only a single explanatory latent
trait in the simulation studies below.

\subsection{Implementation of the estimation}
\label{ss_estimation_implementation}

Software that can do one-step estimation of a model can also be do
two-step estimation of it. We have used Mplus 6.12 software
\citep{muthen+muthen10} for the simulations and application
example of this paper. We have supplemented it with functions in R
\citep{r2022} to manage the process. This automatically
sequences the two steps of estimation, passing the estimated parameter
values from the first step to the code for the second and the estimates
back to R. We have used the \emph{MplusAutomation} package in R
\citep{hallquist+wiley18} to control Mplus from R, and the \emph{brew}
package \citep{horner11} to automatically edit the input files. Examples
of the estimation code are included in supplementary materials for the
paper.

The integrals in the log likelihoods (\ref{loglik1}) and
(\ref{step1loglik}) have no closed form, so numerical integration is
needed to evaluate them. This is the most demanding computational
element of likelihood-based estimation of these models, and it is
important that high-quality software is used to implement and control
it. The computational demands are substantially smaller for two-step
than for one-step estimation, both because step 1 typically involves
only one-dimensional integrals and because step 2 tends to require fewer
iterations (and hence fewer evaluations of the integrals). For example,
most of the computing time in our simulations in Section
\ref{s_simulation} was spent on one-step estimation, and there were
similarly large differences in our substantive application in Section
\ref{s_example}, as discussed further there. It is likely that this is
the most important practical difference between these two methods in
large models, where one-step estimation quickly becomes more and more
demanding. For both approaches, the computational task gets harder with
more latent variables. In our examples we have considered models with up
to three latent variables without problems, but much larger models would
be correspondingly more challenging.

Because the log likelihoods can have multiple local maxima, it is
desirable to use multiple starting values for the estimation. This is a
standard feature in software such as Mplus.

\subsection{Other methods: Non-likelihood-based estimation}
\label{ss_estimation_LS}

The focus of this paper is on a likelihood-based framework, where the
one- and two-step methods are full-information ML and pseudo-ML
estimation respectively. Other methods of estimation are also available
for latent trait models. For example, in the SEM literature it is common
to consider limited-information methods of the kind proposed by
\cite{muthen84} (see also \citealt{muthen+satorra95}) such as Diagonally
Weighted Least Squares (DWLS) estimation. They in effect convert models
with categorical indicators into SEMs for continuous indicators, under
some further assumptions. This is done by presenting the observed
categorical indicators~$\mathbf{Y}$ as coarsened versions of continuous
latent variables $\mathbf{Y}^{*}$ that follow a multivariate normal
distribution given $\mathbf{X}$, using the observed data to estimate the
means and variance matrix of~$\mathbf{Y}^{*}$ given $\mathbf{X}$, and
then treating these as data for estimating the model of interest with
some form of weighted least squares estimation. This produces estimates
that are often nearly as efficient as one-step FIML estimates. These
methods have some limitations of applicability, arising from the
assumption of normality for $\mathbf{Y}^{*}$. For example, they cannot
be used for models with unordered categorical indicators, or with
the logit models for binary or ordinal ones that we use (they imply
probit models instead). Their
handling of missing data is also somewhat inflexible. Their main advantage is
computation, which can be very substantially easier than for
likelihood-based estimation.

With reference to our focus, standard use of limited-information
estimation like this is still one-step estimation, in that both
measurement and structural parameters are estimated together in its last
stage. As such, comparisons between full- and limited-information
estimates are outside the scope of this paper. However, it is worth
noting that the choice between these methods of estimation is
separate from the choice of one or two steps in the estimation. The two-step
idea can be combined with any methods of estimation which satisfy
very general conditions \cite[see][Sections 24.2.4 and
24.2.2]{gourieroux+monfort95}, and it will still give estimates that are
consistent and asymptotically normal. Applied to DWLS, say, this would
mean carrying out its last stage in two steps, i.e.\ in effect applying
``local SAM'' of \cite{rosseel+loh24} to~$\mathbf{Y}^{*}$ given
$\mathbf{X}$. Another example of different possibilities in this spirit
is provided by \cite{levy23} and \cite{levy+mcneish25}, who carry out
both steps using Bayesian estimation.

Another limited-information method that has been developed in the SEM
literature is model-implied instrumental variable (MIIV) estimation. It
was originally proposed for standard SEMs (see \citealt{bollen19} for a
summary) but it can also be used with binary and ordinal indicators
after applying the $\mathbf{Y}$ to $\mathbf{Y}^{*}$ device explained
above (\citealt{jinetal21}; \citealt{fisher+bollen20}). MIIV is a
promising approach, with the particular strength that the estimates are
unsually robust to various misspecifications of the form of the
measurement model. What is interesting about it for our
focus is that while its procedure is ostensibly very different, MIIV has some of
the key conceptual characteristics of two-step estimation. In
particular, it too separates measurement and structural models, in that
the same measurement model (which in MIIV can be left somewhat implicit) can
be combined with different structural models without estimating them
together. This connection with our kind of likelihood-based two-step
estimation would be an interesting topic for future research.

\subsection{Two-step estimation for other types of models}
\label{ss_estimation_extensions}

Two-step estimation could be applied to any instance of the
general latent variable model that was defined at the start of Section
\ref{s_models}. This requires only that the joint model is itself
identified, i.e.\ that one-step estimates could be calculated. To see
this, note first that there are then also identified models for
estimating the measurement model in step 1 (in the extreme, the
redundant choice of the joint model itself could be used for this).
Second, if the structural model is identified when the parameters of the
measurement model are estimated with it, it is also identified when these
parameters are fixed as in the step 2 of two-step estimation.

To provide a clear focus for this paper, however, we concentrate on the
narrower choice of the family of latent trait models that was introduced
in Section~\ref{s_models}. Before proceeding to empirical
investigations of them, we comment briefly on other settings where the
two-step method could be used. Some of them are obvious extensions or
variants of the models considered here, while others could involve more
consequential differences. Implementation and performance of two-step
estimation in many of these other cases remain to be examined, but it
seems plausible that its behaviour would not be fundamentally different
in them.

Considering first the measurement model, some further variants of it
would involve no new issues, just appropriate changes to the model
specification and the likelihood functions. One obvious class of
extensions is produced by models with mixtures of different types of
measurement items. In this paper we focus on categorical items, while
linear structural equation models (as in \citealt{rosseel+loh24}) take
the items to be continuous. But some applications may involve both of
these types, or others such as counts or survival times as measurement
items. The latent variables being measured can also be of different
types and measured by different types of items, such as when categorical
latent variables are measured by categorical items (as in standard
latent class analysis) or by continuous ones (the case known as latent
profile analysis). Similarly, we could use standard specifications to
allow for conditional associations between some measurement indicators,
relaxing the assumption that all items in $\mathbf{Y}_{i}$ are
conditionally independent given the latent
traits~$\boldsymbol{\eta}_{i}$.

Other extensions of the measurement model would require further
modifications of step 1. As discussed in Section
\ref{ss_estimation_twostep}, the general principle is that this should
use the collection of the simplest models that still allow valid
estimation of the measurement parameters. We have already noted that one
simple instance of this is a situation where two latent variables share
part of the measurement model because some items measure both of them
(or, similarly, because some indicators of them are conditionally
associated). Then step 1 has to be carried out for those two variables
together rather than separately. A more involved example of this
principle arises for models which include non-equivalence of
measurement, i.e.\ where $p(\mathbf{Y}_{i}|\boldsymbol{\eta}_{i},
\mathbf{X}_{i};
\boldsymbol{\theta}_{1})=p(\mathbf{Y}_{i}|\boldsymbol{\eta}_{i},
\mathbf{X}_{i}^{*}; \boldsymbol{\theta}_{1})$ for some subset
$\mathbf{X}_{i}^{*}$ of $\mathbf{X}_{i}$. Then the models used for
step~1 should also be conditional on $\mathbf{X}_{i}^{*}$ for those
elements of $\mathbf{Y}_{i}$ that are affected by the non-equivalence
(\citealt{vermunt+magidson21}; \citealt{lyrvalletal25}).

The way to extensions of the structural model is even more obvious.
Whatever estimation procedure (and computer code) would be used for
one-step estimation of a model is also used, now with measurement
parameters fixed, for step 2 of two-step estimation. One simple
extension is to models which include mixtures of different types of
(continuous and categorical) latent variables. Other generalisations of
the structural model could also include such possibilities as
interactions or nonlinear terms of the latent variables, or
non-recursive models. Two-step estimation in these cases has not been
examined. It could behave differently there, if only because such models
are inherently more complex and difficult howsoever they are estimated.

One extension that is easily accommodated is multigroup modelling.
This means models for data on units from different groups, such as
respondents from different countries in cross-national
research. The focus of interest is then on how the parameters of the
structural model do or do not vary between the groups. The simplest
version of this is to include dummy variables for the groups in
$\mathbf{X}_{i}$, thus giving separate intercepts for different groups
in the structural models (\ref{eta_model1})--(\ref{eta_model2}). Beyond
this, latent variable modelling software such as Mplus also have
dedicated multigroup syntax which allows any of the parameters to vary
by the group. This can also be combined with two-step
estimation. In the applied example in Section \ref{s_example} we use
it to fit a model to cross-sectional survey data from 36
countries, where all the regression coefficients and all residual variances
and covariances in the structural model vary between the countries.

%\newpage

\section{Simulations}
\label{s_simulation}

\subsection{First set of simulations}
\label{ss_simulation_1}

We have used simulation studies to examine the properties of two-step
estimates for the types of models considered in this paper. In this section we
describe them for one model setting in detail. In Section
\ref{ss_simulation_2} we then summarise further simulations in several
other settings.

We examine the performance of the point
estimates and accuracy of the estimated standard errors. The main
comparison of interest is with one-step estimates of the same models.
Since they are full maximum likelihood estimates, they serve as a
benchmark whose theoretical properties are well understood. In large
samples, one-step estimates for correctly specified models will be
approximately unbiased and fully efficient, and it is then interesting
to see if two-step estimates do roughly as well. In smaller samples,
either approach could have an advantage. In this section all the models
are correctly specified, but in the next we also include simulations
with a misspecified measurement model in order to examine the robustness
of the estimates in such cases. Another point of comparison is
computational complexity, where two-step estimation should have the
advantage. The simulation results also include (naive) three-step
estimates, but we defer a discussion of results for them to a
separate Section \ref{ss_simulation_3step}.

In these first simulations the structural model for unit
$i=1,\dots,n$ is of the form
\begin{equation}
\eta_{i2} = \beta_{0}+\beta_{1}\eta_{i1} +\epsilon_{i}
\label{sim_model}
\end{equation}
where $\eta_{1i}$ and $\eta_{2i}$ are latent variables
and $\epsilon_{i}\sim N(0,\sigma^{2}_{\epsilon})$ is the residual.
Here $\eta_{1i}\sim N(0,1)$,
$\beta_{0}=0$, and $(\beta_{1}, \sigma^{2}_{\epsilon})$ are such that
marginally $\eta_{2i}\sim N(0,1)$ in all settings.
The value of $\beta_{1}$ is set so that the $R^{2}$ statistic for the response
$\eta_{i2}$ is $R^{2}=0, 0.2$, or $0.4$.
We refer to this as \emph{simulation case A}. Path diagram for it is
shown in Figure A.1 in the supplementary appendix.

The latent variable $\eta_{i1}$ is measured by $p$ binary items
$Y_{i1j}$, coded with values 0 and 1, and $\eta_{i2}$ by a different set
of $p$ binary items similarly, with $p=4$ or $p=8$. All items are
conditionally independent of each other given the latent variables. The
measurement model is of the form $\text{logit}[P(Y_{ikj}=1|\eta_{ik})] =
\tau_{kj} + \lambda_{kj} \eta_{ik}$ for $k=1,2$ and $j=1,\dots,p$. The
parameters $(\tau_{kj},\lambda_{kj})$ are estimated as
distinct parameters for different $k,j$, but their true values
$\tau_{kj}=\tau$ and $\lambda_{kj}=\lambda$ in the data-generating model
are equal for all items in a given simulation. The loading
$\lambda$ is set with reference to the linear model for a
notional continuous latent variable $Y_{ikj}^{*}$ which implies the
logistic model for $Y_{ikj}$. Here $R^{2}$ for $Y_{ikj}^{*}$ given
$\eta_{ik}$ is
$R^{2}_{Y}=\lambda^{2}\var(\eta_{ik})/(\lambda^{2}\var(\eta_{ik})+\pi^{2}/3)$
where $\var(\eta_{ik})=1$, and $\lambda$ is set so that $R^{2}_{Y}=0.4$
or 0.6. The values of $\tau$ are then set so that the marginal
probability $\pi_{Y}=P(Y_{ikj}=1)$ is 0.5 or (approximately) 0.8. In the
estimation, we constrain $\tau_{k1}=0$ and $\lambda_{k1}=1$
for the first item $Y_{ik1}$ of each latent variable, leaving
$\tau_{kj}$ and~$\lambda_{kj}$ for the other $p-1$ items estimable. This
affects the implied scales of the latent variables, so that the value of
$\beta_{1}$ that the estimators should actually be estimating will
depend also on the value of $\lambda$ in the data-generating model.
These true values of $\beta_{1}$ are shown in the results tables.
In each setting we simulate 1000 datasets with $n=200$ or
$n=1000$ independent observations~$i$. All combinations of $n$, $R^{2}$,
$p$, $\pi_{Y}$ and $R^{2}_{Y}$ are considered, resulting in 48 settings.
The two-, one- and naive three-step estimates are obtained using
Mplus, combined with R functions for process management
as discussed in Section \ref{ss_estimation_implementation}.

We focus on the results for the estimates of the regression coefficient
$\beta_{1}$ in the structural model (simulation results for the
measurement model are discussed briefly in Section
\ref{ss_simulation_2}). These results are shown in Table
\ref{t_sim_beta} for the point estimates and Table \ref{t_sim_se} for
the standard errors. For the point estimates, the table includes their
mean bias, root mean squared error (RMSE) and median absolute error
(MAE). To examine the estimated standard errors, we report the standard
deviation of the point estimates of $\beta_{1}$ and the mean of their
estimated standard errors across the simulations, and simulation
coverage of the 95\% confidence intervals. For the two-step estimates we
also report two quantities for the case where we ignore the uncertainty
from step 1 (i.e.\  omit the contribution from $\mathbf{V}_{1}$ to
(\ref{Vmat})) and include only the uncertainty from step~2 in the
standard errors: the average proportion of the full estimated variance
of the two-step estimate of $\beta_{1}$ that this accounts for, and the
coverage of the resulting 95\% confidence interval.

Consider first the point estimates of $\beta_{1}$ in Table
\ref{t_sim_beta}. There is no meaningful difference between two-step and
one-step estimates for the larger sample size of $n=1000$. They also
behave very similarly in almost all settings when $n=200$, except for
some differences  when the number of items is small ($p=4$), the
measurement model is weak ($R^{2}_{Y}=0.4$) and one of the values of the
items is rare ($\pi_{Y}=0.8$). This is the setting where the items
provide the least information about the latent variables. When
$\beta_{1}$ is 0, the two-step estimates have in these cases a slightly
better RMSE and MAE than the one-step estimates. With larger values of
$\beta_{1}$, on the other hand, in a few simulations these cases yield
an extreme value of the two-step estimate, inflating its RMSE. These
extremes are reduced for the one-step estimates, most likely because the
measurement models of $\eta_{1}$ and $\eta_{2}$ stabilise each other
when they are estimated together.

Results for the estimated standard errors of the estimates of
$\beta_{1}$ are reported in Table \ref{t_sim_se}. They are
very satisfactory when $n=1000$: for both the two-step and one-step methods the
standard errors are good estimates of the simulation standard deviations
of estimates of $\beta_{1}$, and coverages of confidence intervals are
close to the nominal 95\%. With the smaller sample size of $n=200$ this
coverage is less correct, ranging from 89\% to 100\% in different
settings. The estimated standard errors are even then reasonably
accurate when the estimated measurement model is sufficiently
informative. In the most difficult settings, with a small sample, weak
measurement model and small number of items, the standard errors tend
to underestimate the true variability.

The last two columns of Table \ref{t_sim_se}
examine what happens if we ignore the uncertainty
from step~1 of two-step estimation, i.e.\ if we calculate its standard
errors using only the term $\mathbf{V}_{2}$ in the variance formula
(\ref{Vmat}). Here this step-2 variance accounts for almost all of the
variance of the estimate of $\beta_{1}$ when the true $\beta_{1}$ is 0,
but much less otherwise, down to less than a third of the overall
variance in the settings with the largest $\beta_{1}$. The coverage of
confidence intervals is correspondingly reduced, down to less than 70\%.
When the association in the structural model is strong, the uncertainty
in its estimates is thus dominated by the uncertainty in the estimated
measurement parameters from step 1.

\clearpage
\newgeometry{top=10mm}
\begin{table}[ht]
\caption{Simulation results for point estimates of structural regression
coefficient $\beta_{1}$ of
a latent covariate $\eta_{1}$ for a conditionally normally distributed
latent response $\eta_{2}$ (simulation case A).}
%\centering
{\small{
\begin{tabular}{|rrrrr|rrr|rrr|rrr|}
  \hline
  & & & &
  & \multicolumn{3}{|c|}{Bias}
  & \multicolumn{3}{|c|}{RMSE}
  & \multicolumn{3}{|c|}{MAE} \\
$p$ & $\pi_{Y}$ & $R^{2}_{Y}$ & $R^{2}_{\eta}$ & $\beta_1$ & 2-step & 1-step & 3-step & 2-step & 1-step & 3-step & 2-step & 1-step & 3-step \\
  \hline \multicolumn{14}{|l|}{$n=200$}\\ \hline
4 & 0.5 & 0.4 & 0.0 & 0.000 & -0.004 & -0.003 & -0.002 & 0.133 & 0.141 & 0.081 & 0.077 & 0.080 & 0.046 \\
  4 & 0.8 & 0.4 & 0.0 & 0.000 & -0.007 & -0.004 & -0.000 & 0.240 & 0.276 & 0.110 & 0.101 & 0.113 & 0.044 \\
  4 & 0.5 & 0.6 & 0.0 & 0.000 & 0.007 & 0.007 & 0.005 & 0.115 & 0.117 & 0.080 & 0.069 & 0.070 & 0.048 \\
  4 & 0.8 & 0.6 & 0.0 & 0.000 & 0.002 & 0.002 & 0.003 & 0.153 & 0.154 & 0.091 & 0.081 & 0.084 & 0.048 \\
  8 & 0.5 & 0.4 & 0.0 & 0.000 & -0.001 & -0.001 & -0.001 & 0.104 & 0.104 & 0.078 & 0.063 & 0.062 & 0.047 \\
  8 & 0.8 & 0.4 & 0.0 & 0.000 & 0.008 & 0.009 & 0.006 & 0.147 & 0.151 & 0.087 & 0.081 & 0.083 & 0.047 \\
  8 & 0.5 & 0.6 & 0.0 & 0.000 & -0.003 & -0.003 & -0.002 & 0.097 & 0.095 & 0.077 & 0.062 & 0.059 & 0.050 \\
  8 & 0.8 & 0.6 & 0.0 & 0.000 & -0.002 & -0.002 & -0.000 & 0.114 & 0.116 & 0.080 & 0.072 & 0.073 & 0.050 \\
  \hline
  4 & 0.5 & 0.4 & 0.2 & 0.447 & 0.016 & 0.035 & -0.162 & 0.236 & 0.237 & 0.221 & 0.132 & 0.132 & 0.202 \\
  4 & 0.8 & 0.4 & 0.2 & 0.447 & 0.031 & 0.070 & -0.208 & 0.404 & 0.385 & 0.300 & 0.183 & 0.177 & 0.267 \\
  4 & 0.5 & 0.6 & 0.2 & 0.447 & 0.019 & 0.025 & -0.116 & 0.182 & 0.189 & 0.176 & 0.117 & 0.116 & 0.148 \\
  4 & 0.8 & 0.6 & 0.2 & 0.447 & 0.030 & 0.039 & -0.144 & 0.257 & 0.244 & 0.225 & 0.135 & 0.134 & 0.183 \\
  8 & 0.5 & 0.4 & 0.2 & 0.447 & 0.013 & 0.010 & -0.104 & 0.177 & 0.188 & 0.170 & 0.109 & 0.110 & 0.137 \\
  8 & 0.8 & 0.4 & 0.2 & 0.447 & 0.023 & 0.040 & -0.152 & 0.249 & 0.250 & 0.222 & 0.144 & 0.136 & 0.197 \\
  8 & 0.5 & 0.6 & 0.2 & 0.447 & 0.005 & -0.008 & -0.081 & 0.143 & 0.159 & 0.143 & 0.092 & 0.093 & 0.116 \\
  8 & 0.8 & 0.6 & 0.2 & 0.447 & 0.024 & 0.028 & -0.100 & 0.183 & 0.181 & 0.170 & 0.114 & 0.111 & 0.137 \\
  \hline
  4 & 0.5 & 0.4 & 0.4 & 0.632 & 0.028 & 0.047 & -0.223 & 0.309 & 0.285 & 0.301 & 0.176 & 0.162 & 0.268 \\
  4 & 0.8 & 0.4 & 0.4 & 0.632 & 0.042 & 0.081 & -0.284 & 0.608 & 0.457 & 0.436 & 0.228 & 0.199 & 0.363 \\
  4 & 0.5 & 0.6 & 0.4 & 0.632 & 0.048 & 0.052 & -0.143 & 0.275 & 0.254 & 0.247 & 0.140 & 0.147 & 0.199 \\
  4 & 0.8 & 0.6 & 0.4 & 0.632 & 0.038 & 0.048 & -0.195 & 0.326 & 0.283 & 0.295 & 0.174 & 0.161 & 0.255 \\
  8 & 0.5 & 0.4 & 0.4 & 0.632 & 0.020 & 0.022 & -0.142 & 0.213 & 0.222 & 0.216 & 0.131 & 0.134 & 0.176 \\
  8 & 0.8 & 0.4 & 0.4 & 0.632 & 0.033 & 0.054 & -0.204 & 0.344 & 0.346 & 0.309 & 0.177 & 0.166 & 0.258 \\
  8 & 0.5 & 0.6 & 0.4 & 0.632 & 0.033 & 0.006 & -0.087 & 0.196 & 0.227 & 0.182 & 0.127 & 0.124 & 0.138 \\
  8 & 0.8 & 0.6 & 0.4 & 0.632 & 0.024 & 0.030 & -0.138 & 0.223 & 0.218 & 0.220 & 0.137 & 0.135 & 0.180 \\
  \hline \multicolumn{14}{|l|}{$n=1000$}\\ \hline
  4 & 0.5 & 0.4 & 0.0 & 0.000 & 0.000 & 0.000 & 0.000 & 0.053 & 0.054 & 0.032 & 0.036 & 0.036 & 0.021 \\
  4 & 0.8 & 0.4 & 0.0 & 0.000 & -0.001 & -0.001 & -0.000 & 0.072 & 0.073 & 0.032 & 0.049 & 0.049 & 0.021 \\
  4 & 0.5 & 0.6 & 0.0 & 0.000 & -0.001 & -0.001 & -0.001 & 0.045 & 0.045 & 0.031 & 0.030 & 0.030 & 0.021 \\
  4 & 0.8 & 0.6 & 0.0 & 0.000 & 0.001 & 0.001 & 0.001 & 0.054 & 0.055 & 0.032 & 0.036 & 0.036 & 0.021 \\
  8 & 0.5 & 0.4 & 0.0 & 0.000 & 0.000 & 0.000 & 0.000 & 0.045 & 0.045 & 0.033 & 0.031 & 0.031 & 0.023 \\
  8 & 0.8 & 0.4 & 0.0 & 0.000 & 0.001 & 0.001 & 0.001 & 0.059 & 0.059 & 0.034 & 0.038 & 0.038 & 0.022 \\
  8 & 0.5 & 0.6 & 0.0 & 0.000 & -0.001 & -0.001 & -0.000 & 0.040 & 0.040 & 0.032 & 0.026 & 0.026 & 0.021 \\
  8 & 0.8 & 0.6 & 0.0 & 0.000 & 0.000 & 0.000 & 0.000 & 0.045 & 0.045 & 0.032 & 0.030 & 0.030 & 0.021 \\
  \hline
  4 & 0.5 & 0.4 & 0.2 & 0.447 & 0.006 & 0.012 & -0.174 & 0.092 & 0.090 & 0.183 & 0.061 & 0.059 & 0.178 \\
  4 & 0.8 & 0.4 & 0.2 & 0.447 & 0.011 & 0.021 & -0.227 & 0.124 & 0.124 & 0.236 & 0.078 & 0.077 & 0.234 \\
  4 & 0.5 & 0.6 & 0.2 & 0.447 & 0.001 & 0.003 & -0.131 & 0.075 & 0.074 & 0.142 & 0.053 & 0.052 & 0.135 \\
  4 & 0.8 & 0.6 & 0.2 & 0.447 & 0.011 & 0.013 & -0.160 & 0.091 & 0.090 & 0.171 & 0.057 & 0.059 & 0.168 \\
  8 & 0.5 & 0.4 & 0.2 & 0.447 & 0.004 & 0.005 & -0.112 & 0.073 & 0.073 & 0.125 & 0.047 & 0.047 & 0.116 \\
  8 & 0.8 & 0.4 & 0.2 & 0.447 & 0.010 & 0.015 & -0.163 & 0.094 & 0.093 & 0.174 & 0.062 & 0.061 & 0.166 \\
  8 & 0.5 & 0.6 & 0.2 & 0.447 & 0.002 & 0.003 & -0.084 & 0.062 & 0.062 & 0.098 & 0.039 & 0.039 & 0.087 \\
  8 & 0.8 & 0.6 & 0.2 & 0.447 & 0.008 & 0.009 & -0.113 & 0.074 & 0.074 & 0.126 & 0.048 & 0.048 & 0.118 \\
  \hline
  4 & 0.5 & 0.4 & 0.4 & 0.632 & 0.009 & 0.014 & -0.244 & 0.118 & 0.112 & 0.255 & 0.076 & 0.076 & 0.250 \\
  4 & 0.8 & 0.4 & 0.4 & 0.632 & 0.003 & 0.014 & -0.319 & 0.156 & 0.150 & 0.329 & 0.103 & 0.099 & 0.329 \\
  4 & 0.5 & 0.6 & 0.4 & 0.632 & 0.006 & 0.007 & -0.177 & 0.093 & 0.090 & 0.190 & 0.061 & 0.058 & 0.182 \\
  4 & 0.8 & 0.6 & 0.4 & 0.632 & 0.009 & 0.010 & -0.218 & 0.117 & 0.109 & 0.232 & 0.074 & 0.071 & 0.223 \\
  8 & 0.5 & 0.4 & 0.4 & 0.632 & 0.003 & 0.005 & -0.158 & 0.088 & 0.086 & 0.172 & 0.059 & 0.057 & 0.164 \\
  8 & 0.8 & 0.4 & 0.4 & 0.632 & -0.001 & 0.007 & -0.231 & 0.114 & 0.111 & 0.244 & 0.078 & 0.072 & 0.237 \\
  8 & 0.5 & 0.6 & 0.4 & 0.632 & 0.007 & 0.008 & -0.110 & 0.080 & 0.079 & 0.128 & 0.054 & 0.053 & 0.112 \\
  8 & 0.8 & 0.6 & 0.4 & 0.632 & 0.006 & 0.008 & -0.152 & 0.088 & 0.086 & 0.167 & 0.060 & 0.058 & 0.156 \\
   \hline
\multicolumn{14}{l}{\emph{Note:} $p$ denotes number of measurement
items $Y_{j}$, $\pi_{Y}$ marginal proportion of $Y_{j}=1$, }\\
\multicolumn{14}{l}{
\hspace*{2em}$R_{Y}^{2}$ and $R^{2}_{\eta}$ the $R^{2}$ statistics in models for
$Y_{j}$ and $\eta_{2}$, and $\beta_{1}$ the true value of $\beta_{1}$.}
\end{tabular}
}}
\label{t_sim_beta}
\end{table}
\clearpage

\thispagestyle{empty}
\begin{table}[ht]
\caption{Simulation results for standard error estimates of estimated structural regression
coefficient $\beta_{1}$ of a latent covariate $\eta_{1}$ for a conditionally
normally distributed latent response~$\eta_{2}$ (simulation case A).}
%\centering
{\small{
\begin{tabular}{|rrrrr|rrrr|rrr|rr|}
  \hline
  &&&&
  & \multicolumn{4}{c|}{Simulation standard deviation}
  &&&
  & \multicolumn{2}{c|}{2-step}
  \\
  &&&&
  & \multicolumn{4}{c|}{vs.\ mean of est.\ std.\ error}
  & \multicolumn{3}{c|}{Coverage of}
  & \multicolumn{2}{c|}{without}
  \\
  &&&&
  & \multicolumn{2}{|c}{2-step}
  & \multicolumn{2}{c|}{1-step}
  & \multicolumn{3}{c|}{95\% conf.\ interval}
  & \multicolumn{2}{c|}{step-1 var.$^{\dagger}$}
\\
$p$ & $\pi_{Y}$ & $R^{2}_{Y}$ & $R^{2}_{\eta}$ & $\beta_1$
& s.d.\ & m(se) & s.d.\ & m(se) & 2-st.\ & 1-st.\ & 3-st.\ & \%var. &
{\footnotesize{cover.}} \\
\hline \multicolumn{14}{|l|}{$n=200$}\\ \hline
4 & 0.5 & 0.4 & 0.0 & 0.000 & 0.133 & 0.137 & 0.141 & 0.138 & 99.5 & 99.3 & 95.6 & 87.3 & 96.0 \\
  4 & 0.8 & 0.4 & 0.0 & 0.000 & 0.240 & 0.224 & 0.276 & 0.235 & 99.9 & 100.0 & 94.7 & 78.0 & 96.0 \\
  4 & 0.5 & 0.6 & 0.0 & 0.000 & 0.115 & 0.114 & 0.116 & 0.114 & 98.9 & 98.5 & 94.3 & 91.4 & 94.9 \\
  4 & 0.8 & 0.6 & 0.0 & 0.000 & 0.153 & 0.146 & 0.154 & 0.145 & 99.2 & 98.6 & 94.2 & 88.2 & 95.1 \\
  8 & 0.5 & 0.4 & 0.0 & 0.000 & 0.105 & 0.105 & 0.105 & 0.102 & 98.5 & 98.5 & 95.7 & 92.5 & 95.8 \\
  8 & 0.8 & 0.4 & 0.0 & 0.000 & 0.147 & 0.143 & 0.151 & 0.143 & 99.3 & 99.0 & 95.0 & 87.1 & 95.9 \\
  8 & 0.5 & 0.6 & 0.0 & 0.000 & 0.097 & 0.095 & 0.095 & 0.090 & 98.4 & 98.1 & 95.5 & 94.3 & 95.7 \\
  8 & 0.8 & 0.6 & 0.0 & 0.000 & 0.115 & 0.112 & 0.116 & 0.112 & 99.1 & 99.0 & 94.5 & 92.1 & 95.0 \\
\hline
  4 & 0.5 & 0.4 & 0.2 & 0.447 & 0.236 & 0.223 & 0.235 & 0.217 & 90.5 & 90.5 & 36.8 & 41.3 & 77.9 \\
  4 & 0.8 & 0.4 & 0.2 & 0.447 & 0.403 & 0.333 & 0.378 & 0.331 & 88.5 & 90.3 & 28.2 & 41.6 & 75.4 \\
  4 & 0.5 & 0.6 & 0.2 & 0.447 & 0.181 & 0.183 & 0.187 & 0.178 & 92.8 & 92.7 & 48.6 & 43.3 & 82.8 \\
  4 & 0.8 & 0.6 & 0.2 & 0.447 & 0.256 & 0.230 & 0.241 & 0.219 & 91.5 & 91.9 & 38.4 & 42.8 & 79.3 \\
  8 & 0.5 & 0.4 & 0.2 & 0.447 & 0.176 & 0.162 & 0.188 & 0.160 & 90.7 & 89.3 & 48.7 & 39.7 & 77.0 \\
  8 & 0.8 & 0.4 & 0.2 & 0.447 & 0.248 & 0.226 & 0.247 & 0.226 & 91.1 & 91.8 & 36.5 & 39.7 & 75.1 \\
  8 & 0.5 & 0.6 & 0.2 & 0.447 & 0.143 & 0.143 & 0.159 & 0.137 & 92.1 & 90.0 & 57.4 & 42.3 & 78.9 \\
  8 & 0.8 & 0.6 & 0.2 & 0.447 & 0.181 & 0.175 & 0.179 & 0.171 & 92.1 & 92.6 & 51.4 & 40.9 & 78.5 \\
\hline
  4 & 0.5 & 0.4 & 0.4 & 0.632 & 0.308 & 0.295 & 0.281 & 0.263 & 91.9 & 92.7 & 28.6 & 27.3 & 72.2 \\
  4 & 0.8 & 0.4 & 0.4 & 0.632 & 0.607 & 0.439 & 0.450 & 0.385 & 89.7 & 92.8 & 19.9 & 27.9 & 69.6 \\
  4 & 0.5 & 0.6 & 0.4 & 0.632 & 0.271 & 0.254 & 0.248 & 0.229 & 94.0 & 94.6 & 36.7 & 28.8 & 73.8 \\
  4 & 0.8 & 0.6 & 0.4 & 0.632 & 0.324 & 0.316 & 0.279 & 0.267 & 93.2 & 94.5 & 29.5 & 28.3 & 71.2 \\
  8 & 0.5 & 0.4 & 0.4 & 0.632 & 0.212 & 0.208 & 0.221 & 0.203 & 92.5 & 92.5 & 40.7 & 24.9 & 66.1 \\
  8 & 0.8 & 0.4 & 0.4 & 0.632 & 0.342 & 0.284 & 0.342 & 0.280 & 91.0 & 92.7 & 27.6 & 24.6 & 65.1 \\
  8 & 0.5 & 0.6 & 0.4 & 0.632 & 0.193 & 0.189 & 0.227 & 0.175 & 94.6 & 90.2 & 47.7 & 25.9 & 66.8 \\
  8 & 0.8 & 0.6 & 0.4 & 0.632 & 0.221 & 0.218 & 0.216 & 0.211 & 93.2 & 94.1 & 38.4 & 26.2 & 68.5 \\
\hline \multicolumn{14}{|l|}{$n=1000$}\\ \hline
  4 & 0.5 & 0.4 & 0.0 & 0.000 & 0.053 & 0.054 & 0.054 & 0.054 & 95.9 & 95.7 & 95.1 & 97.1 & 95.1 \\
  4 & 0.8 & 0.4 & 0.0 & 0.000 & 0.072 & 0.076 & 0.073 & 0.076 & 97.9 & 97.2 & 96.3 & 95.1 & 96.3 \\
  4 & 0.5 & 0.6 & 0.0 & 0.000 & 0.045 & 0.046 & 0.045 & 0.046 & 96.5 & 96.4 & 95.2 & 98.3 & 95.4 \\
  4 & 0.8 & 0.6 & 0.0 & 0.000 & 0.054 & 0.056 & 0.055 & 0.056 & 96.8 & 96.8 & 95.9 & 97.6 & 96.2 \\
  8 & 0.5 & 0.4 & 0.0 & 0.000 & 0.045 & 0.044 & 0.045 & 0.044 & 95.3 & 95.2 & 94.5 & 98.2 & 94.5 \\
  8 & 0.8 & 0.4 & 0.0 & 0.000 & 0.059 & 0.056 & 0.059 & 0.056 & 95.0 & 94.8 & 93.7 & 96.9 & 94.0 \\
  8 & 0.5 & 0.6 & 0.0 & 0.000 & 0.040 & 0.040 & 0.040 & 0.040 & 96.1 & 95.9 & 95.2 & 98.8 & 95.2 \\
  8 & 0.8 & 0.6 & 0.0 & 0.000 & 0.045 & 0.046 & 0.046 & 0.046 & 96.3 & 96.2 & 95.1 & 98.4 & 95.3 \\
\hline
  4 & 0.5 & 0.4 & 0.2 & 0.447 & 0.092 & 0.091 & 0.089 & 0.089 & 94.4 & 95.4 & 5.9 & 41.6 & 79.5 \\
  4 & 0.8 & 0.4 & 0.2 & 0.447 & 0.124 & 0.123 & 0.122 & 0.120 & 93.6 & 94.1 & 3.2 & 42.6 & 80.2 \\
  4 & 0.5 & 0.6 & 0.2 & 0.447 & 0.075 & 0.076 & 0.074 & 0.074 & 94.8 & 94.6 & 12.6 & 43.8 & 79.3 \\
  4 & 0.8 & 0.6 & 0.2 & 0.447 & 0.090 & 0.090 & 0.089 & 0.089 & 96.0 & 95.8 & 8.4 & 43.8 & 81.3 \\
  8 & 0.5 & 0.4 & 0.2 & 0.447 & 0.073 & 0.070 & 0.073 & 0.069 & 93.9 & 94.6 & 18.8 & 39.3 & 77.7 \\
  8 & 0.8 & 0.4 & 0.2 & 0.447 & 0.093 & 0.091 & 0.092 & 0.091 & 93.9 & 94.6 & 8.5 & 38.7 & 76.6 \\
  8 & 0.5 & 0.6 & 0.2 & 0.447 & 0.062 & 0.062 & 0.062 & 0.062 & 94.6 & 94.7 & 28.6 & 41.9 & 79.2 \\
  8 & 0.8 & 0.6 & 0.2 & 0.447 & 0.074 & 0.073 & 0.073 & 0.072 & 95.1 & 94.8 & 17.9 & 41.4 & 80.1 \\
\hline
  4 & 0.5 & 0.4 & 0.4 & 0.632 & 0.118 & 0.118 & 0.112 & 0.109 & 94.3 & 94.9 & 2.3 & 28.9 & 70.9 \\
  4 & 0.8 & 0.4 & 0.4 & 0.632 & 0.156 & 0.155 & 0.149 & 0.143 & 93.4 & 93.8 & 1.6 & 29.1 & 70.4 \\
  4 & 0.5 & 0.6 & 0.4 & 0.632 & 0.093 & 0.097 & 0.089 & 0.092 & 95.7 & 95.9 & 6.8 & 30.5 & 73.9 \\
  4 & 0.8 & 0.6 & 0.4 & 0.632 & 0.116 & 0.115 & 0.109 & 0.108 & 94.7 & 94.7 & 4.0 & 30.4 & 73.9 \\
  8 & 0.5 & 0.4 & 0.4 & 0.632 & 0.088 & 0.089 & 0.086 & 0.087 & 95.5 & 96.2 & 7.9 & 24.6 & 68.0 \\
  8 & 0.8 & 0.4 & 0.4 & 0.632 & 0.114 & 0.114 & 0.110 & 0.111 & 93.3 & 94.5 & 2.6 & 24.5 & 65.3 \\
  8 & 0.5 & 0.6 & 0.4 & 0.632 & 0.080 & 0.080 & 0.078 & 0.077 & 95.1 & 94.6 & 19.8 & 26.3 & 69.2 \\
  8 & 0.8 & 0.6 & 0.4 & 0.632 & 0.088 & 0.092 & 0.086 & 0.089 & 95.5 & 95.8 & 10.5 & 26.3 & 70.8 \\
   \hline
\multicolumn{14}{l}{\emph{Note:} $p$ denotes number of measurement
items $Y_{j}$, $\pi_{Y}$ marginal proportion of $Y_{j}=1$, }\\
\multicolumn{14}{l}{
\hspace*{2em}$R_{Y}^{2}$ and $R^{2}_{\eta}$ the $R^{2}$ statistics in models for
$Y_{j}$ and $\eta_{2}$, and $\beta_{1}$ the true value of $\beta_{1}$.}\\
\multicolumn{14}{l}{
$\dagger$ Average \% of the variance of 2-step estimator accounted
for by step-2 variance alone,}\\
\multicolumn{14}{l}{
\hspace*{1em} and coverage of 95\% confidence interval if only this variance is
included.}\\
\end{tabular}
}}
\label{t_sim_se}
\end{table}
\clearpage
\restoregeometry

\subsection{Further simulations}
\label{ss_simulation_2}

Here we describe some further simulations for different
settings. Path diagrams for these models are shown in Figures A.1--A.2, and
results in Tables A.1--A.25, in the supplementary appendix. The
discussion is short, because we focus on results that extend
or deviate from the conclusions from the first set of simulations in
Section~\ref{ss_simulation_1}.
The following situations are included here:
\begin{itemize}
\item
\emph{Observed covariate, latent response} (\emph{simulation case B}). Structural model is as in case A (equation \ref{sim_model}), except that
$\eta_{i1}$ is replaced by an observed covariate $X_{i}$. Three
distributions for $X_{i}$ are used: $X_{i}\sim N(0,1)$ [\emph{case B1}],
$X_{i} \sim 2\,\text{Bernoulli}(0.5)-1$ [\emph{B2}], and $X_{i}\sim
\text{Gamma}(\alpha,\beta)$ with shape $\alpha=0.06$ and rate $\beta=1$,
scaled to have sample mean 0 and variance 1 [\emph{B3}].
\item \emph{Latent covariate, observed response} (\emph{case C}).
Structural model is as in case A, except
that $\eta_{i2}$ is replaced by an observed response
which is here labelled  $Z_{i}$. Two
conditional distributions for $Z_{i}$ given $\eta_{1i}$ (i.e.\
distributions for $\epsilon_{i}$) are considered,
a normal distribution [\emph{case C1}] and
a skew-normal distribution \citep{azzalini+capitanio14}
set to have a skewness coefficient of just under 1
[\emph{C2}], with $Z_{i}$ having marginal mean of 0 and marginal
variance~1 in all settings.
\item
\emph{Ordinal measurement items} (\emph{case D}). Structural model is
the same as in case A, but each measurement item $Y_{ikj}$ has 4 rather
than 2 categories. Measurement models are again derived from a linear
model for a latent continuous $Y_{ijk}^{*}$,
here resulting in an ordinal logistic
measurement model for each $Y_{ijk}$. Its loading parameter $\lambda$ is
set as in case A. The measurement intercepts are such that the
marginal probabilities of the four categories of each $Y_{ikj}$ are
(0.25,0.25,0.25,0.25) when $R^{2}_{Y}=0.6$ and about
(0.20,0.30,0.30,0.20) when $R^{2}_{Y}=0.4$ (both of these correspond to the
setting $\pi_{Y}=0.5$ in case A).
\item
\emph{Three latent variables} (\emph{case E}). The structural model has two
parts, $\eta_{2i}=\alpha_{0}+\alpha_{1}\eta_{1i}+\epsilon_{2i}$ and
$\eta_{3i}=\beta_{0}+\beta_{1}\eta_{1i}+\beta_{2}\eta_{2i}+\epsilon_{3i}$
for a further latent variable $\eta_{3i}$. The latent variables
are each marginally distributed as $N(0,1)$, and each measured by $p$
separate binary items $Y_{ikj}$. We set $R_{Y}^{2}=0.4$
for both $\eta_{2i}$ and $\eta_{3i}$. What we vary here is the relative
sizes of the coefficients $\beta_{1}$ and $\beta_{2}$, so that the
proportion of the ``indirect effect'' of $\eta_{1i}$ on $\eta_{3i}$ via $\eta_{2i}$
($\alpha_{1}\beta_{2}$)
out of the ``total effect'' ($\beta_{1}+\alpha_{1}\beta_{2}$) is 0, 1/3 or
1/2. Here and for cases F and G we report simulation results for estimates of $\beta_{2}$.
\item
\emph{Measurement model with cross-loadings} (\emph{case F}). Structural
model is the same as in case~E. The difference is that the items
$Y_{i23}$ and $Y_{i24}$, which were previously measures of $\eta_{2i}$
only, are here also measures of $\eta_{1i}$. The measurement loadings
for $Y_{i23}$ and $Y_{i24}$ given $\eta_{1i}$ have the same value
$\lambda$ as all other loadings in the measurement model. In step 1 of
two-step estimation the measurement model for $\eta_{1}$ and $\eta_{2}$
was accordingly estimated for these variables together, including the
cross-loadings.
\item
\emph{Misspecified measurement model}(\emph{case G}). The data-generating model
is as in case F, but the measurement model is estimated as for case E,
i.e.\ incorrectly omitting the cross-loadings from $\eta_{1i}$ to
$Y_{i23}$ and $Y_{i24}$.
\end{itemize}

Cases B and C include the same 48 parameter settings as in case A, but
cases D--F omit some of them for simplicity. There we consider only
settings with the smaller number of $p=4$ measurement items for each
latent variable, and balanced items probabilities with $\pi_{Y}=0.5$.
What is then left to vary are the strength of the measurement models
(with $R^{2}_{Y}=0.4, 0.6$), strengths of the associations in the
structural model (with $R^{2}_{\eta}=0,0.2,0.4$ in case D, and as
described above for cases E--G), and sample sizes ($n=200,1000$). In
case E we also included $n=100$ to further examine the performance of
the estimators in such small samples. The results are again based on
1000 simulated datasets in each setting.

Consider first some of the results for parameters of the measurement
models. Supplementary Table A.1 shows simulation means of the estimated
measurement loadings ($\lambda_{kj}$) and table A.2 means of the
estimated measurement intercepts ($\tau_{kj}$) from one-step estimation
and step 1 of two-step estimation in simulation cases A, B2 and B3.
These means are over both the 1000 simulations and all $p$  or (in case
A) $2p$ loadings or intercepts in the model. The true values of all
loadings and intercepts are 1 and 0 respectively in each case. In case A
these are results for maximum likelihood (ML) estimates of these
parameters for correctly specified models, two one-trait models (for two-step estimation) or one
two-trait model (for one-step estimation). They are, as expected,
approximately unbiased when $n=1000$. With the smaller sample size of
$n=200$, however, the estimates show some bias. The simulation medians
(which are not shown) are even then very close to the true values, so the
observed bias is entirely due to skewness of the sampling distributions
of these estimates in small samples.

Measurement models in cases B2 and B3 are of interest because they are
examples of the situation (as discussed at the end of Section
\ref{ss_estimation_variance}) where step 1 of two-step estimation is
distributionally misspecified because a non-normal covariate $X$ induces
a non-normal marginal distribution of a latent variable (here
$\eta_{2}$). In contrast, the model for one-step estimation (and
two-step when $R^{2}_{\eta}=0$) is correctly specified.
Previous literature on misspecified latent trait models suggests that
this should
be a minor deviation in case B2, because the Bernoulli distribution of
$X$ implies a symmetric two-component mixture model for~$\eta_{2}$.
Case B3 could be more difficult, because there the marginal distribution
of $\eta_{2}$ is skewed. It has a skewness coefficient of about 0.7 when
$R^{2}_{\eta}=0.2$ and about 2.0 (i.e.\ comparable to an exponential
distribution) when $R^{2}_{\eta}=0.4$. However, the results do not show
any new problems in either of these cases: two-step estimates still
behave similarly with the one-step estimates. This
is the case even though case B3 was designed to be so extreme that
it should not really occur in practice: the observed $X$ had a skewness
coefficient of 8.2, which any data analyst would be likely to notice and
address by taking a (log) transformation of $X$. It thus seems that it is difficult
for a non-normal covariate to induce a serious bias in the estimated
measurement parameters from two-step estimation. It should then be even
more difficult for this to lead to bias in the estimated structural
parameters from step 2 --- which are the focus of interest --- and this
is indeed what we find below.

Consider then results for estimates of structural regression
coefficients. They are reported in supplementary Tables A.3--A.25.
First, in cases  B1, C1, D, E and F (and B2 and B3 when
$R^{2}_{\eta}=0$) the estimation model is correctly specified. The broad
conclusion is that most of these results are similar to the results in
case~A. Both estimates perform mostly well, and two-step estimates at
least as well as one-step estimates. These conclusions are thus not
substantially affected when the models are larger than in case~A, in
having ordinal items, more latent variables, or non-simple measurement
models. This is unsurprising, given that the large-sample properties of
the estimators (as discussed in Sections \ref{ss_estimation_variance}
and \ref{sss_estimation_onethree_one}) should not be affected by the
complexity of the model in these (or other) respects.

There are, however, some new findings in these cases. First, in case C1
the estimates perform very similarly even in the most difficult
settings (where the two-step ones had some extreme values in case A). A
possible explanation is that when the latent $\eta_{1}$ is only
an explanatory variable, a weak estimated measurement model for it tends
to attenuate rather than exaggerate the estimate of~$\beta_{1}$,
preventing occasional extreme estimates. Second, in cases with three
latent variables (E and F) there are several settings where the one-step
estimates have a much larger bias and RMSE than the two-step estimates.
This is due to occasional extreme values of the one-step estimates, as
can be seen by observing that the MAEs and simulation medians (the
latter are shown separately in Tables A.18 and A.21) are similar even in these
settings. These extremes occur most prominently with the largest sample
size ($n=1000$), which shows that they are not due to poor
small-sample behaviour. Instead, they indicate
numerical difficulties with one-step estimation, and of the estimation algorithm
failing to converge to (or near) the maximum likelihood estimate. The
code for the simulation was set to increase the number of starting
values for datasets where the initial settings were not enough, but
here the procedure we used for this was not always sufficient. Thus we
could probably have removed some of these extreme estimates by
manually increasing the starting values further. However, we left them
in the results because they illustrate another difference between the
estimators, which is that the two-step approach is computationally
substantially more manageable than the one-step one. For the two-step
estimates there were also some numerical extreme values in cases E (with
$n=100$) and F, but for the estimated standard errors rather than
point estimates of $\beta_{2}$.

In case E we also included a small sample size of $n=100$, where
asymptotic properties of the estimators are least appropriate. This
would arguably be uncomfortably small for substantive applications of
models of this type. Even here the estimators perform mostly similarly
and mostly well, although inevitably less well than with larger samples,
and again with some extreme estimates. The two-step estimates have
consistently the smaller RMSEs and MAEs, suggesting that they may have
somewhat better properties in very small samples.

Consider then cases where the estimation model is misspecified in some
way (B2, B3, C2, and G). There is essentially no difference between C2
and C1, suggesting that misspecification of the distribution of an
observed response $Z$ is inconsequential. In
cases B2 and B3 there is misspecification in step 1 of two-step
estimation, because the distribution of the covariate $X$ is not normal.
As discusssed above, estimated measurement models were
nevertheless not seriously affected. It follows that two-step estimates
of the structural regression coefficient are also unaffected. They
still behave similarly to one-step estimates, for which this model
is correctly specified. In case B2 the results
are essentially similar to the correctly specified B1.
In case B3 the estimates are noticeably biased, at least
with $n=200$. This, however, affects both
estimates --- and is indeed worse for one-step estimates --- so it is
not due to the distributional misspecification in two-step estimation.
Instead, it must be caused by the extreme skeweness of $X$, which hurts
the small-sample performance of both estimates.

Finally, in case F the estimation model is misspecified for both
estimators. Both of them give biased estimates of all the structural
coefficients in this case, with $\beta_{2}$ being overestimated. The
bias is noticeably smaller for two-step estimators, but still
substantial. This is not surprising. Two-step estimation may reduce
bias in estimated \emph{parameters} of the measurement
models because of the way it estimates them separately, but it still
uses the same misspecifified \emph{form} of the model (i.e.\ of which
items measure which latent variables). So we would not in general expect
it to offer much additional protection against bias from model
misspecification.

The conclusions from these simulations are broadly similar to previous
ones for other types of models by \cite{bakk-kuha} and
\cite{rosseel+loh24}. In all of these contexts two-step and one-step
estimates mostly perform very similarly. Differences between them emerge
in difficult situations where the measurement models are weak and sample
sizes are small. Here there is some variation between the findings of
the different studies. In the latent class examples of \cite{bakk-kuha},
two-step estimation does relatively poorly in difficult settings,
possibly because an observed covariate or response provides particularly
useful stabilisation for one-step estimates there. In contrast, in the
simulations of  structural equation models by \cite{rosseel+loh24} it
is one-step estimation that performs much less well in the most
difficult settings. In our simulations the differences in the hardest
situations were somewhat smaller than in these previous studies. It
should be noted that the exact simulation settings are not easily
comparable. For example, \cite{rosseel+loh24} considered models with
even more latent factors (five rather than three), which may be
particularly challenging for the small-sample performance of one-step
estimation (it would also be computationally demanding for it for
latent trait models). They also considered various cases where the
analysis models were misspecified. The conclusions were qualitatively
the same as in our simulations, i.e.\ that both estimates were then
biased but the two-step estimates often somewhat less so.

\subsection{Comments on three-step estimates in the simulations}
\label{ss_simulation_3step}

The simulation tables include also results for naive three-step
estimates which replace the latent variables $\eta_{ik}$ with their
empirical Bayes predictions $\tilde{\eta}_{ik}$. They are typically
badly biased for the true structural parameters. This is to be expected,
based on the general results discussed  in Section
\ref{ss_estimation_onethree}. These naive estimates are clearly an
inadequate way of estimating the structural model for $\eta_{ik}$. As
noted in Section \ref{sss_estimation_onethree_naive3}, it is more
constructive to regard them as estimates for a model for
$\tilde{\eta}_{ik}$, treated as a model of interest in its own right.

An exception here is case C, where the only latent variable $\eta_{1}$
appears as the only explanatory variable in the model. As observed in
Section \ref{sss_estimation_onethree_adjusted3}, naive three-step
estimation which uses the empirical Bayes predictions then becomes a
form of adjusted three-step estimation which avoids measurement error
bias. Accordingly, estimates of $\beta_{1}$ in cases C1 and C2 are
essentially unbiased and very similar to the two-step and one-step
estimates (see supplementary Tables A.11 and A.13). This is not the case
for the other simulations here, and generalising this type of adjustment to
them would not be easy.

The three-step estimates perform relatively well in case G, where the
estimation model is misspecified because of omitted cross-loadings and
where two-step and one-step estimates are clearly biased. The naive
three-step estimates avoid this misspecification bias. This is because
they \emph{only} require valid estimates of the measurement models for
each latent variable individually, and those are here correctly
estimated (even though it includes cross-loadings, the model in case G
implies also a one-trait model for $Y_{21}$--$Y_{24}$ given $\eta_{2}$).
However, the three-step estimates still have their usual measurement
error bias in this case.

\section{Application: Extrinsic and intrinsic work values}
\label{s_example}

Here we apply the methods to a real-data example. We use it,
in particular, to further explore two topics which were not the focus
of the simulations in Section \ref{s_simulation}: the possibility of interpretative
confounding and computational demands of the different estimators.

The example concerns individuals' \emph{work value orientations},
conceptualised in two dimensions: intrinsic and extrinsic work values
(for more information about substantive research in this area, see for
example \cite{gesthuizenetal19} and \cite{bacheretal22}, and references
cited therein). We use data from the fifth wave of the European Values
Study (EVS), conducted in 2017--20 \citep{EVS22}. The survey
includes six items on work values, introduced (in the English-language
version) by \emph{Here are some aspects of a job that people say are
important. Please look at them and tell me which ones you personally
think are important in a job?} The responses are binary, coded as 1 if
the respondent mentioned that aspect and 0 if they did not (and with
``Don't know'' responses coded as missing). Three of the items are
treated as measures of extrinsic work values --- ``Good pay'' (labelled
\emph{Pay} in the tables below), ``Good hours'' (\emph{Hours}), and
``Generous holidays'' (\emph{Holidays}) --- and three of intrinsic work
values --- ``An opportunity to use initiative'' (\emph{Initiative}), ``A
job in which you feel you can achieve something'' (\emph{Achievement}),
and ``A responsible job'' (\emph{Responsibility}). For more discussion
of the measurement of work values in the EVS, see \cite{gesthuizenetal19}.

To provide a focus for this illustrative application, we consider
specifically gender differences in work value orientations between men
and women (in EVS, gender is coded as a binary variable). This question
was also examined, for example, by \cite{bacheretal22}, using data for
Austria from a different survey (The Social Survey of Austria).

We carry out two analyses. The second of them will be a cross-sectional
analysis of all 36 countries in the EVS data. But first we consider just
one country, the Netherlands. In addition to the work value items and
gender, a set of covariates are included: the respondent's age (in
categories: 15--29, 30--49 or 50-- years), whether they are in a
registered partnership and/or live with a partner (yes/no), the age of the youngest
person in the household (0--5, 6-17, or older), respondent's highest
level of education (less than upper secondary, upper secondary, or higher), occupation-based
social class (European Socioeconomic Classification [ESeC08]; grouped by
type of employment regulation of the occupation as Service relationship
[ESeC08 classes 1,2], Mixed [3,6], Labour contract [7,8,9] or
Self-employed [4,5]), whether they are currently in paid employment
(yes/no), and the highest education level of the respondent's parents
(coded as for own education). These parallel, as far as possible with
the EVS data, the covariates used by \cite{bacheretal22}. This analysis
dataset for the Netherlands includes $n=1374$
respondents with observed values of all the covariates and at least
one of the six value items.

In the notation of Section \ref{s_models}, the structural model
considered here is $p(\boldsymbol{\eta}_{i(1)}|\mathbf{X}_{i};
\boldsymbol{\theta}_{21})$, where
$\boldsymbol{\eta}_{i(1)}=(\eta_{i1},\eta_{i2})'$ and $\eta_{i1}$ stands
for extrinsic and $\eta_{i2}$ for intrinsic work values. They are
responses to covariates $\mathbf{X}_{i}$, and there are no observed
response variables. As defined in (\ref{eta_model1}),
the model includes separate linear models for $\eta_{i1}$ and
$\eta_{i2}$, plus their conditional association which we report as the
correlation $\rho=\text{corr}(\eta_{i1},\eta_{i2}|\boldsymbol{X}_{i})$.
The two latent variables are measured by the corresponding 3 items each,
with logistic measurement models (\ref{logit}).

\newpage
We consider three models, with different sets of covariates: Model (1)
which includes only gender, Model (2) which adds respondent's age,
partnership status and age of the youngest person in the household, and Model
(3) which includes all the covariates. Estimates
are shown in Table \ref{t_example_NL}. This includes only the
loading parameters ($\lambda_{kj}$) of the measurement models, the
regression coefficient of gender (as dummy variable for men) and the
conditional correlation $\rho$. The full sets of structural coefficients
are shown in supplementary Tables B.1 and B.2.

\begin{table}
\caption{Models for extrinsic and intrinsic work values given gender,
controlling for different
sets of covariates,
estimated from EVS2017 data for Netherlands. The table shows two-step
(``2-st.''), one-step
(``1-st.'')
and naive three-step
(``3-st.'')
estimates. ``Step 1'' denotes the
estimated measurement loadings from step 1 of two-step estimation, which
are used for two-step and (for latent trait scores of) three-step estimation of all three models.
}
{\small{\hspace*{-3em}
\begin{tabular}{|l|r|rrr|rrr|rrr|}\hline
& &
\multicolumn{3}{|c|}{Model (1)$^{*}$} &
\multicolumn{3}{|c|}{Model (2)$^{*}$} &
\multicolumn{3}{|c|}{Model (3)$^{*}$} \\
 & Step 1 &
2-st.\ & 1-st.\ & 3-st. &
2-st.\ & 1-st.\ & 3-st. &
2-st.\ & 1-st.\ & 3-st. \\ \hline
\multicolumn{11}{|l|}{\emph{Model for Extrinsic work values}} \\ \hline
\multicolumn{11}{|l|}{Measurement loadings:} \\[.3ex]
\emph{Pay}      & 1.000 &  & 1.000 &  &  & 1.000 &  &  & 1.000 &  \\
\emph{Hours}    & 1.161 &  & 1.101 &  &  & 1.148 &  &  & 1.190 &  \\
\emph{Holidays} & 3.080 &  & 3.123 &  &  & 2.586 &  &  & 2.573 &  \\
\hline
\multicolumn{11}{|l|}{Structural model: coefficient of Gender (man)} \\[.3ex]
  Male &  & -0.042 & -0.033 & -0.033 & 0.025 & 0.022 & 0.024 & 0.018 & 0.009 & 0.018 \\
  (s.e.) &  & (0.103) & (0.104) & (0.082) & (0.101) & (0.104) &
  (0.080) &
  (0.102) & (0.104) & (0.081) \\ \hline\hline
\multicolumn{11}{|l|}{\emph{Model for Intrinsic work values}} \\ \hline
\multicolumn{11}{|l|}{Measurement loadings:} \\[.3ex]
\emph{Initiative} & 1.000 &  & 1.000 &  &  & 1.000 &  &  & 1.000 &  \\
\emph{Achievement} & 0.431 &  & 0.453 &  &  & 0.500 &  &  & 0.556 &  \\
\emph{Responsibility} & 0.563 &  & 0.589 &  &  & 0.631 &  &  & 0.804 &  \\
\hline
\multicolumn{11}{|l|}{Structural model: coefficient of Gender (man)} \\[.3ex]
  Male &  & 0.607 & 0.603 & 0.460 & 0.657 & 0.638 & 0.500 & 0.516 & 0.506 & 0.400 \\
  (s.e.) &  & (0.193) & (0.186) & (0.142) & (0.194) & (0.176) & (0.142)
  & (0.188) & (0.151) & (0.139) \\
\hline\hline
\multicolumn{11}{|l|}{\emph{Correlation of Extrinsic and
Intrinsic work values given covariates}} \\[.3ex]
  &  & 0.550 & 0.551 & 0.345 & 0.543 & 0.549 & 0.334 & 0.559 & 0.567 & 0.334 \\
\hline
\multicolumn{11}{l}{\footnotesize{* Model (1) includes gender as
covariate, Model (2) also
respondent's age, partnership status and age of youngest person
}}\\
\multicolumn{11}{l}{\footnotesize{%\hspace*{1em}
in the household, and Model (3) also
respondent's education, social class and work status,
and their parents'
education.}} \\
\multicolumn{11}{l}{\footnotesize{
The full sets of estimated coefficients of the structural model are
shown in the supplementary appendix.}}
\end{tabular}
}}
\label{t_example_NL}
\end{table}

Consider first the estimated measurement models in Table
\ref{t_example_NL}. The column ``Step 1'' shows the measurement loadings
that are obtained from step 1 of the two-step method. The measurement
parameters from it are used for step 2 for all choices of the set of
covariates $\mathbf{X}_{i}$. They are also used to calculate the trait
scores that are used for three-step estimation of all the models.

In contrast, the one-step method estimates the measurement model anew
whenever the structural model changes; these estimated loadings are
shown under the ``1-st.'' columns of the table. This raises the
possibility of interpretational confounding, where the definition of a
latent variable is affected not only by its indicators but also by its
predictors. Previously, \cite{bakk-kuha} examined this for latent class
models. In one of their examples, different choices of covariates
flipped the measurement model between two qualitatively quite different
configurations of the classes, so that comparisons of one-step estimates
between different structural models became effectively meaningless. The
differences are not similarly dramatic in our example here. All of the
one-step estimates of the loadings are positive, indicating that for
both the extrinsic and intrinsic factors higher values correspond to a
respondent placing higher importance on that dimension of work values,
and the loadings of different items are in the same order of size in
each case. This is perhaps not surprising, in that we might expect that
for continuous latent variables such changes in the measurement model
will be quantitative shifts rather than qualitative jumps. Nevertheless,
in this more subtle way the one-step estimates for the three models in
Table \ref{t_example_NL} are still for three slightly different response
variables which differ both from each other and from the response in all
of the two-step estimates.

Consider then the estimated structural models for the Netherlands.
In terms of the settings of the simulations in Section
\ref{s_simulation}, these data have $n=1374$ and $R^{2}_{\eta}$
around 0.1 for both latent factors, each of them measured by $p=3$
items with $R^{2}_{Y}$ of around 0.3--0.8 and $\pi_{Y}$ of 0.4--0.7
(different for different items). These place this example around the
middle of the simulation settings in terms of the strength of the
measurement model, but with a somewhat larger sample size than even the
larger $n=1000$ considered there. The simulations suggest that
the two-step and one-step estimates should both behave well.
Here they are also similar to each other, with somewhat higher standard errors for
the two-step estimates. For the naive three-step estimates, for better
comparability we first re-scaled the factor scores so that their
variances equal the estimated marginal variances of the corresponding
latent variables from step 1 of two-step estimation. Even after this,
the three-step estimates of regression coefficients are attenuated toward zero, as would be
expected given the theoretical and simulation results above.

The results in Table \ref{t_example_NL} indicate that individuals'
intrinsic and extrinsic work values are strongly positively correlated
even given covariates. Between genders, there is essentially no
difference in extrinsic values, but for intrinsic ones men indicate on
average significantly (at the 5\% level) higher level of importance on
them. Some other significant associations with covariates are also
found, as shown by the estimated coefficients in supplementary Tables
B.1 and B.2. The importance of extrinsic rewards of a job (pay, hours
and holidays) tends to be higher for respondents who are younger, have
young children at home, or are currently in paid employment. For
intrinsic values (related to the scope of initiative, achievement and
responsibility), the expected levels are higher for individuals who have
(or whose parents have) higher levels education and who are in
professional, administrative or managerial occupations.

Summarising previous literature on gender differences in work values,
\cite{bacheretal22} observed that ``[...]intrinsic work values,
especially altruistic or social work values, seem still to be more
important for women than men, at least in some countries and/or in
combination with other factors[...]'' and that ``Gender differences in
extrinsic work value orientations are absent or smaller than those for
intrinsic work value orientations''. In their analysis in Austria,
they also found higher levels of intrinsic values for women. Our
conclusion from the Netherlands is the opposite. This may be because the
three intrinsic items in EVS do not include ones of ``altruistic or
social work'' kind. But it could also reflect variation
between different country contexts. To examine this further, we can use
the EVS to carry out a comparative cross-national analysis. Here we
include as covariates the respondents' gender, age, education, social
class and work status; the other covariates were omitted from this
illustrative example because they had large proportions of missing data
in some countries. The list of the 36 countries and their sample sizes
in this analysis are shown in supplementary Tables B.3 and B.4.

\begin{figure}[t]
\caption{Estimates of the coefficient (with 95\% confidence
intervals) of
dummy variable for men in models for intrinsic work values (circles and
solid lines) and extrinsic work values (triangles and dashed lines).
Two-step estimates obtained separately for each country in the 2017
European Values Survey, but with the same measurement model for each
country estimated from pooled data in step 1.}

\includegraphics[width=.95\textwidth]{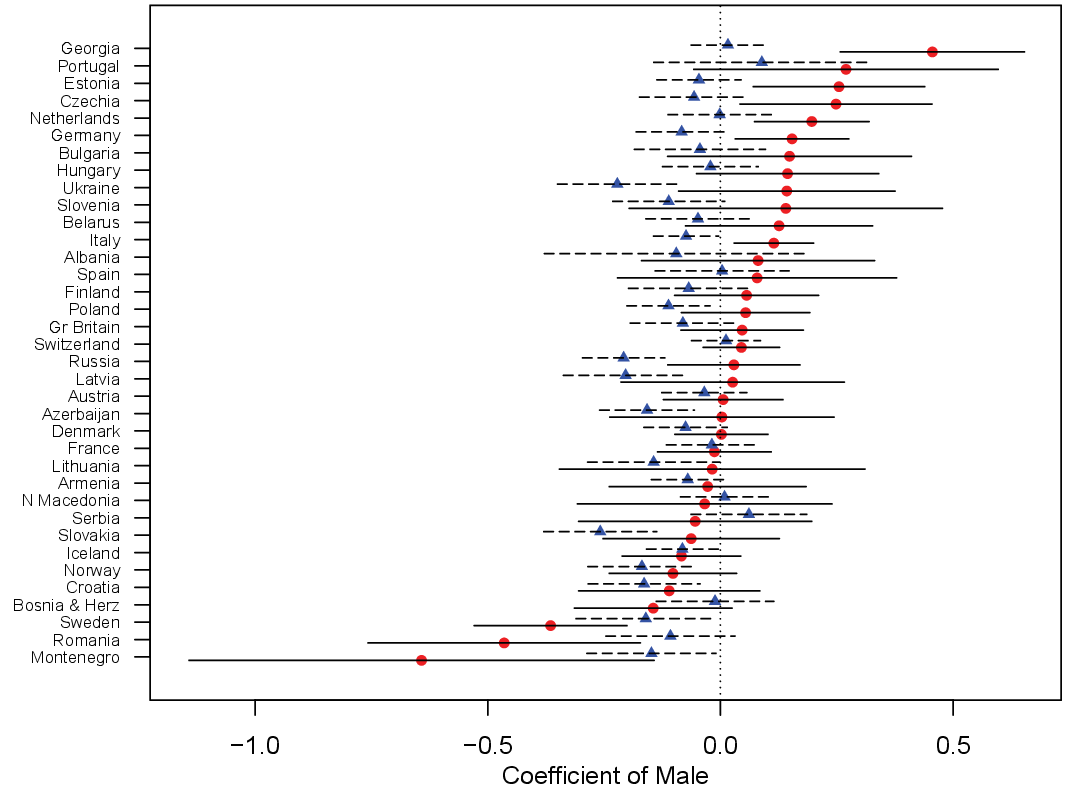}

\vspace*{-1ex}
{\scriptsize{\emph{Note}:
The coefficients are expressed on a scale where the residual
variance of both factors for the Netherlands is 1.
The models also include as covariates the respondent's age, education,
social class and employment status (working
vs.\ not).
}}
\label{f_countries}
\end{figure}

We allow all parameters of the structural model (regression coefficients
and residual variances and covariance) to have different values in each
country. Every measurement parameter, in contrast, is
constrained to have the same value in all of them, to specify
measurement equivalence across the countries. We note that because the
step-1 estimates are thus based on much more data than the step-2
estimates (with ca.\ 59,000 observations in the pooled data but
1000--2000 in each country), step 1 contributes essentially nothing to
the standard errors of the estimated structural coefficients. So in this
example we could safely treat the measurement parameters from step 1 as
known and omit the term for them from the estimated standard errors.

Estimates of gender differences in the work values in each country are
shown in supplementary Tables B.3 and B.4. We focus here on the two-step
estimates, which are also shown in Figure~\ref{f_countries}. For
extrinsic values, most of the point estimates indicate higher average
importance of them for women, but these differences are small and mostly
not statistically significant (at the 5\% level). For intrinsic values
there are a few more countries with significant differences, but they
include ones in both directions and have no very clear geographic
regularities in the variation across countries. In other words, these results
do not show strong evidence of consistent gender differences in work
value orientations, suggesting at best small differences with much
cross-national variation in them.

As mentioned at the start of this this section, this application example
also allowed us to examine two topics in the relative performance of
two-step estimation. For the analysis for the Netherlands
above, we discussed the level of interpretative confounding in
one-step estimates. The second, cross-national analysis provides an
extreme illustration of the difference in computational convenience
between two- and one-step estimation. The model has 836 estimable
parameters. One-step estimates have to be calculated for all of them at
once, from the pooled dataset of all the countries. On our computer this
took 4.5 hours, using just one starting value. Two-step estimation, in
contrast, splits the task in a way which considerably simplifies it. In
its step 1, the measurement models are estimated from the pooled
dataset, as two one-trait models for three binary items each and
without any covariates. Step 2 can then be done separately for each
country, because their structural parameters are distinct. Even with
multiple starting values (\texttt{Starts=30 10} in Mplus syntax) this
took just 1-2 minutes for both step 1 and each country in step 2, for a
total of a little over 0.5 hours for the whole model.

\section{Conclusions and discussion}
\label{s_discussion}

Two-step estimation is applicable to any
types of latent variable models where the measurement and structural
parameters are distinct and the focus of interest is on the structural
model. In this paper we have described it for the broad family of latent
trait models, where continuous latent variables are measured by observed
categorical indicators. In our simulations and examples, two-step
estimation performed very well for them, as was also expected on
theoretical grounds. We compared it with one-step estimation which
estimates all model parameters at once. The two approaches behaved
generally similarly, with two-step estimates losing next to no
efficiency in large samples and often having somewhat better properties
in small samples. This is in line with previous findings on two-step
estimation for other families of models, in particular latent class
analysis of categorical variables and structural equation modelling of
continuous variables. Together these studies provide increasingly strong
evidence and reassurance that the two-step method is a useful and
convenient approach to estimation of latent variable models, and an
attractive alternative to the existing one-step and three-step methods.

We also argue that two-step estimation has two major advantages in
general. One of them is practical and computational
convenience compared to one-step estimation.
This may be particularly important for latent trait
models, because they are more often used with complex structural models
than are latent class models, and are also inherently more
computationally demanding than structural equation models. Another
important characteristic of the two-step approach is that it avoids
interpretational confounding, where the effective definition of the
latent variables is affected by the specification of the structural
model. This may be more consequential for latent class models, where the
interpretation of a categorical latent class variable can change in a
discontinuous way, than for latent trait models and structural equation
models which have continuous latent variables.

To build on these promising results so far, different directions of
further research on different elements of two-step estimation could be
pursued. The accessibility of the approach would be much improved by
integrated implementation of it (including calculation of correct standard
errors of the estimates) in general-purpose software for latent variable
models, including latent trait models. This would
extend and complement the procedures that are already available
for latent class and structural equation modelling. For better
understanding of the scope of two-step estimation, its implementation
and performance could be further studied for other categories of models,
for example ones which combine different types of latent variables and
measurement items, for extensions such as multilevel models and ones
with non-equivalent measurement models, or for other methods of
estimation than likelihood-based estimation. These and other areas of
further research remain to be explored.

\vspace*{1ex}
\textbf{\Large{Declarations}}

Funding: No funding was received for conducting this study.\\
Competing interests: The authors have no relevant financial or non-financial interests to disclose.

%\bibliographystyle{chicago}
%\bibliography{irt}

\vspace*{-1ex}

%\begin{comment}

\newpage

\setcounter{page}{1}
\renewcommand{\thepage}{A.\arabic{page}}

\appendix

\begin{center}
{\huge{
\textbf{Two-step estimation of latent trait models}

\textbf{Supplementary appendices}
}}
\end{center}

\vspace*{3ex}

These supplementary materials provide the following information:

\section{Results of the simulations (Section 4 of the paper)}
\label{s_a_sims}

\begin{itemize}
\item
Path diagrams that represent the cases that
are considered in the simulations of Section~4 of the paper are shown in
Figures A.1 and A.2:
\begin{itemize}
\item
Cases with 1 or 2 latent variables (simulation cases A--D) in Figure
A.1.
\item
Cases with 3 latent variables (simulation cases E--G) in Figure
A.2.
\end{itemize}
\item
Results of the simulations for point estimates of measurement parameters in
simulation cases A, B2 and B3 are given in Tables A.1 (for the
measurement loadings) and A.2 (for the measurement intercepts):
\item
Results of the simulations for estimates of structural regression
coefficients are given in Tables A.3--A.25:
\begin{itemize}
\item Case A (latent covariate, latent response)
in Tables A.2 (point estimates) and A.3 (standard errors and
confidence intervals). These tables are also shown in Section
4 of the main paper, as Tables 1 and 2 respectively.
\item Case B1, B2 and B3 (observed covariate, latent response)
in Tables A.5, A.7, and A.9 (point estimates)and A.6, A.8, and A.10 (standard errors and
confidence intervals).
\item Cases C1 and C2 (latent covariate, observed response)
in Tables A.11 and A.13 (point estimates) and A.12 and A.14 (standard errors and
confidence intervals).
\item Case D (like case A, but with ordinal measurement items)
in Tables A.15 (point estimates) and A.16 (standard errors and
confidence intervals).
\item
Case E (three latent variables) in
in Tables A.17 (and with medians rather than means of the estimates in A.18)
and A.19 (standard errors and confidence intervals).
\item
Case F (like case E, but with cross-loadings in the measurement model) in
in Tables A.20 (and with medians rather than means of the estimates in
A.21) and A.22 (standard errors and confidence intervals).
\item
Case G (misspecified model: true model as in case F, but estimation model as in case E) in
in Tables A.23 (and with medians rather than means of the estimates in
A.24) and A.25 (standard errors and confidence intervals).
\end{itemize}
\end{itemize}

\newpage
\section{Results of the example (Section 5 of the paper)}
\label{s_a_example}

Further results of the applied example discussed in Section 5
are given Tables B.1--B.4:

\begin{itemize}
\item
Full sets of coefficients for the structural models for
Netherlands in Tables B.1--B.2.
\begin{itemize}
\item
Of these, the results for the coefficient of the dummy variable for Male
are also shown in Section
5, as part of Table 3.
\end{itemize}
\item
The list of countries in the cross-national analysis, their sample
sizes, and estimated coefficients of the dummy variable for Male in each
of them in Tables B.3--B.4.
\begin{itemize}
\item
Of these, the two-step estimates of the coefficients and their 95\%
confidence intervals are also displayed in Figure 1 in Section 5
(with the further
standardisation that there the coefficients are expressed on a scale where the residual
variance of both intrinsic and extrinsic values for the Netherlands is 1).
\end{itemize}
\end{itemize}

\newpage
\setcounter{figure}{0}
\renewcommand{\thefigure}{A.\arabic{figure}}

\begin{figure}
\caption{Path diagrams representing cases A--D that are considered
in the simulation studies of Section 4 of the paper. Here the dashed
arrows represent the measurement model that is estimated in step 1 of
two-step estimation, and the solid arrow the structural model
(regression model for
$\eta_{2}$ given $\eta_{1}$,
$\eta_{2}$ given $X$, or
$Z$ given $\eta_{1}$)
that is estimated in its step 2.
%Cases A and D differ in that the items $Y$ are binary in A but ordinal in D.
}
\begin{center}
\vspace*{3ex}
\includegraphics[width=.7\textwidth]{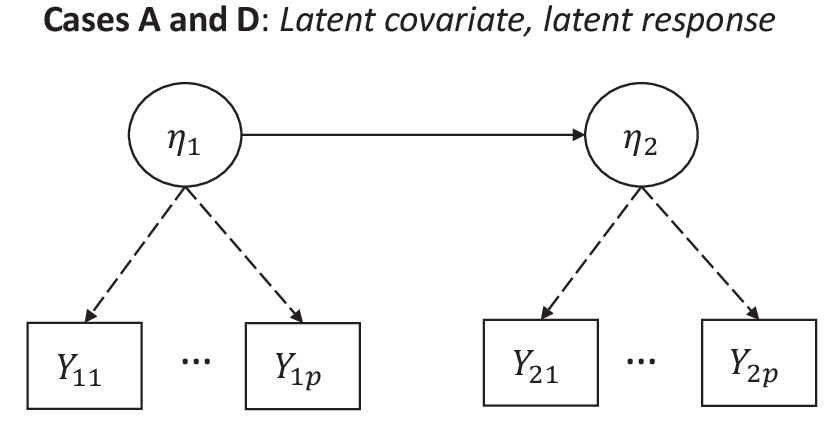}

\vspace*{3ex}
\includegraphics[width=.7\textwidth]{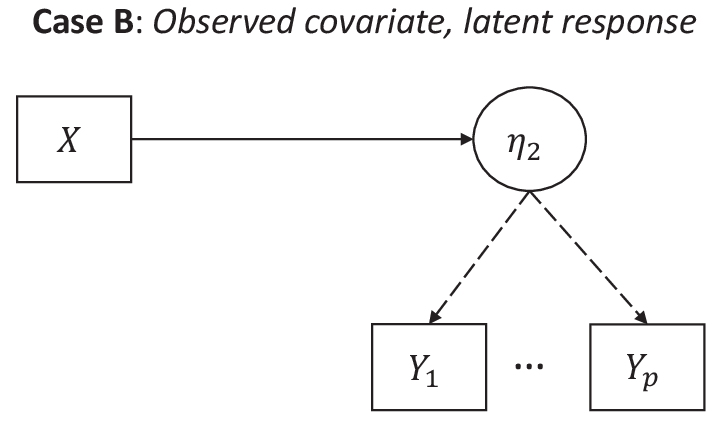}

\vspace*{3ex}
\includegraphics[width=.7\textwidth]{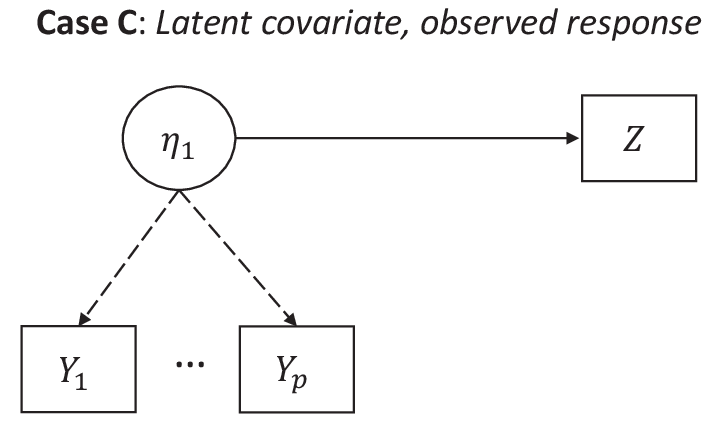}
\end{center}
\label{f_a_cases}
\end{figure}

\begin{figure}
\caption{Path diagrams representing cases E--G that are considered
in the simulation studies of Section 4 of the paper. Here the dashed and
dotted arrows represent the measurement model that is estimated in step 1 of
two-step estimation, and the solid arrows the structural model
(regression models for $\eta_{2}$ given $\eta_{1}$ and
for $\eta_{3}$ given $(\eta_{1},\eta_{2}))$
that is estimated in its step 2.
%Cases A and D differ in that the items $Y$ are binary in A but ordinal in D.
}
\begin{center}
\vspace*{-1ex}
\includegraphics[width=.9\textwidth]{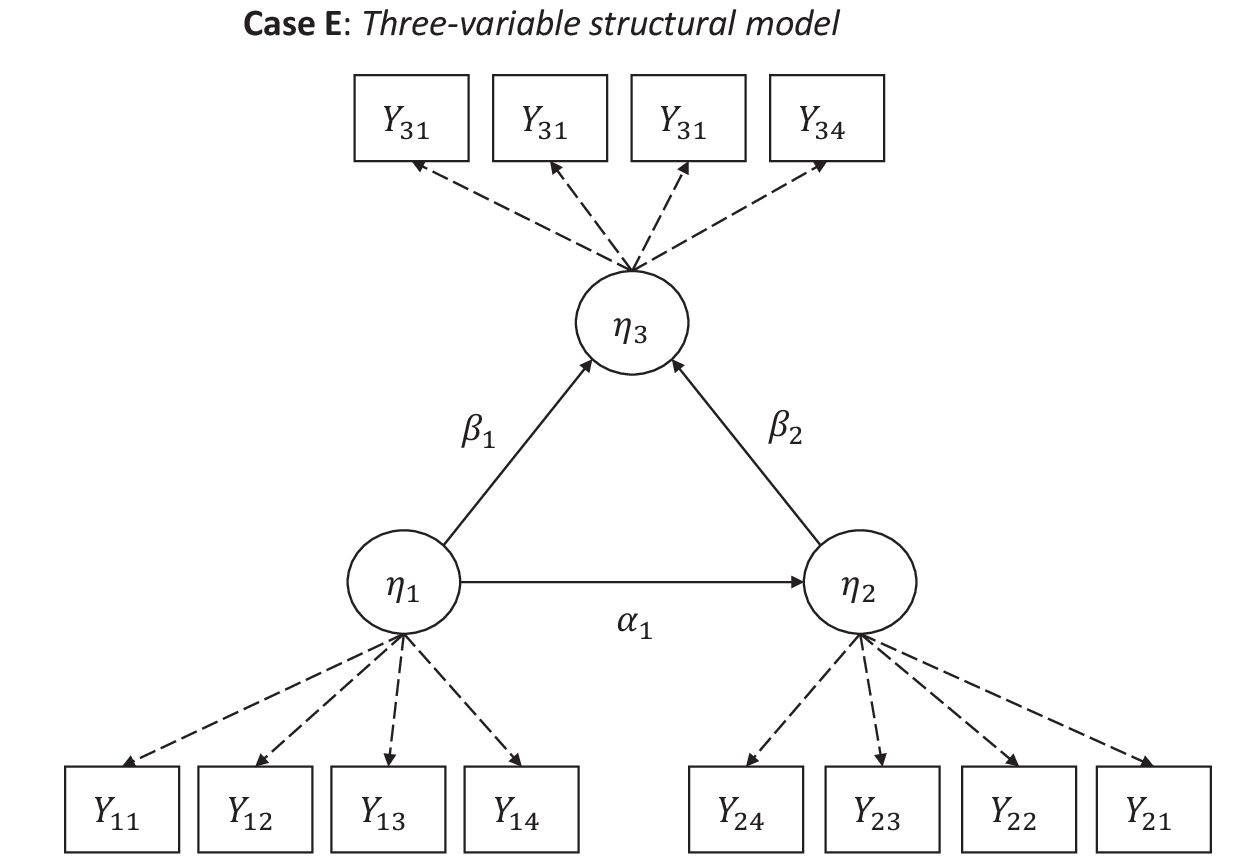}

\vspace*{3ex}
\includegraphics[width=.9\textwidth]{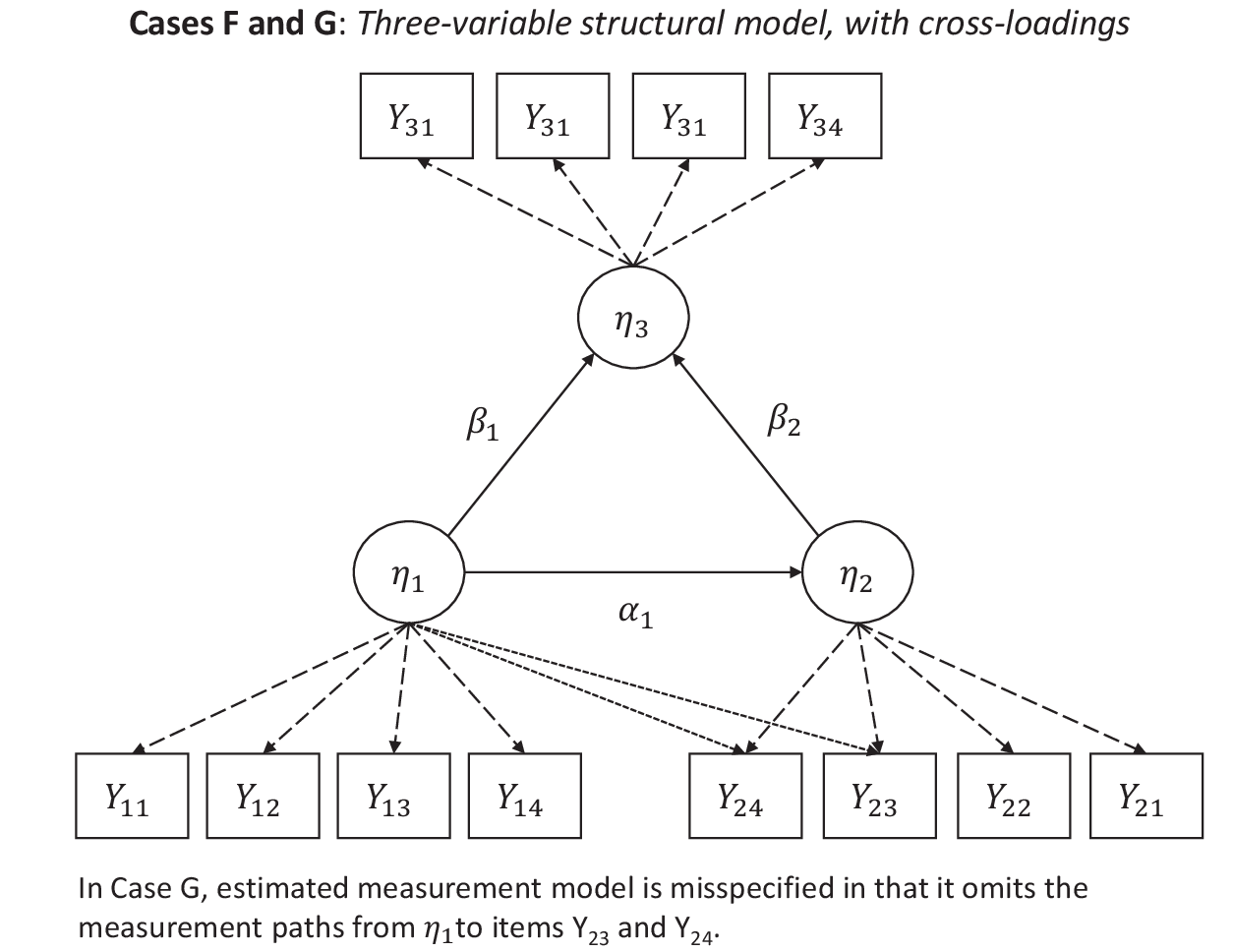}
\end{center}
\label{f_b_cases}
\end{figure}

\clearpage
\newpage
\setcounter{table}{0}
\renewcommand{\thetable}{A.\arabic{table}}

\thispagestyle{empty}
\newgeometry{top=10mm}

\begin{table}[ht]
\caption{Simulation means of two-step and one-step point estimates of
all loadings parameters ($\lambda_{kj}$) of the measurement models
in simulation cases A, B2 and B3.
In each case the true value of all the loadings is 1.}
\centering
{\small{
\begin{tabular}{|rrrr|rr|rr|rr|}
  \hline
  & & &
  & \multicolumn{2}{|c|}{Case A}
  & \multicolumn{2}{|c|}{Case B2}
  & \multicolumn{2}{|c|}{Case B3} \\
$p$ & $\pi_{Y}$ & $R^{2}_{Y}$ & $R^{2}_{\eta}$  & 2-step & 1-step & 2-step & 1-step & 2-step & 1-step\\
  \hline \multicolumn{10}{|l|}{$n=200$}\\ \hline
4 & 0.5 & 0.4 & 0.0 & 1.095 & 1.046 & 1.116 & 1.118 & 1.118 & 1.120 \\
  4 & 0.8 & 0.4 & 0.0 & 1.345 & 1.632 & 1.349 & 1.358 & 1.331 & 1.397 \\
  4 & 0.5 & 0.6 & 0.0 & 1.068 & 0.916 & 1.081 & 1.082 & 1.080 & 1.081 \\
  4 & 0.8 & 0.6 & 0.0 & 1.107 & 1.172 & 1.117 & 1.115 & 1.139 & 1.141 \\
  8 & 0.5 & 0.4 & 0.0 & 1.045 & -0.597 & 1.035 & 1.036 & 1.049 & 1.049 \\
\end{tabular}
}}
\label{meas_loadings}
\end{table}

\clearpage
\thispagestyle{empty}

\begin{table}[ht]
\caption{Simulation means of two-step and one-step point estimates of
all intercept parameters ($\tau_{kj}$) of the measurement models
in simulation cases A, B2 and B3.
In each case the true value of all the intercepts is 0.}
\centering
{\small{
\begin{tabular}{|rrrr|rr|rr|rr|}
  \hline
  & & &
  & \multicolumn{2}{|c|}{Case A}
  & \multicolumn{2}{|c|}{Case B2}
  & \multicolumn{2}{|c|}{Case B3} \\
$p$ & $\pi_{Y}$ & $R^{2}_{Y}$ & $R^{2}_{\eta}$  & 2-step & 1-step & 2-step & 1-step & 2-step & 1-step\\
  \hline \multicolumn{10}{|l|}{$n=200$}\\ \hline
4 & 0.5 & 0.4 & 0.0 & -0.013 & -0.008 & -0.002 & -0.002 & 0.001 & 0.002 \\
  4 & 0.8 & 0.4 & 0.0 & 0.462 & 0.743 & 0.442 & 0.452 & 0.403 & 0.490 \\
  4 & 0.5 & 0.6 & 0.0 & 0.005 & 0.020 & -0.003 & -0.003 & -0.001 & -0.001 \\
  4 & 0.8 & 0.6 & 0.0 & 0.085 & 0.148 & 0.092 & 0.090 & 0.136 & 0.138 \\
  8 & 0.5 & 0.4 & 0.0 & 0.009 & 0.023 & 0.001 & 0.001 & -0.006 & -0.006 \\
\end{tabular}
}}
\label{meas_intercepts}
\end{table}

\clearpage
\thispagestyle{empty}
\begin{table}[ht]
\caption{Simulation results for point estimates of structural regression
coefficient $\beta_{1}$ of
a latent covariate $\eta_{1}$ for a conditionally normally distributed
latent response $\eta_{2}$ (Case~A).}
\centering
{\small{
\begin{tabular}{|rrrrr|rrr|rrr|rrr|}
  \hline
  & & & &
  & \multicolumn{3}{|c|}{Bias}
  & \multicolumn{3}{|c|}{RMSE}
  & \multicolumn{3}{|c|}{MAE} \\
$p$ & $\pi_{Y}$ & $R^{2}_{Y}$ & $R^{2}_{\eta}$ & $\beta_1$ & 2-step & 1-step & 3-step & 2-step & 1-step & 3-step & 2-step & 1-step & 3-step \\
  \hline \multicolumn{14}{|l|}{$n=200$}\\ \hline
4 & 0.5 & 0.4 & 0.0 & 0.000 & -0.004 & -0.003 & -0.002 & 0.133 & 0.141 & 0.081 & 0.077 & 0.080 & 0.046 \\
  4 & 0.8 & 0.4 & 0.0 & 0.000 & -0.007 & -0.004 & -0.000 & 0.240 & 0.276 & 0.110 & 0.101 & 0.113 & 0.044 \\
  4 & 0.5 & 0.6 & 0.0 & 0.000 & 0.007 & 0.007 & 0.005 & 0.115 & 0.117 & 0.080 & 0.069 & 0.070 & 0.048 \\
  4 & 0.8 & 0.6 & 0.0 & 0.000 & 0.002 & 0.002 & 0.003 & 0.153 & 0.154 & 0.091 & 0.081 & 0.084 & 0.048 \\
  8 & 0.5 & 0.4 & 0.0 & 0.000 & -0.001 & -0.001 & -0.001 & 0.104 & 0.104 & 0.078 & 0.063 & 0.062 & 0.047 \\
\end{tabular}
}}
\label{sres_301_348}
\end{table}

\clearpage
\thispagestyle{empty}
\begin{table}[ht]
\caption{Simulation results for standard error estimates of structural regression
coefficient $\beta_{1}$ of a latent covariate $\eta_{1}$ for a conditionally
normally distributed latent response~$\eta_{2}$ (Case A).}
\centering
{\small{
\begin{tabular}{|rrrrr|rrrr|rrr|rr|}
  \hline
  &&&&
  & \multicolumn{4}{c|}{Simulation standard deviation}
  &&&
  & \multicolumn{2}{c|}{2-step}
  \\
  &&&&
  & \multicolumn{4}{c|}{vs.\ mean of est.\ std.\ error}
  & \multicolumn{3}{c|}{Coverage of}
  & \multicolumn{2}{c|}{without}
  \\
  &&&&
  & \multicolumn{2}{|c}{2-step}
  & \multicolumn{2}{c|}{1-step}
  & \multicolumn{3}{c|}{95\% conf.\ interval}
  & \multicolumn{2}{c|}{step-1 var.$^{\dagger}$}
\\
$p$ & $\pi_{Y}$ & $R^{2}_{Y}$ & $R^{2}_{\eta}$ & $\beta_1$
& s.d.\ & m(se) & s.d.\ & m(se) & 2-st.\ & 1-st.\ & 3-st.\ & \%var. &
{\footnotesize{cover.}} \\
\hline \multicolumn{14}{|l|}{$n=200$}\\ \hline
4 & 0.5 & 0.4 & 0.0 & 0.000 & 0.133 & 0.137 & 0.141 & 0.138 & 99.5 & 99.3 & 95.6 & 87.3 & 96.0 \\
  4 & 0.8 & 0.4 & 0.0 & 0.000 & 0.240 & 0.224 & 0.276 & 0.235 & 99.9 & 100.0 & 94.7 & 78.0 & 96.0 \\
  4 & 0.5 & 0.6 & 0.0 & 0.000 & 0.115 & 0.114 & 0.116 & 0.114 & 98.9 & 98.5 & 94.3 & 91.4 & 94.9 \\
\end{tabular}
}}
\label{sres_301_348_se}
\end{table}

\clearpage

\begin{table}[ht]
\caption{Simulation results for point estimates of structural regression
coefficient $\beta_{1}$ of a normally distributed
covariate $X$ for a latent response $\eta_{2}$
(Case B1)
.}
\centering
{\small{
\begin{tabular}{|rrrrr|rrr|rrr|rrr|}
  \hline
  & & & &
  & \multicolumn{3}{|c|}{Bias}
  & \multicolumn{3}{|c|}{RMSE}
  & \multicolumn{3}{|c|}{MAE} \\
$p$ & $\pi_{Y}$ & $R^{2}_{Y}$ & $R^{2}_{\eta}$ & $\beta_1$ & 2-step & 1-step & 3-step & 2-step & 1-step & 3-step & 2-step & 1-step & 3-step \\
  \hline \multicolumn{14}{|l|}{$n=200$}\\ \hline
4 & 0.5 & 0.4 & 0.0 & 0.000 & -0.003 & -0.004 & -0.002 & 0.146 & 0.151 & 0.090 & 0.084 & 0.086 & 0.052 \\
  4 & 0.8 & 0.4 & 0.0 & 0.000 & 0.007 & 0.006 & 0.003 & 0.196 & 0.208 & 0.094 & 0.100 & 0.102 & 0.045 \\
  4 & 0.5 & 0.6 & 0.0 & 0.000 & 0.008 & 0.008 & 0.006 & 0.207 & 0.208 & 0.145 & 0.125 & 0.126 & 0.086 \\
  4 & 0.8 & 0.6 & 0.0 & 0.000 & 0.002 & 0.002 & 0.002 & 0.251 & 0.253 & 0.151 & 0.144 & 0.145 & 0.085 \\
  8 & 0.5 & 0.4 & 0.0 & 0.000 & -0.005 & -0.005 & -0.004 & 0.132 & 0.132 & 0.098 & 0.085 & 0.085 & 0.063 \\
\end{tabular}
}}
\label{sres_1_48}
\end{table}

\clearpage
\thispagestyle{empty}
\begin{table}[ht]
\caption{Simulation results for standard error estimates of structural regression
coefficient $\beta_{1}$ of a normally distributed
covariate $X$ for a latent response $\eta_{2}$
(Case B1).}
\centering
{\small{
\begin{tabular}{|rrrrr|rrrr|rrr|rr|}
  \hline
  &&&&
  & \multicolumn{4}{c|}{Simulation standard deviation}
  &&&
  & \multicolumn{2}{c|}{2-step}
  \\
  &&&&
  & \multicolumn{4}{c|}{vs.\ mean of est.\ std.\ error}
  & \multicolumn{3}{c|}{Coverage of}
  & \multicolumn{2}{c|}{without}
  \\
  &&&&
  & \multicolumn{2}{|c}{2-step}
  & \multicolumn{2}{c|}{1-step}
  & \multicolumn{3}{c|}{95\% conf.\ interval}
  & \multicolumn{2}{c|}{step-1 var.$^{\dagger}$}
\\
$p$ & $\pi_{Y}$ & $R^{2}_{Y}$ & $R^{2}_{\eta}$ & $\beta_1$
& s.d.\ & m(se) & s.d.\ & m(se) & 2-st.\ & 1-st.\ & 3-st.\ & \%var. &
{\footnotesize{cover.}} \\
\hline \multicolumn{14}{|l|}{$n=200$}\\ \hline
4 & 0.5 & 0.4 & 0.0 & 0.000 & 0.146 & 0.149 & 0.151 & 0.149 & 98.4 & 98.0 & 95.3 & 92.6 & 95.3 \\
  4 & 0.8 & 0.4 & 0.0 & 0.000 & 0.196 & 0.204 & 0.208 & 0.201 & 99.2 & 98.1 & 94.8 & 86.5 & 95.0 \\
  4 & 0.5 & 0.6 & 0.0 & 0.000 & 0.207 & 0.205 & 0.208 & 0.204 & 96.1 & 95.8 & 93.2 & 95.2 & 93.4 \\
\end{tabular}
}}
\label{sres_1_48_se}
\end{table}

\clearpage
\thispagestyle{empty}
\begin{table}[ht]
\caption{Simulation results for point estimates of structural regression
coefficient $\beta_{1}$ of a binary
covariate $X$ for a latent response $\eta_{2}$
(Case B2).}
\centering
{\small{
\begin{tabular}{|rrrrr|rrr|rrr|rrr|}
  \hline
  & & & &
  & \multicolumn{3}{|c|}{Bias}
  & \multicolumn{3}{|c|}{RMSE}
  & \multicolumn{3}{|c|}{MAE} \\
$p$ & $\pi_{Y}$ & $R^{2}_{Y}$ & $R^{2}_{\eta}$ & $\beta_1$ & 2-step & 1-step & 3-step & 2-step & 1-step & 3-step & 2-step & 1-step & 3-step \\
  \hline \multicolumn{14}{|l|}{$n=200$}\\ \hline
4 & 0.5 & 0.4 & 0.0 & 0.000 & 0.001 & 0.001 & 0.000 & 0.144 & 0.147 & 0.090 & 0.090 & 0.092 & 0.055 \\
  4 & 0.8 & 0.4 & 0.0 & 0.000 & 0.007 & 0.008 & 0.003 & 0.191 & 0.213 & 0.091 & 0.099 & 0.106 & 0.046 \\
  4 & 0.5 & 0.6 & 0.0 & 0.000 & -0.001 & -0.002 & -0.001 & 0.209 & 0.210 & 0.147 & 0.129 & 0.129 & 0.091 \\
  4 & 0.8 & 0.6 & 0.0 & 0.000 & 0.002 & 0.001 & 0.001 & 0.259 & 0.254 & 0.160 & 0.138 & 0.141 & 0.081 \\
  8 & 0.5 & 0.4 & 0.0 & 0.000 & -0.002 & -0.002 & -0.002 & 0.125 & 0.126 & 0.093 & 0.086 & 0.086 & 0.063 \\
\end{tabular}
}}
\label{sres_201_132}
\end{table}

\clearpage
\thispagestyle{empty}
\begin{table}[ht]
\caption{Simulation results for standard error estimates of structural regression
coefficient $\beta_{1}$ of a binary covariate $X$ for a latent response
$\eta_{2}$ (Case B2).}
\centering
{\small{
\begin{tabular}{|rrrrr|rrrr|rrr|rr|}
  \hline
  &&&&
  & \multicolumn{4}{c|}{Simulation standard deviation}
  &&&
  & \multicolumn{2}{c|}{2-step}
  \\
  &&&&
  & \multicolumn{4}{c|}{vs.\ mean of est.\ std.\ error}
  & \multicolumn{3}{c|}{Coverage of}
  & \multicolumn{2}{c|}{without}
  \\
  &&&&
  & \multicolumn{2}{|c}{2-step}
  & \multicolumn{2}{c|}{1-step}
  & \multicolumn{3}{c|}{95\% conf.\ interval}
  & \multicolumn{2}{c|}{step-1 var.$^{\dagger}$}
\\
$p$ & $\pi_{Y}$ & $R^{2}_{Y}$ & $R^{2}_{\eta}$ & $\beta_1$
& s.d.\ & m(se) & s.d.\ & m(se) & 2-st.\ & 1-st.\ & 3-st.\ & \%var. &
{\footnotesize{cover.}} \\
\hline \multicolumn{14}{|l|}{$n=200$}\\ \hline
4 & 0.5 & 0.4 & 0.0 & 0.000 & 0.144 & 0.147 & 0.147 & 0.147 & 97.9 & 97.7 & 95.5 & 92.7 & 95.6 \\
  4 & 0.8 & 0.4 & 0.0 & 0.000 & 0.191 & 0.239 & 0.213 & 0.214 & 99.2 & 98.8 & 94.4 & 85.2 & 94.7 \\
  4 & 0.5 & 0.6 & 0.0 & 0.000 & 0.209 & 0.204 & 0.210 & 0.204 & 95.8 & 95.6 & 93.7 & 95.2 & 93.7 \\
\end{tabular}
}}
\label{sres_201_132_se}
\end{table}

\clearpage
\thispagestyle{empty}
\begin{table}[ht]
\caption{Simulation results for point estimates of structural regression
coefficient $\beta_{1}$ of a
covariate $X$ with a skewed distribution for a latent response $\eta_{2}$
(Case B3).}
\centering
{\small{
\begin{tabular}{|rrrrr|rrr|rrr|rrr|}
  \hline
  & & & &
  & \multicolumn{3}{|c|}{Bias}
  & \multicolumn{3}{|c|}{RMSE}
  & \multicolumn{3}{|c|}{MAE} \\
$p$ & $\pi_{Y}$ & $R^{2}_{Y}$ & $R^{2}_{\eta}$ & $\beta_1$ & 2-step & 1-step & 3-step & 2-step & 1-step & 3-step & 2-step & 1-step & 3-step \\
  \hline \multicolumn{14}{|l|}{$n=200$}\\ \hline
4 & 0.5 & 0.4 & 0.0 & 0.000 & -0.008 & -0.008 & -0.004 & 0.226 & 0.232 & 0.119 & 0.105 & 0.108 & 0.062 \\
  4 & 0.8 & 0.4 & 0.0 & 0.000 & 0.077 & 0.085 & 0.002 & 0.441 & 0.545 & 0.115 & 0.119 & 0.124 & 0.049 \\
  4 & 0.5 & 0.6 & 0.0 & 0.000 & 0.003 & 0.004 & 0.002 & 0.278 & 0.281 & 0.173 & 0.144 & 0.148 & 0.100 \\
  4 & 0.8 & 0.6 & 0.0 & 0.000 & 0.047 & 0.049 & -0.005 & 0.347 & 0.350 & 0.172 & 0.165 & 0.166 & 0.092 \\
  8 & 0.5 & 0.4 & 0.0 & 0.000 & 0.003 & 0.003 & 0.002 & 0.173 & 0.174 & 0.119 & 0.090 & 0.090 & 0.066 \\
\end{tabular}
}}
\label{sres_901_948}
\end{table}

\clearpage
\thispagestyle{empty}
\begin{table}[ht]
\caption{Simulation results for standard error estimates of structural regression
coefficient $\beta_{1}$ of a
covariate $X$ with a skewed distribution for a latent response $\eta_{2}$
(Case B3).}
\centering
{\small{
\begin{tabular}{|rrrrr|rrrr|rrr|rr|}
  \hline
  &&&&
  & \multicolumn{4}{c|}{Simulation standard deviation}
  &&&
  & \multicolumn{2}{c|}{2-step}
  \\
  &&&&
  & \multicolumn{4}{c|}{vs.\ mean of est.\ std.\ error}
  & \multicolumn{3}{c|}{Coverage of}
  & \multicolumn{2}{c|}{without}
  \\
  &&&&
  & \multicolumn{2}{|c}{2-step}
  & \multicolumn{2}{c|}{1-step}
  & \multicolumn{3}{c|}{95\% conf.\ interval}
  & \multicolumn{2}{c|}{step-1 var.$^{\dagger}$}
\\
$p$ & $\pi_{Y}$ & $R^{2}_{Y}$ & $R^{2}_{\eta}$ & $\beta_1$
& s.d.\ & m(se) & s.d.\ & m(se) & 2-st.\ & 1-st.\ & 3-st.\ & \%var. &
{\footnotesize{cover.}} \\
\hline \multicolumn{14}{|l|}{$n=200$}\\ \hline
4 & 0.5 & 0.4 & 0.0 & 0.000 & 0.226 & 0.179 & 0.232 & 0.179 & 97.9 & 97.5 & 94.4 & 92.2 & 94.5 \\
  4 & 0.8 & 0.4 & 0.0 & 0.000 & 0.434 & 0.254 & 0.539 & 0.281 & 99.4 & 98.9 & 96.0 & 86.7 & 96.6 \\
  4 & 0.5 & 0.6 & 0.0 & 0.000 & 0.278 & 0.247 & 0.281 & 0.247 & 97.8 & 97.8 & 94.6 & 95.3 & 95.3 \\
\end{tabular}
}}
\label{sres_901_948_se}
\end{table}

\clearpage
\thispagestyle{empty}
\begin{table}[ht]
\caption{Simulation results for point estimates of structural regression
coefficient $\beta_{1}$ of
a latent covariate $\eta_{1}$ for a conditionally normally distributed
response $Z$
(Case C1).}
\centering
{\small{
\begin{tabular}{|rrrrr|rrr|rrr|rrr|}
  \hline
  & & & &
  & \multicolumn{3}{|c|}{Bias}
  & \multicolumn{3}{|c|}{RMSE}
  & \multicolumn{3}{|c|}{MAE} \\
$p$ & $\pi_{Y}$ & $R^{2}_{Y}$ & $R^{2}_{Z}$ & $\beta_1$ & 2-step & 1-step & 3-step & 2-step & 1-step & 3-step & 2-step & 1-step & 3-step \\
  \hline \multicolumn{14}{|l|}{$n=200$}\\ \hline
4 & 0.5 & 0.4 & 0.0 & 0.000 & -0.003 & -0.003 & -0.003 & 0.067 & 0.070 & 0.067 & 0.041 & 0.042 & 0.040 \\
  4 & 0.8 & 0.4 & 0.0 & 0.000 & -0.003 & -0.003 & -0.003 & 0.089 & 0.104 & 0.088 & 0.050 & 0.057 & 0.050 \\
  4 & 0.5 & 0.6 & 0.0 & 0.000 & -0.000 & -0.000 & -0.000 & 0.039 & 0.040 & 0.039 & 0.024 & 0.025 & 0.024 \\
  4 & 0.8 & 0.6 & 0.0 & 0.000 & -0.000 & -0.000 & -0.000 & 0.044 & 0.046 & 0.044 & 0.027 & 0.029 & 0.027 \\
  8 & 0.5 & 0.4 & 0.0 & 0.000 & 0.001 & 0.001 & 0.001 & 0.059 & 0.060 & 0.059 & 0.038 & 0.038 & 0.038 \\
\end{tabular}
}}
\label{sres_49_96}
\end{table}

\clearpage
\thispagestyle{empty}
\begin{table}[ht]
\caption{Simulation results for standard error estimates of structural regression
coefficient $\beta_{1}$ of a latent covariate $\eta_{1}$ for a conditionally
normally distributed response $Z$ (Case C1).}
\centering
{\small{
\begin{tabular}{|rrrrr|rrrr|rrr|rr|}
  \hline
  &&&&
  & \multicolumn{4}{c|}{Simulation standard deviation}
  &&&
  & \multicolumn{2}{c|}{2-step}
  \\
  &&&&
  & \multicolumn{4}{c|}{vs.\ mean of est.\ std.\ error}
  & \multicolumn{3}{c|}{Coverage of}
  & \multicolumn{2}{c|}{without}
  \\
  &&&&
  & \multicolumn{2}{|c}{2-step}
  & \multicolumn{2}{c|}{1-step}
  & \multicolumn{3}{c|}{95\% conf.\ interval}
  & \multicolumn{2}{c|}{step-1 var.$^{\dagger}$}
\\
$p$ & $\pi_{Y}$ & $R^{2}_{Y}$ & $R^{2}_{Z}$ & $\beta_1$
& s.d.\ & m(se) & s.d.\ & m(se) & 2-st.\ & 1-st.\ & 3-st.\ & \%var. &
{\footnotesize{cover.}} \\
\hline \multicolumn{14}{|l|}{$n=200$}\\ \hline
4 & 0.5 & 0.4 & 0.0 & 0.000 & 0.067 & 0.065 & 0.070 & 0.066 & 97.9 & 97.6 & 94.7 & 92.5 & 95.2 \\
  4 & 0.8 & 0.4 & 0.0 & 0.000 & 0.089 & 0.085 & 0.105 & 0.093 & 98.6 & 98.4 & 91.8 & 86.4 & 93.3 \\
  4 & 0.5 & 0.6 & 0.0 & 0.000 & 0.039 & 0.039 & 0.040 & 0.040 & 97.7 & 97.7 & 96.1 & 95.7 & 96.3 \\
\end{tabular}
}}
\label{sres_49_96_se}
\end{table}

\clearpage
\thispagestyle{empty}
\begin{table}[ht]
\caption{Simulation results for point estimates of structural regression
coefficient $\beta_{1}$ of
a latent covariate $\eta_{1}$ for a conditionally skew-normally distributed
response $Z$ (Case~C2).}
\centering
{\small{
\begin{tabular}{|rrrrr|rrr|rrr|rrr|}
  \hline
  & & & &
  & \multicolumn{3}{|c|}{Bias}
  & \multicolumn{3}{|c|}{RMSE}
  & \multicolumn{3}{|c|}{MAE} \\
$p$ & $\pi_{Y}$ & $R^{2}_{Y}$ & $R^{2}_{Z}$ & $\beta_1$ & 2-step & 1-step & 3-step & 2-step & 1-step & 3-step & 2-step & 1-step & 3-step \\
  \hline \multicolumn{14}{|l|}{$n=200$}\\ \hline
4 & 0.5 & 0.4 & 0.0 & 0.000 & 0.003 & 0.003 & 0.003 & 0.068 & 0.070 & 0.067 & 0.040 & 0.042 & 0.040 \\
  4 & 0.8 & 0.4 & 0.0 & 0.000 & 0.005 & 0.008 & 0.003 & 0.085 & 0.113 & 0.084 & 0.043 & 0.048 & 0.043 \\
  4 & 0.5 & 0.6 & 0.0 & 0.000 & 0.000 & 0.000 & 0.000 & 0.039 & 0.040 & 0.039 & 0.025 & 0.026 & 0.025 \\
  4 & 0.8 & 0.6 & 0.0 & 0.000 & 0.001 & 0.001 & -0.000 & 0.044 & 0.046 & 0.044 & 0.026 & 0.027 & 0.026 \\
  8 & 0.5 & 0.4 & 0.0 & 0.000 & 0.002 & 0.002 & 0.002 & 0.056 & 0.057 & 0.056 & 0.037 & 0.038 & 0.037 \\
\end{tabular}
}}
\label{sres_133_180}
\end{table}

\clearpage
\thispagestyle{empty}
\begin{table}[ht]
\caption{Simulation results for standard error estimates of structural regression
coefficient $\beta_{1}$ of a latent covariate $\eta_{1}$ for a conditionally
skew-normally distributed response~$Z$ (Case C2).}
\centering
{\small{
\begin{tabular}{|rrrrr|rrrr|rrr|rr|}
  \hline
  &&&&
  & \multicolumn{4}{c|}{Simulation standard deviation}
  &&&
  & \multicolumn{2}{c|}{2-step}
  \\
  &&&&
  & \multicolumn{4}{c|}{vs.\ mean of est.\ std.\ error}
  & \multicolumn{3}{c|}{Coverage of}
  & \multicolumn{2}{c|}{without}
  \\
  &&&&
  & \multicolumn{2}{|c}{2-step}
  & \multicolumn{2}{c|}{1-step}
  & \multicolumn{3}{c|}{95\% conf.\ interval}
  & \multicolumn{2}{c|}{step-1 var.$^{\dagger}$}
\\
$p$ & $\pi_{Y}$ & $R^{2}_{Y}$ & $R^{2}_{Z}$ & $\beta_1$
& s.d.\ & m(se) & s.d.\ & m(se) & 2-st.\ & 1-st.\ & 3-st.\ & \%var. &
{\footnotesize{cover.}} \\
\hline \multicolumn{14}{|l|}{$n=200$}\\ \hline
4 & 0.5 & 0.4 & 0.0 & 0.000 & 0.068 & 0.066 & 0.070 & 0.067 & 97.9 & 97.6 & 95.7 & 93.1 & 95.9 \\
  4 & 0.8 & 0.4 & 0.0 & 0.000 & 0.085 & 0.085 & 0.113 & 0.101 & 99.4 & 99.2 & 95.7 & 87.8 & 96.6 \\
  4 & 0.5 & 0.6 & 0.0 & 0.000 & 0.039 & 0.039 & 0.040 & 0.040 & 96.5 & 96.3 & 94.9 & 95.6 & 95.0 \\
\end{tabular}
}}
\label{sres_133_180_se}
\end{table}

\clearpage
\thispagestyle{empty}
\begin{table}[ht]
\caption{\small{Simulation results for point estimates of structural regression
coefficient $\beta_{1}$ of latent covariate $\eta_{1}$ for a
conditionally normally distributed
latent response $\eta_{2}$, when each latent variable is measured by 4
measurement items with 4 ordered categories (Case~D).}}

%\vspace*{1ex}
\centering
{\small{
\begin{tabular}{|rrr|rrr|rrr|rrr|}
  \hline
  & &
  & \multicolumn{3}{|c|}{Bias}
  & \multicolumn{3}{|c|}{RMSE}
  & \multicolumn{3}{|c|}{MAE} \\
$R^{2}_{Y}$ & $R^{2}_{\eta}$ & $\beta_1$ & 2-step & 1-step & 3-step & 2-step & 1-step & 3-step & 2-step & 1-step & 3-step \\
  \hline \multicolumn{12}{|l|}{$n=200$}\\ \hline
0.4 & 0.0 & 0.000 & 0.001 & 0.001 & 0.001 & 0.110 & 0.113 & 0.079 & 0.068 & 0.070 & 0.048 \\
  0.6 & 0.0 & 0.000 & -0.001 & 0.005 & -0.001 & 0.096 & 0.204 & 0.079 & 0.063 & 0.063 & 0.051 \\
  \hline
  0.4 & 0.2 & 0.447 & 0.003 & 0.014 & -0.125 & 0.152 & 0.154 & 0.169 & 0.096 & 0.093 & 0.146 \\
  0.6 & 0.2 & 0.447 & 0.002 & 0.006 & -0.078 & 0.129 & 0.130 & 0.134 & 0.087 & 0.085 & 0.106 \\
  \hline
  0.4 & 0.4 & 0.632 & 0.003 & 0.018 & -0.179 & 0.198 & 0.192 & 0.233 & 0.126 & 0.120 & 0.212 \\
  0.6 & 0.4 & 0.632 & 0.010 & 0.016 & -0.103 & 0.157 & 0.154 & 0.169 & 0.102 & 0.098 & 0.134 \\
  \hline \multicolumn{12}{|l|}{$n=1000$}\\ \hline
  0.4 & 0.0 & 0.000 & 0.000 & 0.000 & 0.000 & 0.046 & 0.046 & 0.033 & 0.031 & 0.031 & 0.022 \\
  0.6 & 0.0 & 0.000 & 0.000 & 0.000 & 0.000 & 0.039 & 0.039 & 0.032 & 0.027 & 0.027 & 0.022 \\
  \hline
  0.4 & 0.2 & 0.447 & -0.001 & 0.002 & -0.132 & 0.065 & 0.065 & 0.140 & 0.043 & 0.043 & 0.135 \\
  0.6 & 0.2 & 0.447 & 0.000 & 0.002 & -0.081 & 0.055 & 0.055 & 0.094 & 0.036 & 0.036 & 0.084 \\
  \hline
  0.4 & 0.4 & 0.632 & -0.005 & -0.001 & -0.189 & 0.080 & 0.079 & 0.198 & 0.054 & 0.051 & 0.195 \\
  0.6 & 0.4 & 0.632 & 0.002 & 0.004 & -0.112 & 0.065 & 0.062 & 0.125 & 0.045 & 0.044 & 0.113 \\
   \hline
\multicolumn{12}{l}{\emph{Note:}
$R_{Y}^{2}$ and $R^{2}_{\eta}$ denote the $R^{2}$ statistics in models for
$Y_{j}$ and $\eta_{2}$, and $\beta_{1}$ true value of $\beta_{1}$.}
\end{tabular}
}}
\label{sres_b1_ord}
\end{table}

\begin{table}[hb]
\caption{Simulation results for standard error estimates
of the estimates considered in Table \ref{sres_b1_ord}.}

\vspace*{1ex}

%of structural regression
%coefficient $\beta_{1}$ of latent covariate $\eta_{1}$ for a
%conditionally normally distributed
%latent response $\eta_{2}$, when each latent variable is measured by 4
%measurement items with 4 ordered categories.}

%\vspace*{1ex}
\centering
{\small{
\begin{tabular}{|rrr|rrrr|rrr|rr|}
  \hline
  &&
  & \multicolumn{4}{c|}{Simulation standard deviation}
  &&&
  & \multicolumn{2}{c|}{2-step}
  \\
  &&
  & \multicolumn{4}{c|}{vs.\ mean of est.\ std.\ error}
  & \multicolumn{3}{c|}{Coverage of}
  & \multicolumn{2}{c|}{without}
  \\
  &&
  & \multicolumn{2}{|c}{2-step}
  & \multicolumn{2}{c|}{1-step}
  & \multicolumn{3}{c|}{95\% conf.\ interval}
  & \multicolumn{2}{c|}{step-1 var.$^{\dagger}$}
\\
$R^{2}_{Y}$ & $R^{2}_{\eta}$ & $\beta_1$
& s.d.\ & m(se) & s.d.\ & m(se) & 2-st.\ & 1-st.\ & 3-st.\ & \%var. &
{\footnotesize{cover.}} \\
\hline \multicolumn{12}{|l|}{$n=200$}\\ \hline
0.4 & 0.0 & 0.000 & 0.110 & 0.107 & 0.113 & 0.108 & 97.2 & 97.1 & 95.0 & 92.7 & 95.0 \\
  0.6 & 0.0 & 0.000 & 0.096 & 0.092 & 0.204 & 0.111 & 96.8 & 96.6 & 95.3 & 95.5 & 95.3 \\
  \hline
  0.4 & 0.2 & 0.447 & 0.152 & 0.152 & 0.154 & 0.151 & 93.1 & 93.6 & 46.1 & 45.8 & 80.9 \\
\end{tabular}
}}
\label{sres_se_ord}
\end{table}

\clearpage
\begin{table}[ht]
\caption{\small{Simulation results for point estimates of structural
regression coefficient $\beta_{2}$ of latent covariate $\eta_{2}$ for a
latent response $\eta_{3}$, in a model which also includes another
latent covariate $\eta_{1}$ (which has coefficient $\beta_{1}$)
(Case~E).}}
\vspace*{1ex}
\centering
{\small{
\begin{tabular}{|rrr|rrr|rrr|rrr|}
  \hline
  & &
  & \multicolumn{3}{|c|}{Bias}
  & \multicolumn{3}{|c|}{RMSE}
  & \multicolumn{3}{|c|}{MAE} \\
$R^{2}_{Y}$ & $\beta_1$ & $\beta_2$ & 2-step & 1-step & 3-step & 2-step & 1-step & 3-step & 2-step & 1-step & 3-step \\
\hline \multicolumn{12}{|l|}{$n=100$}\\ \hline
  0.4 & 0.390 & 0.309 & 0.036 & 0.044 & -0.037 & 0.451 & 0.756 & 0.235 & 0.198 & 0.214 & 0.150 \\
  0.6 & 0.390 & 0.309 & 0.034 & 0.036 & -0.021 & 0.302 & 0.326 & 0.195 & 0.158 & 0.161 & 0.131 \\
  0.4 & 0.273 & 0.423 & 0.007 & 0.049 & -0.120 & 0.524 & 0.623 & 0.291 & 0.230 & 0.249 & 0.223 \\
  0.6 & 0.273 & 0.423 & 0.053 & 0.074 & -0.068 & 0.341 & 0.427 & 0.227 & 0.186 & 0.195 & 0.165 \\
0.4 & 0.000 & 0.632 & 0.026 & 0.154 & -0.257 & 0.514 & 0.954 & 0.367 & 0.284 & 0.302 & 0.336 \\
\end{tabular}
}}
\label{sres_601_618}
\end{table}

\clearpage
\begin{table}[ht]
\caption{\small{Simulation results for point estimates of structural
regression coefficient $\beta_{2}$ of latent covariate $\eta_{2}$ for a
latent response $\eta_{3}$, in a model which also includes another
latent covariate $\eta_{1}$ (which has coefficient $\beta_{1}$)
(Case~E).
\textbf{Note}: Here bias of the estimates is quantified by
the difference between their median over the simulations
and the true value of $\beta_{2}$, whereas in Table \ref{sres_601_618}} the simulation mean rather than the median was used for
this. Other columns of this table identical to those of Table
\ref{sres_601_618}.
}
\vspace*{1ex}
\centering
{\small{
\begin{tabular}{|rrr|rrr|rrr|rrr|}
  \hline
  & &
  & \multicolumn{3}{|c|}{Bias (median)}
  & \multicolumn{3}{|c|}{RMSE}
  & \multicolumn{3}{|c|}{MAE} \\
$R^{2}_{Y}$ & $\beta_1$ & $\beta_2$ & 2-step & 1-step & 3-step & 2-step & 1-step & 3-step & 2-step & 1-step & 3-step \\
\hline \multicolumn{12}{|l|}{$n=100$}\\ \hline
  0.4 & 0.390 & 0.309 & -0.045 & -0.033 & -0.097 & 0.451 & 0.756 & 0.235 & 0.198 & 0.214 & 0.150 \\
  0.6 & 0.390 & 0.309 & -0.020 & -0.015 & -0.066 & 0.302 & 0.326 & 0.195 & 0.158 & 0.161 & 0.131 \\
  0.4 & 0.273 & 0.423 & -0.089 & -0.059 & -0.183 & 0.524 & 0.623 & 0.291 & 0.230 & 0.249 & 0.223 \\
  0.6 & 0.273 & 0.423 & -0.006 & -0.002 & -0.109 & 0.341 & 0.427 & 0.227 & 0.186 & 0.195 & 0.165 \\
0.4 & 0.000 & 0.632 & -0.090 & 0.000 & -0.317 & 0.514 & 0.954 & 0.367 & 0.284 & 0.302 & 0.336 \\
  0.6 & 0.000 & 0.632 & -0.035 & 0.011 & -0.230 & 0.427 & 0.486 & 0.319 & 0.226 & 0.232 & 0.261 \\
\hline \multicolumn{12}{|l|}{$n=200$}\\ \hline
  0.4 & 0.390 & 0.309 & -0.025 & -0.013 & -0.091 & 0.276 & 0.962 & 0.167 & 0.144 & 0.152 & 0.120 \\
  0.6 & 0.390 & 0.309 & -0.003 & -0.004 & -0.057 & 0.184 & 1.183 & 0.124 & 0.109 & 0.114 & 0.089 \\
  0.4 & 0.273 & 0.423 & -0.021 & -0.009 & -0.151 & 0.260 & 0.272 & 0.193 & 0.161 & 0.151 & 0.164 \\
  0.6 & 0.273 & 0.423 & -0.022 & -0.011 & -0.120 & 0.212 & 0.487 & 0.170 & 0.127 & 0.133 & 0.135 \\
  0.4 & 0.000 & 0.632 & -0.050 & -0.008 & -0.304 & 0.346 & 0.350 & 0.323 & 0.211 & 0.195 & 0.311 \\
  0.6 & 0.000 & 0.632 & -0.026 & -0.019 & -0.233 & 0.258 & 0.490 & 0.266 & 0.163 & 0.158 & 0.236 \\
\hline \multicolumn{12}{|l|}{$n=1000$}\\ \hline
  0.4 & 0.390 & 0.309 & -0.002 & -0.003 & -0.077 & 0.093 & 0.092 & 0.089 & 0.059 & 0.058 & 0.078 \\
  0.6 & 0.390 & 0.309 & -0.005 & -0.005 & -0.054 & 0.077 & 3.529 & 0.070 & 0.050 & 0.052 & 0.057 \\
  0.4 & 0.273 & 0.423 & -0.012 & -0.008 & -0.154 & 0.106 & 2.134 & 0.159 & 0.070 & 0.064 & 0.154 \\
  0.6 & 0.273 & 0.423 & -0.002 & 0.003 & -0.110 & 0.085 & 5.133 & 0.120 & 0.058 & 0.059 & 0.110 \\
  0.4 & 0.000 & 0.632 & -0.007 & 0.001 & -0.292 & 0.141 & 0.133 & 0.297 & 0.094 & 0.091 & 0.292 \\
  0.6 & 0.000 & 0.632 & -0.004 & -0.002 & -0.227 & 0.110 & 8.196 & 0.233 & 0.072 & 0.070 & 0.227 \\
   \hline
\multicolumn{12}{l}{\emph{Note:}
$R_{Y}^{2}$ denotes the $R^{2}$ statistic in measurement models for
items $Y_{j}$.}\\
\multicolumn{12}{l}{\hspace*{2em}
Results for 1-step estimates exclude simulations where the estimation
did not converge.
}\\
\multicolumn{12}{l}{\hspace*{2em}
This occurred in 0--9 of the 1000 simulations in different settings.
}\\
\end{tabular}
}}
\label{sres_601_618B}
\end{table}

\clearpage
\begin{table}[ht]
\caption{\small{Simulation results for
standard error estimates of structural
regression coefficient $\beta_{2}$ of latent covariate $\eta_{2}$ for a
latent response $\eta_{3}$, in a model which also includes another
latent covariate $\eta_{1}$ (which has coefficient $\beta_{1}$)
(Case~E).}}
\vspace*{1ex}
\centering
{\small{
\begin{tabular}{|rrr|rrrr|rrr|rr|}
  \hline
  &&
  & \multicolumn{4}{c|}{Simulation standard deviation}
  &&&
  & \multicolumn{2}{c|}{2-step}
  \\
  &&
  & \multicolumn{4}{c|}{vs.\ mean of est.\ std.\ error}
  & \multicolumn{3}{c|}{Coverage of}
  & \multicolumn{2}{c|}{without}
  \\
  &&
  & \multicolumn{2}{|c}{2-step}
  & \multicolumn{2}{c|}{1-step}
  & \multicolumn{3}{c|}{95\% conf.\ interval}
  & \multicolumn{2}{c|}{step-1 var.$^{\dagger}$}
\\
$R^{2}_{Y}$ & $\beta_1$ & $\beta_2$
& s.d.\ & m(se) & s.d.\ & m(se) & 2-st.\ & 1-st.\ & 3-st.\ & \%var. &
{\footnotesize{cover.}} \\
\hline \multicolumn{12}{|l|}{$n=100$}\\ \hline
  0.4 & 0.390 & 0.309 & 0.450 & 1.937 & 0.755 & 2.350 & 93.2 & 94.2 & 63.8 & 58.9 & 87.8 \\
  0.6 & 0.390 & 0.309 & 0.300 & 0.299 & 0.324 & 0.303 & 93.6 & 93.1 & 68.7 & 64.1 & 89.2 \\
  0.4 & 0.273 & 0.423 & 0.524 & 0.581 & 0.621 & 0.513 & 88.6 & 91.1 & 49.2 & 53.4 & 80.3 \\
  0.6 & 0.273 & 0.423 & 0.337 & 0.332 & 0.421 & 0.343 & 92.7 & 92.7 & 60.1 & 56.3 & 85.9 \\
  0.4 & 0.000 & 0.632 & 0.514 & 1.893 & 0.942 & 0.657 & 88.4 & 90.3 & 34.9 & 42.9 & 76.8 \\
  0.6 & 0.000 & 0.632 & 0.423 & 0.427 & 0.474 & 0.429 & 89.4 & 91.1 & 44.2 & 43.8 & 79.4 \\
\hline \multicolumn{12}{|l|}{$n=200$}\\ \hline
  0.4 & 0.390 & 0.309 & 0.276 & 0.255 & 0.962 & 0.486 & 95.2 & 95.2 & 56.6 & 64.2 & 89.3 \\
  0.6 & 0.390 & 0.309 & 0.183 & 0.181 & 1.184 & 0.304 & 94.0 & 94.3 & 70.3 & 69.7 & 89.9 \\
  0.4 & 0.273 & 0.423 & 0.259 & 0.277 & 0.269 & 0.268 & 93.3 & 94.5 & 45.8 & 56.4 & 86.0 \\
  0.6 & 0.273 & 0.423 & 0.212 & 0.203 & 0.487 & 0.225 & 91.6 & 91.9 & 52.0 & 59.7 & 84.3 \\
  0.4 & 0.000 & 0.632 & 0.345 & 0.354 & 0.344 & 0.330 & 91.0 & 93.6 & 22.8 & 44.0 & 79.8 \\
  0.6 & 0.000 & 0.632 & 0.258 & 0.256 & 0.488 & 0.368 & 92.6 & 93.1 & 29.1 & 45.4 & 80.8 \\
\hline \multicolumn{12}{|l|}{$n=1000$}\\ \hline
  0.4 & 0.390 & 0.309 & 0.093 & 0.095 & 0.092 & 0.092 & 95.4 & 94.8 & 36.7 & 73.1 & 90.8 \\
  0.6 & 0.390 & 0.309 & 0.077 & 0.074 & 3.523 & 0.134 & 94.7 & 94.5 & 54.9 & 74.6 & 89.7 \\
  0.4 & 0.273 & 0.423 & 0.106 & 0.106 & 2.133 & 0.146 & 93.4 & 93.2 & 10.8 & 62.0 & 87.4 \\
  0.6 & 0.273 & 0.423 & 0.085 & 0.083 & 5.123 & 0.155 & 93.8 & 93.8 & 22.6 & 62.3 & 87.1 \\
  0.4 & 0.000 & 0.632 & 0.141 & 0.137 & 0.132 & 0.127 & 93.2 & 94.2 & 0.9 & 47.2 & 81.0 \\
  0.6 & 0.000 & 0.632 & 0.110 & 0.107 & 8.181 & 0.195 & 94.0 & 94.6 & 2.6 & 46.5 & 80.6 \\
   \hline
\multicolumn{12}{l}{
$\dagger$ Average \% of the variance of 2-step estimator accounted
for by step-2 variance alone,}\\
\multicolumn{12}{l}{
\hspace*{1em} and coverage of 95\% confidence interval if only this variance is
included.}\\
\multicolumn{12}{l}{
Results for 1-step estimates exclude simulations where the estimation
did not converge.
}\\
\multicolumn{12}{l}{
This occurred in 0--9 of the 1000 simulations in different settings.
}\\
\end{tabular}
}}
\label{sres_601_618_se}
\end{table}

\clearpage
%\restoregeometry

\begin{table}[ht]
\caption{\small{Simulation results for point estimates of structural
regression coefficient $\beta_{2}$ of latent covariate $\eta_{2}$ for a
latent response $\eta_{3}$, in a model which also includes another
latent covariate $\eta_{1}$
(which has coefficient $\beta_{1}$)
and where the measurement model includes
cross-loadings from $\eta_{1}$ to measures of $\eta_{2}$
(Case~F).
}}
%\vspace*{1ex}
\centering
{\small{
\begin{tabular}{|rrr|rrr|rrr|rrr|}
  \hline
  & &
  & \multicolumn{3}{|c|}{Bias}
  & \multicolumn{3}{|c|}{RMSE}
  & \multicolumn{3}{|c|}{MAE} \\
$R^{2}_{Y}$ & $\beta_1$ & $\beta_2$ & 2-step & 1-step & 3-step & 2-step & 1-step & 3-step & 2-step & 1-step & 3-step \\
\hline \multicolumn{12}{|l|}{$n=200$}\\ \hline
  0.4 & 0.390 & 0.309 & 0.019 & 0.073 & -0.099 & 0.341 & 0.935 & 0.230 & 0.158 & 0.176 & 0.161 \\
  0.6 & 0.390 & 0.309 & 0.025 & 0.046 & -0.067 & 0.216 & 0.346 & 0.177 & 0.123 & 0.127 & 0.127 \\
  0.4 & 0.273 & 0.423 & 0.026 & 0.068 & -0.134 & 0.470 & 0.921 & 0.303 & 0.173 & 0.181 & 0.196 \\
  0.6 & 0.273 & 0.423 & 0.007 & -0.202 & -0.108 & 0.231 & 5.275 & 0.207 & 0.139 & 0.147 & 0.163 \\
0.4 & 0.000 & 0.632 & 0.033 & 0.152 & 0.014 & 0.528 & 0.700 & 7.030 & 0.236 & 0.234 & 0.300 \\
  0.6 & 0.000 & 0.632 & 0.009 & 0.044 & -0.159 & 0.291 & 0.313 & 0.272 & 0.176 & 0.177 & 0.226 \\
\hline \multicolumn{12}{|l|}{$n=1000$}\\ \hline
  0.4 & 0.390 & 0.309 & 0.002 & 0.010 & -0.116 & 0.192 & 0.112 & 0.135 & 0.072 & 0.070 & 0.125 \\
  0.6 & 0.390 & 0.309 & 0.002 & -0.046 & -0.087 & 0.082 & 1.613 & 0.106 & 0.055 & 0.055 & 0.093 \\
  0.4 & 0.273 & 0.423 & -0.012 & 0.005 & -0.163 & 0.230 & 0.124 & 0.181 & 0.081 & 0.083 & 0.173 \\
  0.6 & 0.273 & 0.423 & 0.001 & 0.006 & -0.115 & 0.091 & 0.092 & 0.134 & 0.059 & 0.058 & 0.121 \\
  0.4 & 0.000 & 0.632 & 0.007 & 0.017 & -0.239 & 0.160 & 0.155 & 0.260 & 0.099 & 0.095 & 0.251 \\
  0.6 & 0.000 & 0.632 & 0.005 & -0.103 & -0.169 & 0.122 & 3.527 & 0.192 & 0.080 & 0.083 & 0.176 \\
   \hline
\multicolumn{12}{l}{\emph{Note:}
$R_{Y}^{2}$ denotes the $R^{2}$ statistic in measurement models for
items $Y_{j}$.}\\
\multicolumn{12}{l}{
Results for 1-step estimates exclude simulations where the estimation
did not converge.
}\\
\multicolumn{12}{l}{
This occurred in 0--1 of the 1000 simulations in different settings.}
\end{tabular}
}}
\label{sres_701_718}
\end{table}

\begin{table}[hb]
\caption{\small{Simulation results for point estimates of structural
regression coefficient $\beta_{2}$ of latent covariate $\eta_{2}$ for a
latent response $\eta_{3}$, in a model which also includes another
latent covariate $\eta_{1}$
(which has coefficient $\beta_{1}$)
and where the measurement model includes
cross-loadings from $\eta_{1}$ to measures of $\eta_{2}$
(Case~F).
\textbf{Note}: Here bias of the estimates is quantified by
the difference between their median over the simulations
and the true value of $\beta_{2}$, whereas in Table \ref{sres_701_718}
the simulation mean rather than the median was used for
this. Other columns of this table identical to those of Table
\ref{sres_701_718}.
}}
%\vspace*{1ex}
\centering
{\small{
\begin{tabular}{|rrr|rrr|rrr|rrr|}
  \hline
  & &
  & \multicolumn{3}{|c|}{Bias (median)}
  & \multicolumn{3}{|c|}{RMSE}
  & \multicolumn{3}{|c|}{MAE} \\
$R^{2}_{Y}$ & $\beta_1$ & $\beta_2$ & 2-step & 1-step & 3-step & 2-step & 1-step & 3-step & 2-step & 1-step & 3-step \\
\hline \multicolumn{12}{|l|}{$n=200$}\\ \hline
  0.4 & 0.390 & 0.309 & -0.023 & -0.006 & -0.132 & 0.341 & 0.935 & 0.230 & 0.158 & 0.176 & 0.161 \\
  0.6 & 0.390 & 0.309 & -0.002 & 0.011 & -0.088 & 0.216 & 0.346 & 0.177 & 0.123 & 0.127 & 0.127 \\
  0.4 & 0.273 & 0.423 & -0.031 & 0.014 & -0.178 & 0.470 & 0.921 & 0.303 & 0.173 & 0.181 & 0.196 \\
  0.6 & 0.273 & 0.423 & -0.029 & -0.007 & -0.136 & 0.231 & 5.275 & 0.207 & 0.139 & 0.147 & 0.163 \\
0.4 & 0.000 & 0.632 & -0.076 & -0.004 & -0.279 & 0.528 & 0.700 & 7.030 & 0.236 & 0.234 & 0.300 \\
  0.6 & 0.000 & 0.632 & -0.041 & -0.013 & -0.200 & 0.291 & 0.313 & 0.272 & 0.176 & 0.177 & 0.226 \\
\hline \multicolumn{12}{|l|}{$n=1000$}\\ \hline
  0.4 & 0.390 & 0.309 & -0.004 & -0.001 & -0.125 & 0.192 & 0.112 & 0.135 & 0.072 & 0.070 & 0.125 \\
  0.6 & 0.390 & 0.309 & -0.004 & 0.000 & -0.091 & 0.082 & 1.613 & 0.106 & 0.055 & 0.055 & 0.093 \\
  0.4 & 0.273 & 0.423 & -0.015 & -0.009 & -0.173 & 0.230 & 0.124 & 0.181 & 0.081 & 0.083 & 0.173 \\
  0.6 & 0.273 & 0.423 & -0.004 & -0.001 & -0.120 & 0.091 & 0.092 & 0.134 & 0.059 & 0.058 & 0.121 \\
  0.4 & 0.000 & 0.632 & -0.013 & 0.005 & -0.251 & 0.160 & 0.155 & 0.260 & 0.099 & 0.095 & 0.251 \\
  0.6 & 0.000 & 0.632 & -0.005 & -0.002 & -0.176 & 0.122 & 3.527 & 0.192 & 0.080 & 0.083 & 0.176 \\
   \hline
\multicolumn{12}{l}{\emph{Note:}
$R_{Y}^{2}$ denotes the $R^{2}$ statistic in measurement models for
items $Y_{j}$.}\\
\multicolumn{12}{l}{
Results for 1-step estimates exclude simulations where the estimation
did not converge.
}\\
\multicolumn{12}{l}{
This occurred in 0--1 of the 1000 simulations in different settings.}
\end{tabular}
}}
\label{sres_701_718B}
\end{table}

\begin{table}[ht]
\caption{\small{Simulation results for
standard error estimates of structural
regression coefficient $\beta_{2}$ of latent covariate $\eta_{2}$ for a
latent response $\eta_{3}$, in a model which also includes another
latent covariate $\eta_{1}$
(which has coefficient $\beta_{1}$)
and where the measurement model includes
cross-loadings from $\eta_{1}$ to measures of $\eta_{2}$
(Case~F).
}}
%\vspace*{1ex}
\centering
{\small{
\begin{tabular}{|rrr|rrrr|rrr|rr|}
  \hline
  &&
  & \multicolumn{4}{c|}{Simulation standard deviation}
  &&&
  & \multicolumn{2}{c|}{2-step}
  \\
  &&
  & \multicolumn{4}{c|}{vs.\ mean of est.\ std.\ error}
  & \multicolumn{3}{c|}{Coverage of}
  & \multicolumn{2}{c|}{without}
  \\
  &&
  & \multicolumn{2}{|c}{2-step}
  & \multicolumn{2}{c|}{1-step}
  & \multicolumn{3}{c|}{95\% conf.\ interval}
  & \multicolumn{2}{c|}{step-1 var.$^{\dagger}$}
\\
$R^{2}_{Y}$ & $\beta_1$ & $\beta_2$
& s.d.\ & m(se) & s.d.\ & m(se) & 2-st.\ & 1-st.\ & 3-st.\ & \%var. &
{\footnotesize{cover.}} \\
\hline \multicolumn{12}{|l|}{$n=200$}\\ \hline
  0.4 & 0.390 & 0.309 & 0.340 & 0.624 & 0.933 & 0.533 & 94.7 & 95.0 & 66.7 & 61.9 & 90.1 \\
  0.6 & 0.390 & 0.309 & 0.214 & 0.220 & 0.343 & 0.233 & 94.8 & 94.4 & 73.6 & 64.0 & 89.5 \\
  0.4 & 0.273 & 0.423 & 0.470 & 2.922 & 0.919 & 0.464 & 94.9 & 95.6 & 56.9 & 54.6 & 88.9 \\
  0.6 & 0.273 & 0.423 & 0.232 & 0.234 & 5.273 & 0.334 & 92.4 & 93.0 & 59.7 & 55.9 & 85.7 \\
0.4 & 0.000 & 0.632 & 0.527 & 2.240 & 0.684 & 0.506 & 88.8 & 90.7 & 43.6 & 46.2 & 79.1 \\
  0.6 & 0.000 & 0.632 & 0.291 & 0.301 & 0.310 & 0.296 & 90.5 & 91.7 & 50.1 & 44.6 & 79.3 \\
\hline \multicolumn{12}{|l|}{$n=1000$}\\ \hline
  0.4 & 0.390 & 0.309 & 0.192 & 0.164 & 0.111 & 0.109 & 95.1 & 94.9 & 38.9 & 70.8 & 90.2 \\
  0.6 & 0.390 & 0.309 & 0.082 & 0.081 & 1.613 & 0.094 & 94.8 & 94.4 & 48.4 & 70.5 & 90.5 \\
  0.4 & 0.273 & 0.423 & 0.230 & 2.564 & 0.124 & 0.121 & 94.1 & 94.5 & 24.1 & 61.8 & 85.9 \\
  0.6 & 0.273 & 0.423 & 0.091 & 0.092 & 0.092 & 0.093 & 95.1 & 95.2 & 34.2 & 60.1 & 87.2 \\
  0.4 & 0.000 & 0.632 & 0.160 & 0.159 & 0.154 & 0.151 & 93.3 & 94.3 & 12.1 & 50.5 & 83.9 \\
  0.6 & 0.000 & 0.632 & 0.122 & 0.120 & 3.527 & 0.162 & 94.2 & 92.7 & 23.0 & 47.2 & 80.5 \\
   \hline
\multicolumn{12}{l}{
$\dagger$ Average \% of the variance of 2-step estimator accounted
for by step-2 variance alone,}\\
\multicolumn{12}{l}{
\hspace*{1em} and coverage of 95\% confidence interval if only this variance is
included.}\\
\multicolumn{12}{l}{
Results for 1-step estimates exclude simulations where the estimation
did not converge.
}\\
\multicolumn{12}{l}{
This occurred in 0--1 of the 1000 simulations in different settings.}
\end{tabular}
}}
\label{sres_701_718_se}
\end{table}

\clearpage
%\restoregeometry

\begin{table}[ht]
\caption{\small{Simulation results for point estimates of structural
regression coefficient $\beta_{2}$ of latent covariate $\eta_{2}$ for a
latent response $\eta_{3}$, in a model which also includes another
latent covariate $\eta_{1}$
(which has coefficient $\beta_{1}$)
and where the measurement model includes
cross-loadings from $\eta_{1}$ to measures of $\eta_{2}$ (as in case G),
but when the measurement model is estimated without the cross-loadings
(as in case F) (Case~G).
}}
%\vspace*{1ex}

\centering
{\small{
\begin{tabular}{|rrr|rrr|rrr|rrr|}
  \hline
  & &
  & \multicolumn{3}{|c|}{Bias}
  & \multicolumn{3}{|c|}{RMSE}
  & \multicolumn{3}{|c|}{MAE} \\
$R^{2}_{Y}$ & $\beta_1$ & $\beta_2$ & 2-step & 1-step & 3-step & 2-step & 1-step & 3-step & 2-step & 1-step & 3-step \\
\hline \multicolumn{12}{|l|}{$n=200$}\\ \hline
  0.4 & 0.390 & 0.309 & 0.246 & 0.314 & 0.026 & 0.634 & 1.064 & 0.174 & 0.263 & 0.371 & 0.095 \\
  0.6 & 0.390 & 0.309 & 0.252 & 0.320 & 0.074 & 0.461 & 1.022 & 0.178 & 0.247 & 0.341 & 0.095 \\
  0.4 & 0.273 & 0.423 & 0.369 & 0.542 & -0.033 & 0.787 & 1.199 & 0.198 & 0.306 & 0.442 & 0.123 \\
  0.6 & 0.273 & 0.423 & 0.402 & 0.552 & 0.038 & 0.601 & 0.808 & 0.187 & 0.337 & 0.460 & 0.112 \\
0.4 & 0.000 & 0.632 & 0.491 & 0.830 & -0.217 & 0.864 & 1.356 & 0.285 & 0.345 & 0.543 & 0.262 \\
  0.6 & 0.000 & 0.632 & 0.550 & 0.796 & -0.092 & 0.774 & 1.076 & 0.222 & 0.456 & 0.659 & 0.167 \\
\hline \multicolumn{12}{|l|}{$n=1000$}\\ \hline
  0.4 & 0.390 & 0.309 & 0.215 & 0.278 & 0.007 & 0.294 & 0.443 & 0.065 & 0.196 & 0.251 & 0.044 \\
  0.6 & 0.390 & 0.309 & 0.234 & 0.230 & 0.061 & 0.278 & 2.043 & 0.089 & 0.225 & 0.305 & 0.061 \\
  0.4 & 0.273 & 0.423 & 0.288 & 0.418 & -0.072 & 0.352 & 0.912 & 0.098 & 0.274 & 0.345 & 0.078 \\
  0.6 & 0.273 & 0.423 & 0.325 & 0.468 & 0.003 & 0.365 & 0.888 & 0.072 & 0.309 & 0.416 & 0.048 \\
  0.4 & 0.000 & 0.632 & 0.450 & 0.714 & -0.238 & 0.527 & 1.496 & 0.249 & 0.408 & 0.538 & 0.244 \\
  0.6 & 0.000 & 0.632 & 0.482 & 0.572 & -0.124 & 0.524 & 3.318 & 0.146 & 0.466 & 0.613 & 0.128 \\
   \hline
\multicolumn{12}{l}{\emph{Note:}
$R_{Y}^{2}$ denotes the $R^{2}$ statistic in measurement models for
items $Y_{j}$.}\\
\multicolumn{12}{l}{
Results for 1-step estimates exclude simulations where the estimation
did not converge.
}\\
\multicolumn{12}{l}{
This occurred in 1 of the 1000 simulations in one setting.}
\end{tabular}
}}
\label{sres_801_818}
\end{table}

\begin{table}[hb]
\caption{\small{Simulation results for point estimates of structural
regression coefficient $\beta_{2}$ of latent covariate $\eta_{2}$ for a
latent response $\eta_{3}$, in a model which also includes another
latent covariate $\eta_{1}$
(which has coefficient $\beta_{1}$)
and where the measurement model includes
cross-loadings from $\eta_{1}$ to measures of $\eta_{2}$ (as in case G),
but when the measurement model is estimated without the cross-loadings
(as in case F) (Case~G).
\textbf{Note}: Here bias of the estimates is quantified by
the difference between their median over the simulations
and the true value of $\beta_{2}$, whereas in Table \ref{sres_801_818}
the simulation mean rather than the median was used for
this. Other columns of this table identical to those of Table
\ref{sres_801_818}.
}}
%\vspace*{1ex}

\centering
{\small{
\begin{tabular}{|rrr|rrr|rrr|rrr|}
  \hline
  & &
  & \multicolumn{3}{|c|}{Bias (median)}
  & \multicolumn{3}{|c|}{RMSE}
  & \multicolumn{3}{|c|}{MAE} \\
$R^{2}_{Y}$ & $\beta_1$ & $\beta_2$ & 2-step & 1-step & 3-step & 2-step & 1-step & 3-step & 2-step & 1-step & 3-step \\
\hline \multicolumn{12}{|l|}{$n=200$}\\ \hline
  0.4 & 0.390 & 0.309 & 0.180 & 0.228 & -0.008 & 0.634 & 1.064 & 0.174 & 0.263 & 0.371 & 0.095 \\
  0.6 & 0.390 & 0.309 & 0.201 & 0.273 & 0.049 & 0.461 & 1.022 & 0.178 & 0.247 & 0.341 & 0.095 \\
  0.4 & 0.273 & 0.423 & 0.263 & 0.380 & -0.068 & 0.787 & 1.199 & 0.198 & 0.306 & 0.442 & 0.123 \\
  0.6 & 0.273 & 0.423 & 0.324 & 0.454 & 0.011 & 0.601 & 0.808 & 0.187 & 0.337 & 0.460 & 0.112 \\
  0.4 & 0.000 & 0.632 & 0.305 & 0.526 & -0.251 & 0.864 & 1.356 & 0.285 & 0.345 & 0.543 & 0.262 \\
  0.6 & 0.000 & 0.632 & 0.456 & 0.656 & -0.126 & 0.774 & 1.076 & 0.222 & 0.456 & 0.659 & 0.167 \\
\hline \multicolumn{12}{|l|}{$n=1000$}\\ \hline
  0.4 & 0.390 & 0.309 & 0.192 & 0.249 & 0.003 & 0.294 & 0.443 & 0.065 & 0.196 & 0.251 & 0.044 \\
  0.6 & 0.390 & 0.309 & 0.225 & 0.304 & 0.058 & 0.278 & 2.043 & 0.089 & 0.225 & 0.305 & 0.061 \\
  0.4 & 0.273 & 0.423 & 0.274 & 0.345 & -0.074 & 0.352 & 0.912 & 0.098 & 0.274 & 0.345 & 0.078 \\
  0.6 & 0.273 & 0.423 & 0.309 & 0.416 & -0.003 & 0.365 & 0.888 & 0.072 & 0.309 & 0.416 & 0.048 \\
  0.4 & 0.000 & 0.632 & 0.408 & 0.537 & -0.244 & 0.527 & 1.496 & 0.249 & 0.408 & 0.538 & 0.244 \\
  0.6 & 0.000 & 0.632 & 0.466 & 0.612 & -0.128 & 0.524 & 3.318 & 0.146 & 0.466 & 0.613 & 0.128 \\
   \hline
\multicolumn{12}{l}{\emph{Note:}
$R_{Y}^{2}$ denotes the $R^{2}$ statistic in measurement models for
items $Y_{j}$.}\\
\multicolumn{12}{l}{
Results for 1-step estimates exclude simulations where the estimation
did not converge.
}\\
\multicolumn{12}{l}{
This occurred in 1 of the 1000 simulations in one setting.}
\end{tabular}
}}
\label{sres_801_818B}
\end{table}

\begin{table}[ht]
\caption{\small{Simulation results for
standard error estimates of structural
regression coefficient $\beta_{2}$ of latent covariate $\eta_{2}$ for a
latent response $\eta_{3}$, in a model which also includes another
latent covariate $\eta_{1}$
(which has coefficient $\beta_{1}$)
and where the measurement model includes
cross-loadings from $\eta_{1}$ to measures of $\eta_{2}$ (as in case G),
but when the measurement model is estimated without the cross-loadings
(as in case F) (Case~G).
}}
%\vspace*{1ex}

\centering
{\small{
\begin{tabular}{|rrr|rrrr|rrr|rr|}
  \hline
  &&
  & \multicolumn{4}{c|}{Simulation standard deviation}
  &&&
  & \multicolumn{2}{c|}{2-step}
  \\
  &&
  & \multicolumn{4}{c|}{vs.\ mean of est.\ std.\ error}
  & \multicolumn{3}{c|}{Coverage of}
  & \multicolumn{2}{c|}{without}
  \\
  &&
  & \multicolumn{2}{|c}{2-step}
  & \multicolumn{2}{c|}{1-step}
  & \multicolumn{3}{c|}{95\% conf.\ interval}
  & \multicolumn{2}{c|}{step-1 var.$^{\dagger}$}
\\
$R^{2}_{Y}$ & $\beta_1$ & $\beta_2$
& s.d.\ & m(se) & s.d.\ & m(se) & 2-st.\ & 1-st.\ & 3-st.\ & \%var. &
{\footnotesize{cover.}} \\
\hline \multicolumn{12}{|l|}{$n=200$}\\ \hline
  0.4 & 0.390 & 0.309 & 0.585 & 1.555 & 1.017 & 1.639 & 99.5 & 99.5 & 73.9 & 64.9 & 95.0 \\
  0.6 & 0.390 & 0.309 & 0.386 & 1.111 & 0.971 & 0.519 & 98.0 & 98.4 & 78.5 & 70.4 & 92.1 \\
  0.4 & 0.273 & 0.423 & 0.695 & 1.499 & 1.071 & 4.246 & 98.7 & 98.7 & 63.9 & 60.9 & 92.3 \\
  0.6 & 0.273 & 0.423 & 0.446 & 0.460 & 0.590 & 0.545 & 96.7 & 97.0 & 73.7 & 68.1 & 86.2 \\
0.4 & 0.000 & 0.632 & 0.712 & 1.156 & 1.073 & 1.398 & 99.2 & 97.9 & 32.5 & 55.4 & 93.9 \\
  0.6 & 0.000 & 0.632 & 0.544 & 0.557 & 0.723 & 0.671 & 97.2 & 97.7 & 57.4 & 62.9 & 83.1 \\
\hline \multicolumn{12}{|l|}{$n=1000$}\\ \hline
  0.4 & 0.390 & 0.309 & 0.202 & 0.205 & 0.344 & 0.245 & 85.0 & 81.6 & 73.9 & 79.3 & 78.4 \\
  0.6 & 0.390 & 0.309 & 0.150 & 0.148 & 2.031 & 0.204 & 66.0 & 56.6 & 64.3 & 84.6 & 59.1 \\
  0.4 & 0.273 & 0.423 & 0.202 & 0.210 & 0.811 & 0.331 & 77.2 & 72.1 & 47.6 & 79.1 & 67.6 \\
  0.6 & 0.273 & 0.423 & 0.167 & 0.152 & 0.755 & 0.351 & 44.2 & 33.9 & 75.0 & 84.8 & 37.2 \\
  0.4 & 0.000 & 0.632 & 0.273 & 0.263 & 1.315 & 1.311 & 67.6 & 55.2 & 3.9 & 74.4 & 48.3 \\
  0.6 & 0.000 & 0.632 & 0.205 & 0.181 & 3.270 & 0.309 & 20.1 & 12.5 & 29.1 & 81.5 & 14.2 \\
   \hline
\multicolumn{12}{l}{
$\dagger$ Average \% of the variance of 2-step estimator accounted
for by step-2 variance alone,}\\
\multicolumn{12}{l}{
\hspace*{1em} and coverage of 95\% confidence interval if only this variance is
included.}\\
\multicolumn{12}{l}{
Results for 1-step estimates exclude simulations where the estimation
did not converge.
}\\
\multicolumn{12}{l}{
This occurred in 1 of the 1000 simulations in one settings.}
\end{tabular}
}}
\label{sres_801_818_se}
\end{table}

\clearpage
\restoregeometry
\setcounter{table}{0}
\renewcommand{\thetable}{B.\arabic{table}}
\begin{table}[ht]
\caption{
Estimated coefficients of structural models for extrinsic work values given
different sets of covariates (with standard errors in parentheses), estimated for EVS2017 data for Netherlands. The table shows two-step
(``2-st.''), one-step (``1-st.'') and naive three-step (``3-st.'') estimates.
This example is considered in Section 5 of the paper, and the
coefficients of gender of respondent (dummy variable for men)
are
also shown in Table 3 there.
}

\vspace*{1ex}
\centering
{\small{
\begin{tabular}{|l|rrr|rrr|rrr|}
  \hline
 &
\multicolumn{3}{|c|}{Model (1)} &
\multicolumn{3}{|c|}{Model (2)} &
\multicolumn{3}{|c|}{Model (3)} \\
&
2-st.\ & 1-st.\ & 3-st. &
2-st.\ & 1-st.\ & 3-st. &
2-st.\ & 1-st.\ & 3-st. \\ \hline
Intercept & 1.403 & 1.398 & 1.699 & 1.385 & 1.404 & 1.694 & 1.155 & 1.160 & 1.518 \\
    & (0.113) & (0.116) & (0.057) & (0.231) & (0.238) & (0.172) &
    (0.287) & (0.294) &    (0.224) \\[1ex]
Man
       & -0.042 & -0.033 & -0.033 & 0.025 & 0.022 & 0.024 & 0.018 & 0.009 & 0.018 \\
      & (0.103) & (0.104) & (0.082) & (0.101) & (0.104) & (0.080) &
      (0.102) & (0.104) & (0.081) \\[1ex]
\multicolumn{4}{|l|}{Age (vs.\ 30--49)} & & & & & & \\
\hspace*{2em}15--29
&  &  &  & 0.472 & 0.489 & 0.380 & 0.448 & 0.462 & 0.364 \\
      &  &  &  & (0.207) & (0.214) & (0.166) & (0.208) & (0.213) &
      (0.167) \\[.5ex]
      \hspace*{2em}50--
          &  &  &  & -0.391 & -0.408 & -0.326 & -0.191 & -0.196 & -0.162 \\
      &  &  &  & (0.150) & (0.155) & (0.119) & (0.152) & (0.155) &
      (0.123) \\[1ex]
Has partner
   &  &  &  & -0.019 & -0.022 & -0.017 & -0.052 & -0.056 & -0.048 \\
      &  &  &  & (0.163) & (0.169) & (0.133) & (0.163) & (0.167) &
      (0.133) \\[1ex]
\multicolumn{7}{|l|}{Age of youngest person in household (vs.\ older)} &&&\\
\hspace*{2em}0--5
&  &  &  & 0.542 & 0.570 & 0.437 & 0.477 & 0.499 & 0.390 \\
      &  &  &  & (0.194) & (0.201) & (0.151) & (0.191) & (0.195) &
      (0.150) \\[.5ex]
\hspace*{2em}6--17
   &  &  &  & 0.367 & 0.381 & 0.299 & 0.272 & 0.279 & 0.225 \\
      &  &  &  & (0.150) & (0.156) & (0.120) & (0.149) & (0.153) &
      (0.121) \\[1ex]
\multicolumn{4}{|l|}{Education level (vs.\ upper secondary)} &&&&&&\\
\hspace*{2em}  Lower
  &  &  &  &  &  &  & -0.104 & -0.105 & -0.080 \\
      &  &  &  &  &  &  & (0.144) & (0.147) & (0.117) \\[.5ex]
\hspace*{2em}  Higher
  &  &  &  &  &  &  & 0.034 & 0.031 & 0.039 \\
      &  &  &  &  &  &  & (0.133) & (0.135) & (0.107) \\[1ex]
\multicolumn{7}{|l|}{Occupation-based social class (vs.\
Managerial/service contract)} &&&\\
\hspace*{1em}Mixed
   &  &  &  &  &  &  & -0.074 & -0.074 & -0.069 \\
      &  &  &  &  &  &  & (0.144) & (0.147) & (0.116) \\[.5ex]
\hspace*{1em}Self-empl.\
 &  &  &  &  &  &  & -0.350 & -0.358 & -0.273 \\
       &  &  &  &  &  &  & (0.243) & (0.248) & (0.192) \\[.5ex]
\hspace*{1em}Labour
  &  &  &  &  &  &  & 0.231 & 0.229 & 0.185 \\
       &  &  &  &  &  &  & (0.144) & (0.147) & (0.116) \\[1ex]
\multicolumn{4}{|l|}{Currently in paid employment}
&&&& 0.449 & 0.468 & 0.358 \\
       &  &  &  &  &  &  & (0.123) & (0.125) & (0.092) \\[1ex]
\multicolumn{7}{|l|}{Parents' education level (highest, vs.\ upper secondary)} &&&\\
\hspace*{2em}Lower
   &  &  &  &  &  &  & -0.122 & -0.124 & -0.108 \\
       &  &  &  &  &  &  & (0.146) & (0.149) & (0.118) \\[.5ex]
\hspace*{2em}Higher
   &  &  &  &  &  &  & -0.092 & -0.099 & -0.092 \\
       &  &  &  &  &  &  & (0.159) & (0.162) & (0.128) \\
   \hline
\end{tabular}
}}
\label{NL_mods_ext}
\end{table}

\clearpage

\clearpage
\begin{table}[ht]
\caption{
Estimated coefficients of structural models for intrinsic work values given
different sets of covariates (with standard errors in parentheses), estimated for EVS2017 data for Netherlands. The table shows two-step
(``2-st.''), one-step (``1-st.'') and naive three-step (``3-st.'') estimates.
This example is considered in Section 5 of the paper, and the
coefficients
coefficients of gender of respondent (dummy variable for men)
are also shown in Table 3 there.
}

\vspace*{1ex}
\centering
{\small{
\begin{tabular}{|l|rrr|rrr|rrr|}
  \hline
 &
\multicolumn{3}{|c|}{Model (1)} &
\multicolumn{3}{|c|}{Model (2)} &
\multicolumn{3}{|c|}{Model (3)} \\
&
2-st.\ & 1-st.\ & 3-st. &
2-st.\ & 1-st.\ & 3-st. &
2-st.\ & 1-st.\ & 3-st. \\ \hline
Intercept & 0.624 & 0.611 & 0.955 & 0.474 & 0.449 & 0.847 & 1.311 & 1.087 & 1.480 \\
    & (0.190) & (0.165) & (0.099) & (0.415) & (0.376) & (0.305) &
    (0.569) & (0.430) &    (0.387) \\[1ex]
  Man
  & 0.607 & 0.603 & 0.460 & 0.657 & 0.638 & 0.500 & 0.516 & 0.506 & 0.400 \\
      & (0.193) & (0.186) & (0.142) & (0.194) & (0.176) & (0.142) &
      (0.188) & (0.151) &      (0.139) \\[1ex]
\multicolumn{4}{|l|}{Age (vs.\ 30--49)} & & & & & & \\
\hspace*{2em}15--29
 &  &  &  & 0.783 & 0.749 & 0.573 & 0.568 & 0.507 & 0.417 \\
      &  &  &  & (0.405) & (0.365) & (0.293) & (0.390) & (0.312) &
      (0.288) \\[.5ex]
      \hspace*{2em}50--
  &  &  &  & -0.584 & -0.545 & -0.452 & -0.183 & -0.162 & -0.159 \\
      &  &  &  & (0.293) & (0.261) & (0.210) & (0.283) & (0.227) &
      (0.213) \\[1ex]
Has partner
 &  &  &  & 0.412 & 0.371 & 0.311 & 0.168 & 0.108 & 0.132 \\
      &  &  &  & (0.328) & (0.293) & (0.235) & (0.311) & (0.248) &
      (0.229) \\[1ex]
\multicolumn{7}{|l|}{Age of youngest person in household (vs.\ older)} &&&\\
\hspace*{2em}0--5
    & &  &  & -0.081 & -0.072 & -0.067 & -0.276 & -0.221 & -0.228 \\
      &  &  &  & (0.355) & (0.324) & (0.267) & (0.349) & (0.278) &
      (0.259) \\[.5ex]
\hspace*{2em}6--17
   &  &  &  & 0.028 & 0.010 & 0.029 & -0.132 & -0.130 & -0.098 \\
      &  &  &  & (0.282) & (0.257) & (0.213) & (0.276) & (0.221) &
      (0.208) \\[1ex]
\multicolumn{4}{|l|}{Education level (vs.\ upper secondary)} &&&&&&\\
\hspace*{2em}  Lower
   &  &  &  &  &  &  & -0.459 & -0.374 & -0.378 \\
      &  &  &  &  &  &  & (0.279) & (0.218) & (0.202) \\[.5ex]
\hspace*{2em}  Higher
   &  &  &  &  &  &  & 0.446 & 0.365 & 0.342 \\
      &  &  &  &  &  &  & (0.257) & (0.201) & (0.185) \\[1ex]
\multicolumn{7}{|l|}{Occupation-based social class (vs.\
Managerial/service contract)} &&&\\
\hspace*{1em}Mixed
   &  &  &  &  &  &  & -0.588 & -0.503 & -0.454 \\
      &  &  &  &  &  &  & (0.279) & (0.216) & (0.200) \\[.5ex]
\hspace*{1em}Self-empl.\
   &  &  &  &  &  &  & -0.452 & -0.402 & -0.337 \\
       &  &  &  &  &  &  & (0.444) & (0.355) & (0.332) \\[.5ex]
\hspace*{1em}Labour
   &  &  &  &  &  &  & -0.857 & -0.748 & -0.652 \\
       &  &  &  &  &  &  & (0.289) & (0.220) & (0.200) \\[1ex]
\multicolumn{4}{|l|}{Currently in paid employment}
  &  &  &  & 0.304 & 0.252 & 0.237 \\
       &  &  &  &  &  &  & (0.219) & (0.172) & (0.160) \\[1ex]
\multicolumn{7}{|l|}{Parents' education level (highest, vs.\ upper secondary)} &&&\\
\hspace*{2em}Lower
   &  &  &  &  &  &  & -1.009 & -0.796 & -0.758 \\
       &  &  &  &  &  &  & (0.344) & (0.239) & (0.203) \\[.5ex]
\hspace*{2em}Higher
   &  &  &  &  &  &  & -0.238 & -0.180 & -0.167 \\
       &  &  &  &  &  &  & (0.306) & (0.242) & (0.221) \\
   \hline
\end{tabular}
}}
\label{NL_mods_int}
\end{table}

\clearpage

\begin{table}[ht]
\caption{
Estimated coefficients of gender (as dummy variable for men) in
structural models for extrinsic work values, for each country
in EVS2017 data. The table shows two-step, one-step and naive three-step estimates.
The models also include as covariates the respondent's age, education,
occupation-based social class and current employment status (working
vs.\ not).
All the coefficients of the structrural model are estimated separately
for each country, but the measurement model for work values is the same
in all countries.
The coefficients and confidence intervals from two-step estimation are
also also shown in Figure 1 in Section 5 of the paper
(with the further
standardisation that they are expressed on a scale where the residual
variance of extrinsic values for the Netherlands is 1).
}

\vspace*{2ex}
\centering
{\small{
\begin{tabular}{|llrrrrrr|}
  \hline
country & $n$&\hspace*{2em} 2-step & (s.e.) & 1-step & (s.e.) & 3-step & (s.e.) \\
  \hline
Albania    &1054& -0.180 & (0.276) & -0.195 & (0.320) & -0.016 & (0.069) \\
  Armenia  &1157& -0.133 & (0.076) & -0.149 & (0.089) & -0.124 & (0.074) \\
  Austria  &1450& -0.066 & (0.089) & -0.045 & (0.103) & -0.056 & (0.079) \\
Azerbaijan &1205& -0.300 & (0.099) & -0.346 & (0.116) & -0.240 & (0.081) \\
  Belarus  &1452& -0.092 & (0.109) & -0.096 & (0.127) & -0.069 & (0.078) \\
  Bosnia \& Herzegovina
           &1002& -0.022 & (0.123) & -0.023 & (0.142) & -0.019 & (0.098) \\
  Bulgaria &1327& -0.085 & (0.137) & -0.091 & (0.156) & -0.046 & (0.081) \\
  Croatia  &1328& -0.313 & (0.117) & -0.352 & (0.135) & -0.214 & (0.086) \\
  Czechia  &1546& -0.108 & (0.114) & -0.113 & (0.131) & -0.078 & (0.076) \\
  Denmark  &3024& -0.143 & (0.087) & -0.118 & (0.100) & -0.106 & (0.063) \\
  Estonia  &1272& -0.088 & (0.088) & -0.081 & (0.104) & -0.076 & (0.084) \\
  Finland  &1035& -0.130 & (0.127) & -0.111 & (0.146) & -0.104 & (0.099) \\
  France   &1729& -0.035 & (0.095) & -0.027 & (0.106) & -0.026 & (0.078) \\
  Georgia  &1756& 0.032  & (0.077) & 0.051  & (0.090) & 0.035  & (0.072) \\
  Germany  &1890& -0.159 & (0.095) & -0.155 & (0.109) & -0.128 & (0.076) \\
Great Britain
           &1673& -0.154 & (0.110) & -0.154 & (0.127) & -0.105 & (0.082) \\
  Hungary  &1360& -0.041 & (0.100) & -0.041 & (0.117) & -0.023 & (0.078) \\
  Iceland  &1547& -0.156 & (0.075) & -0.172 & (0.089) & -0.139 & (0.072) \\
  Italy    &1782& -0.141 & (0.068) & -0.134 & (0.077) & -0.123 & (0.064) \\
  Latvia   &1204& -0.389 & (0.130) & -0.441 & (0.149) & -0.292 & (0.091) \\
 Lithuania &1232& -0.274 & (0.138) & -0.313 & (0.159) & -0.142 & (0.069) \\
Montenegro &\hspace*{.5em}716& -0.283 & (0.135) & -0.320 & (0.153) & -0.237 & (0.112) \\
Netherlands
           &2068& -0.003 & (0.108) & 0.032  & (0.123) & -0.010 & (0.077) \\
North Macedonia
           &\hspace*{.5em}748& 0.017  & (0.092) & 0.022  & (0.109) & 0.014  & (0.081) \\
  Norway   &1079& -0.322 & (0.113) & -0.334 & (0.129) & -0.266 & (0.095) \\
  Poland   &1200& -0.213 & (0.087) & -0.244 & (0.098) & -0.202 & (0.083) \\
  Portugal &1092& 0.169  & (0.226) & 0.215  & (0.258) & 0.077  & (0.097) \\
  Romania  &1098& -0.205 & (0.135) & -0.232 & (0.158) & -0.139 & (0.080) \\
  Russia   &1582& -0.397 & (0.086) & -0.435 & (0.097) & -0.363 & (0.078) \\
  Serbia   &1168& 0.117  & (0.121) & 0.133  & (0.139) & 0.082  & (0.085) \\
  Slovakia &1280& -0.493 & (0.118) & -0.560 & (0.134) & -0.391 & (0.093) \\
  Slovenia &\hspace*{.5em}954& -0.212 & (0.117) & -0.247 & (0.140) & -0.169 & (0.088) \\
  Spain    &1002&  0.007 & (0.140) & 0.020  & (0.164) & -0.003 & (0.084) \\
  Sweden   &1053& -0.307 & (0.146) & -0.340 & (0.169) & -0.225 & (0.106) \\
Switzerland&2841& 0.023  & (0.072) & 0.066  & (0.082) & 0.029  & (0.057) \\
  Ukraine  &1455& -0.423 & (0.125) & -0.476 & (0.143) & -0.253 & (0.073) \\
   \hline
\end{tabular}
}}
\label{country_mods_ext}
\end{table}

\clearpage

\begin{table}[ht]
\caption{
Estimated coefficients of gender (as dummy variable for men) in
structural models for intrinsic work values, for each country
in EVS2017 data. The table shows two-step, one-step and naive three-step estimates.
The models also include as covariates the respondent's age, education,
occupation-based social class and current employment status (working
vs.\ not).
All the coefficients of the structural model are estimated separately
for each country, but the measurement model for work values is the same
in all countries.
The coefficients and confidence intervals from two-step estimation are
also also shown in Figure 1 in Section 5 of the paper
(with the further
standardisation that they are expressed on a scale where the residual
variance of intrinsic values for the Netherlands is 1).
}

\vspace*{2ex}
\centering
{\small{
\begin{tabular}{|llrrrrrr|}
  \hline
country & $n$&\hspace*{2em} 2-step & (s.e.) & 1-step & (s.e.) & 3-step & (s.e.) \\
  \hline
Albania    &1054&
0.152 & (0.237) & 0.141 & (0.235) & 0.092 & (0.122) \\
  Armenia  &1157&
-0.051 & (0.201) & -0.052 & (0.196) & -0.039 & (0.142) \\
  Austria  &1450&
  0.010 & (0.122) & 0.014 & (0.119) & 0.006 & (0.119) \\
Azerbaijan &1205&
  0.010 & (0.229) & 0.003 & (0.222) & 0.024 & (0.160) \\
  Belarus  &1452&
  0.235 & (0.191) & 0.239 & (0.184) & 0.164 & (0.141) \\
Bosnia \& Herzegovina &1002&
  -0.269 & (0.161) & -0.268 & (0.153) & -0.229 & (0.145) \\
  Bulgaria &1327&
  0.277 & (0.249) & 0.269 & (0.242) & 0.185 & (0.137) \\
  Croatia  &1328&
  -0.204 & (0.184) & -0.204 & (0.178) & -0.145 & (0.137) \\
  Czechia  &1546&
  0.463 & (0.196) & 0.454 & (0.190) & 0.295 & (0.130) \\
  Denmark  &3024&
  0.004 & (0.095) & -0.007 & (0.092) & 0.017 & (0.079) \\
  Estonia  &1272&
  0.474 & (0.175) & 0.449 & (0.169) & 0.407 & (0.146) \\
  Finland  &1035&
  0.105 & (0.147) & 0.096 & (0.143) & 0.112 & (0.136) \\
  France   &1729&
  -0.024 & (0.116) & -0.035 & (0.113) & -0.026 & (0.107) \\
  Georgia  &1756&
  0.854 & (0.188) & 0.813 & (0.181) & 0.558 & (0.128) \\
  Germany  &1890&
  0.287 & (0.116) & 0.284 & (0.114) & 0.252 & (0.101) \\
Great Britain &1673&
  0.087 & (0.125) & 0.077 & (0.121) & 0.076 & (0.110) \\
  Hungary  &1360&
  0.268 & (0.186) & 0.253 & (0.181) & 0.203 & (0.134) \\
  Iceland  &1547&
  -0.155 & (0.121) & -0.165 & (0.118) & -0.137 & (0.109) \\
  Italy    &1782&
  0.214 & (0.081) & 0.192 & (0.077) & 0.248 & (0.093) \\
  Latvia   &1204&
  0.049 & (0.228) & 0.058 & (0.219) & 0.033 & (0.156) \\
 Lithuania &1232&
  -0.033 & (0.312) & -0.036 & (0.304) & -0.047 & (0.147) \\
Montenegro &\hspace*{.5em}716&
  -1.196 & (0.475) & -1.146 & (0.454) & -0.556 & (0.213) \\
Netherlands &2068&
  0.366 & (0.117) & 0.355 & (0.114) & 0.334 & (0.105) \\
North Macedonia &\hspace*{.5em}748&
  -0.063 & (0.260) & -0.068 & (0.254) & -0.007 & (0.148) \\
  Norway   &1079&
  -0.190 & (0.130) & -0.196 & (0.126) & -0.174 & (0.126) \\
  Poland   &1200&
  0.101 & (0.131) & 0.084 & (0.125) & 0.098 & (0.131) \\
  Portugal &1092&
  0.501 & (0.311) & 0.478 & (0.302) & 0.239 & (0.147) \\
  Romania  &1098&
  -0.865 & (0.277) & -0.849 & (0.271) & -0.462 & (0.134) \\
  Russia   &1582&
  0.054 & (0.135) & 0.048 & (0.128) & 0.039 & (0.107) \\
  Serbia   &1168&
  -0.101 & (0.238) & -0.101 & (0.230) & -0.078 & (0.156) \\
  Slovakia &1280&
  -0.117 & (0.180) & -0.105 & (0.173) & -0.104 & (0.145) \\
  Slovenia &\hspace*{.5em}954&
  0.262 & (0.320) & 0.253 & (0.312) & 0.103 & (0.106) \\
  Spain    &1002&
  0.147 & (0.285) & 0.134 & (0.277) & 0.085 & (0.159) \\
  Sweden   &1053&
  -0.681 & (0.156) & -0.689 & (0.153) & -0.495 & (0.116) \\
Switzerland&2841&
  0.084 & (0.078) & 0.077 & (0.076) & 0.086 & (0.077) \\
  Ukraine  &1455&
  0.266 & (0.220) & 0.262 & (0.213) & 0.169 & (0.138) \\
   \hline
\end{tabular}
}}
\label{country_mods_int}
\end{table}

\end{document}